\def\dir{./}
\documentclass[useAMS,usenatbib]{\dir mn2e}
\usepackage[fleqn]{amsmath}
\usepackage{amssymb}
\usepackage[super]{nth}
\usepackage{natbib}
\usepackage{graphicx}
\usepackage{float}
\usepackage{mathrsfs}
\usepackage{mathtools}
\usepackage{multirow}
\usepackage[usenames,dvipsnames]{color}
\usepackage{\dir aas_macros}
\usepackage{comment}
\usepackage{subfigure}
\usepackage{hyperref}
\usepackage{mathrsfs}
\usepackage{epstopdf, dcolumn}
\DeclareGraphicsExtensions{.pdf}
\usepackage{wasysym}
\usepackage{dashrule}
\usepackage{xcolor}
\usepackage{arydshln}
\usepackage{threeparttable}

\newlength\replength
\newcommand\repfrac{.33}

\newcommand\rulewidth{.6pt}
\newcommand\tdashfill[1][\repfrac]{\cleaders\hbox to \replength{%
  \smash{\rule[\arraystretch\ht\strutbox]{\repfrac\replength}{\rulewidth}}}\hfill}

\newcommand\tdotfill[1][\repfrac]{\cleaders\hbox to \replength{%
  \smash{\raisebox{\arraystretch\dimexpr\ht\strutbox-.1ex\relax}{.}}}\hfill}

\newcommand{\appropto}{\mathrel{\vcentre{
			\offinterlineskip\halign{\hfil$##$\cr
				\propto\cr\noalign{\kern2pt}\sim\cr\noalign{\kern-2pt}}}}}

\newcommand{\Rom}[1]{\uppercase\expandafter{\romannumeral #1}}
\newcommand{\rom}[1]{\lowercase\expandafter{\romannumeral #1}}
\newcommand{\HII}{\textrm{H}\,\textsc{ii}}

\definecolor{colorM16}{RGB}{228,26,28}
\definecolor{colorfiducial}{RGB}{55,126,184}
\definecolor{colorstrongRadio}{RGB}{77,175,74}
\definecolor{colorstrongQuasar}{RGB}{152,78,163}
\definecolor{colorQuasarReion}{RGB}{255,127,0}
\definecolor{colorQuasarReionnodust}{RGB}{166,86,40}
\definecolor{colorMaxStellar}{RGB}{166,86,40}

\defcitealias{Mutch2016}{M16}
\defcitealias{Becker2013}{BB13}
\defcitealias{Giallongo2015}{G15}
\defcitealias{PlanckCollaboration2015}{Planck15}
\defcitealias{PlanckCollaboration2016}{Planck16}

\title[DRAGONS \Rom{10}: quasars to reionization]{\parbox{1.05\textwidth}{Dark-ages~Reionization~and~Galaxy~Formation~Simulation~-~\Rom{10}.} The small contribution of quasars to reionization}

\author[Qin et al.]{Yuxiang Qin$^{1}$\thanks{E-mail: Yuxiang.L.Qin@gmail.com},
Simon J. Mutch$^1$, Gregory B. Poole$^1$, Chuanwu Liu$^1$, Paul W. Angel$^1$,
\newauthor Alan R. Duffy$^2$, Paul M. Geil$^1$, Andrei Mesinger$^3$ and J. Stuart B. Wyithe$^1$\thanks{E-mail: swyithe@unimelb.edu.au}\\
$^{1}$School of Physics, University of Melbourne, Parkville, VIC 3010, Australia\\
$^{2}$Centre for Astrophysics and Supercomputing, Swinburne University of Technology, PO Box 218, Hawthorn VIC 3122, Australia\\
$^{3}$Scuola Normale Superiore, Piazza dei Cavalieri 7, I-56126 Pisa, Italy}

\begin{document}
\date{Accepted 2017 July 25. Received 2017 July 23; in original form 2017 March 13}
\pagerange{\pageref{firstpage}--\pageref{lastpage}} \pubyear{2017}
\maketitle
\label{firstpage}

\begin{abstract}
Motivated by recent measurements of the number density of faint AGN at high redshift, we investigate the contribution of quasars to reionization by tracking the growth of central supermassive black holes in an update of the {\sc Meraxes} semi-analytic model. The model is calibrated against the observed stellar mass function at $z\sim0.6-7$, the black hole mass function at $z\lesssim0.5$, the global ionizing emissivity at $z\sim2-5$ and the Thomson scattering optical depth. The model reproduces a Magorrian relation in agreement with observations at $z<0.5$ and predicts a decreasing black hole mass towards higher redshifts at fixed total stellar mass. With the implementation of an opening angle of $80$ deg for quasar radiation, corresponding to an observable fraction of ${\sim}23.4$ per cent due to obscuration by dust, the model is able to reproduce the observed quasar luminosity function at $z\sim0.6-6$. The stellar light from galaxies hosting faint AGN contributes a significant or dominant fraction of the UV flux. At high redshift, the model is consistent with the bright end quasar luminosity function and suggests that the recent faint $z\sim4$ AGN sample compiled by \citet{Giallongo2015} includes a significant fraction of stellar light. Direct application of this luminosity function to the calculation of AGN ionizing emissivity consequently overestimates the number of ionizing photons produced by quasars by a factor of 3 at $z\sim6$. We conclude that quasars are unlikely to make a significant contribution to reionization.
\end{abstract}


\section{Introduction}
The epoch of reionization (EoR) is the phase of the Universe when neutral hydrogen in the intergalactic medium (IGM) was reionized. Star-forming galaxies at high redshift are believed to be one of the dominant sources of ionizing UV photons provided one assumes a high average escape fraction of Lyman continuum radiation ($f_\mathrm{esc,*}{\gtrsim}10$ per cent) and extends the UV luminosity function to faint dwarf galaxies \citep{Kuhlen2012,duffy2014low,Feng2016,Mesinger2016}. However, the value of $f_\mathrm{esc, *}$ is very uncertain. Observations of star-forming galaxies at low redshift usually indicate a much lower escape fraction. For example, by measuring the ratio of Ly$\alpha$ to H$\beta$ line emission, \citet{Ciardullo2014} derived an escape fraction of 4.4 per cent while \citet{Matthee2016} measured a median escape fraction of 1.6 per cent using stacking of H$\alpha$-selected galaxies. Moreover, some theoretical works also suggest a low $f_\mathrm{esc,*}$ \citep{Gnedin2007,Hassan2016,Sun2015}. 

In order to reconcile the difference between low-redshift observations and the photon budget at high redshift, some propose a rapid increase of $f_\mathrm{esc,*}$ with redshift \citep{Haardt2012,Khaire2016,Price2016,Sharma2016} and with decreasing mass \citep{Paardekooper2013,Kimm2014,Wise2014}. This is supported by identifying local analogues of high-redshift galaxies and extrapolating the observed $f_\mathrm{esc,*}$ using indicators such as the [O{\sc iii}]/[O{\sc ii}] ratio to high redshift (\citealt{Faisst2016} and references therein). On the other hand, additional contributors to reionization may also be present. For example, \citet{Ma2016} included a binary population into a set of radiative transfer cosmological simulations and found that they produced significantly more ionizing photons among the old stellar population compared to a model without binaries, reducing the requirement of high escape fraction. In addition, these photons produced at later times can escape from galaxies more easily since the local feedback efficiently clears out nearby gas, leading to a lower required escape fraction on average. With a high escape fraction ($f_\mathrm{esc,q}\sim1$, \citealt{Barkana2000}), quasars (AGN) are potential contributors to reionization \citep{Volonteri2009,Fontanot2014,Madau2015,Mitra2015}, despite their relatively low number. Recently, \citet{Giallongo2015} identified faint AGN candidates in the CANDELS GOODS-South field and suggested that there is a high number density of faint AGN at $z=4-6$. These faint quasars provide a new source of reionization \citep{Madau2015}. However, it is still debated whether there are enough luminous quasars at high redshift to make a significant contribution \citep{Bouwens2015,Weigel2015,Manti2017,Parsa2017} and whether the escape fraction of high-redshift low-luminosity AGN is of order of unity \citep{Cristiani2016,Micheva2016}.

To explore the consequences for galaxy formation and reionization from theses faint quasars, this paper describes the addition of a population of evolving black holes to the {\sc Meraxes}\footnote{\href{http://dragons.ph.unimelb.edu.au/}{http://dragons.ph.unimelb.edu.au}} semi-analytic model of galaxy formation and reionization \citep{Mutch2016}. This new model enables a detailed exploration of the relative role of quasars during the EoR. The paper is organized as follows. We begin with a description of the semi-analytic model in Section \ref{sec:model}, in which the black hole growth model is introduced in detail. We present the black hole properties in Section \ref{sec:properties} and explore reionization from quasars in Sections \ref{sec:reionization} and \ref{sec:discussion}. Conclusions are given in Section \ref{sec:conclusions}. In this work, we adopt cosmological parameters from the Planck 2015 results ($\Omega_{\mathrm{m}}, \Omega_{\mathrm{b}}, \Omega_{\mathrm{\Lambda}}, h, \sigma_8, n_\mathrm{s} $ = 0.308, 0.0484, 0.692, 0.678, 0.815, 0.968; \citealt{PlanckCollaboration2015}). 

\section{Modelling Black hole growth}\label{sec:model}

Built on halo\footnote{Note that in this work, haloes are defined as the substructures of {\sc fof} groups. Central haloes (the most massive halo in a {\sc fof} group) and satellites co-evolve in {\sc Meraxes} and both contribute to reionization.} merger trees constructed from the \textit{Tiamat} collisionless \textit{N}-body simulation \citep{Poole2015}, the {\sc Meraxes} semi-analytic model \citep{Mutch2016} was specifically designed to study galaxy formation and reionization at high redshift. The model computes galaxy properties according to different astrophysical processes including gas infall, cooling, star formation, supernova feedback, metal enrichment, mergers and reionization. In order to properly track the evolution of galaxies during reionization, \textit{Tiamat} provide 100 snapshots between $z=35$ and 5 with a time interval of ${\sim}11.1$ Myr and 64 additional snapshots between $z=5$ and 2 separated equally in units of Hubble time. The mass resolution of \textit{Tiamat} is ${\sim} 2.64\times10^6h^{-1}\mathrm{M}_\odot$ and the box size is 67.8$h^{-1}$Mpc. Additionally, in order to obtain information of more massive objects and lower redshifts, we also take advantages of the dark matter halo merger trees generated from the \textit{Tiamat-125-HR} simulation (Poople et al. in prep). The \textit{Tiamat-125-HR} simulation shares identical cosmology with \textit{Tiamat} but has a lower mass resolution of $0.12\times10^9h^{-1}\mathrm{M}_\odot$ in a bigger simulation volume, with side lengths equal to $125h^{-1}\mathrm{Mpc}$. The \textit{Tiamat-125-HR} trees are constructed down to $z=0.56$ with the same snapshot separation strategy as the \textit{Tiamat} trees. 

In the following subsections, we briefly describe the galaxy formation in {\sc Meraxes} and introduce the new implementation of black hole growth and feedback in detail. More details about the implemented galaxy formation physics can be found in \citet{Mutch2016}.

\subsection{Galaxy evolution}\label{sec:galaxy_evolution}

In the {\sc Meraxes} semi-analytic model, cooling and star formation are assumed to be negligible in haloes below the atomic cooling mass threshold, ${\sim}10^8 \mathrm{M}_\odot$. Thus, once a halo grows larger than the atomic cooling limit it is designated as a galaxy, which has three baryonic components: gas, stars and a central black hole. During one time step, $\Delta t$, additional gas falls into the hot gas component of a galaxy from the IGM when the mass fraction of baryons in the halo is lower than the cosmic mean value, $f_\mathrm{b}{=}\Omega_\mathrm{b}/\Omega_\mathrm{m}$:
\begin{equation}\label{eq:minfall}
\Delta m_{\mathrm{hot}}{=}{\max}\left[0, \chi_\mathrm{r}f_\mathrm{b}M_\mathrm{vir}{-}\left(m_{\mathrm{\star}}{+}m_{\mathrm{cold}}{+}m_{\mathrm{hot}}{+}m_{\mathrm{eject}}\right)\right],
\end{equation}
where $m_{\mathrm{\star}},\ m_{\mathrm{cold}},\ m_{\mathrm{hot}}$ and $m_{\mathrm{eject}}$ are the masses of the stellar component, cold gas, hot gas and ejected gas, respectively, and $M_\mathrm{vir}$ is the virial mass of the host halo in which the galaxy forms.\footnote{A central halo consist of the majority of its {\sc fof} particles and dominates the gravitational potential of the entire system. In the code, the infall gas is added into the central halo based on the baryon fraction of its {\sc fof} group while the satellites do not get any fresh gas (see more details in \citealt{Mutch2016}).} $\chi_\mathrm{r}$ is a baryon fraction modifier to take account of reionization feedback and will be introduced later in Section \ref{sec:reionization model}.

Some of the hot gas, $m_\mathrm{cool}$, cools and collapses on to the cold disc. Assuming the cooling process is in quasi-static thermal equilibrium, one can calculate the cooling time by
\begin{equation}\label{eq:t_cool}
t_{\mathrm{cool}}\left(r\right) = \dfrac{3\bar{\mu}m_\mathrm{p}kT_\mathrm{hot}}{2\rho_\mathrm{hot}\left(r\right)\Lambda\left(T_\mathrm{hot},Z_\mathrm{hot}\right)},
\end{equation}
where $\mu=0.59$, $m_\mathrm{p}$ and $k$ are the mean molecular weight for fully ionized gas, the mass of a proton and the Boltzmann constant, respectively. $T_\mathrm{hot}$ and $\rho_\mathrm{hot}$ are the temperature and density of the hot gas. $\Lambda$ is the cooling function \citep{Sutherland1993} determined by the temperature and metallicity, $Z_\mathrm{hot}$. Assuming the hot gas shares the same temperature as the host halo ($T_\mathrm{hot}=T_\mathrm{vir}$) due to shock heating, and follows a singular isothermal sphere density profile, equation (\ref{eq:t_cool}) becomes
\begin{equation}\label{eq:t_cool_2}
t_{\mathrm{cool}}\left(r\right) = \dfrac{6\pi\bar{\mu}m_\mathrm{p}kT_\mathrm{vir} R_\mathrm{vir}r^2}{m_\mathrm{hot}\Lambda\left(T_\mathrm{vir},Z_\mathrm{hot}\right)},
\end{equation}
where $R_\mathrm{vir}$ is the virial radius. Following \citet{croton2006many}, we calculate the cooling radius, $r_\mathrm{cool}$, at which the cooling time is equal to the halo dynamical time, through
\begin{equation}
r_\mathrm{cool} = \sqrt{\dfrac{m_\mathrm{hot}\Lambda\left(T_\mathrm{vir},Z_\mathrm{hot}\right)}{6\pi \bar{\mu}m_\mathrm{p}kT_\mathrm{vir}V_\mathrm{vir}}},
\end{equation}
where $V_\mathrm{vir}$ is the virial velocity. Cooling is sufficient within $r_\mathrm{cool}$ and we estimate the cooling mass by
\begin{equation}\label{eq:mcool}
m_{\mathrm{cool}} \equiv
m_{\mathrm{hot}}\times\min\left[1, \mathrm{min}\left(1,\dfrac{r_{\mathrm{cool}}}{R_{\mathrm{vir}}}\right)\times\dfrac{\Delta t}{t_{\mathrm{cool}}}\right],
\end{equation}
which is removed from the hot gas reservoir and redistributed into the cold gas disc, $\Delta m_{\mathrm{hot}}=-\Delta m_{\mathrm{cold}}{=}-m_{\mathrm{cool}}$. From equation (\ref{eq:mcool}), we see that depending on the ratio of $r_\mathrm{cool}$ to $R_\mathrm{vir}$, cooling is separated into two regimes: \textit{static hot halo} ($r_\mathrm{cool}>R_\mathrm{vir}$) and \textit{rapid cooling} ($r_\mathrm{cool}<R_\mathrm{vir}$, see more in \citealt{croton2006many}). In the case of a static hot halo, cooling is in thermal equilibrium and the cooling rate is determined by the continuity equation. On the other hand, if the cooling time is shorter than the dynamical time this leads to a rapid cooling process, with a free-fall of all the hot gas on to the cold gas disc. We note that $m_{\mathrm{cool}}$ is set to be zero when the mass of the host halo drops below the atomic cooling due to stripping.

When the galaxy collects enough cold gas, $m_\mathrm{cold}{>}m_{\mathrm{crit}}$, based on \citet{kennicutt1998global} and \citet{kauffmann1996disc}, it forms new stars, $\Delta m_{\star}$, through a burst
\begin{equation}\label{eq:sfr}  
\Delta m_{\star}{=} \min\left[\dfrac{\alpha_{\mathrm{sf}}\mathrm{max}\left(0,m_{\mathrm{cold}}{-}m_{\mathrm{crit}}\right)\Delta t}{t_{\mathrm{dyn,disc}}},m_\mathrm{cold}\right],
\end{equation}
where $t_{\mathrm{dyn,disc}}= 1.5\sqrt{2}\lambda R_\mathrm{vir}/V_\mathrm{vir}$ is the dynamical time of the cold gas disc ($\lambda$ is the spin parameter of the host halo) and $\alpha_{\mathrm{sf}}$ is a free parameter corresponding to the star formation efficiency. This mass is removed from the cold gas reservoir, $\Delta m_{\mathrm{cold}}=-\Delta m_{\star}$. 

The new stellar mass is assumed to form following a \citet{Salpeter1955ApJ...121..161S} initial mass function (IMF). Ultimately some of these stars, $\eta_\mathrm{SN{\Rom{2}}}\Delta m_{\star}$, recycle their mass back to the interstellar medium (ISM) through type-\Rom{2} supernovae explosion.\footnote{{\sc Meraxes} assumes the Salpeter IMF with a mass range of $0.1-120\mathrm{M}_\odot$. We use the lifetime-mass relation of stars \citep{Portinari1997} to calculate the fraction ($\eta_\mathrm{SN{\Rom{2}}}$) of stellar mass in a single stellar population that have reached the supernova stage after a certain period of time. For example, assuming stars more massive than $8\mathrm{M}_\odot$ will reach the type-\Rom{2} supernova stage after ${\sim}40$ Myr, $\eta_\mathrm{SN{\Rom{2}}} = 7.432\times10^{-3}$.} The energy produced by these supernovae heats the ISM and converts some of the cold gas to the hot phase or, if there is sufficient energy, even ejects a fraction of hot gas from the galaxy. Assuming that the efficiency for supernovae energy coupling with the ISM scales with mass and is in proportion to a free parameter \citep{guo2011dwarf}, $\alpha_{\mathrm{energy}}$, then the total energy released by supernovae that couples to the ISM is
\begin{equation}\label{eq:esn}
\begin{split}
e_{\mathrm{total}} {=} &\mathrm{min}\left\{ \alpha_{\mathrm{energy}}\left[0.5{+}\left(\dfrac{V_\mathrm{max}}{70 \mathrm{km\ s^{{-}1}}}\right)^{-2.0}\right],1\right\} \\
&\times10^{51}\mathrm{erg}\times\eta_\mathrm{SN{\Rom{2}}}\Delta m_{\star},
\end{split}
\end{equation}
where $V_\mathrm{max}$ is the maximum circular velocity of the host halo. Since a massive star (${\gtrsim}8\mathrm{M}_\odot$) takes ${\sim}40$ million years, or 4 snapshots, before reaching its type-\Rom{2} supernova stage, {\sc Meraxes} accounts for supernovae not only from the current snapshot, $j$, but also from the stars formed in the previous 4 snapshots. Therefore, the total energy released during one snapshot is 
\begin{equation}\label{eq:Esn}
    E_{\mathrm{total}} = \mathrm{\sum}^{i=j}_{i=j-4} e_{\mathrm{total},i}\left(\Delta m_{\star,i}, V_\mathrm{max,i}, \eta_\mathrm{SNII,i}\right).
\end{equation}
Since the hot gas shares the same virial temperature as the host halo, assuming the mass loading factor for reheating cold gas is a free parameter, $\alpha_{\mathrm{mass}}$, the energy utilized in gas heating can be calculated by
\begin{equation}\label{eq:E_reheat}
\begin{split}
E_{\mathrm{reheat}} = \dfrac{1}{2}\alpha_{\mathrm{mass}}\Delta m_{\star}V_\mathrm{vir}^{2},
\end{split}
\end{equation}
where $V_\mathrm{vir}$ is the virial velocity. Depending on the available energy, $E_{\mathrm{total}}$ and the required energy for re-heating, $E_{\mathrm{reheat}}$, the reheated mass, $m_{\mathrm{reheat}}$, is
\begin{equation}\label{eq:Delta m_reheat}
m_{\mathrm{reheat}} {=} \mathrm{min}\left[m_\mathrm{cold}, \dfrac{\mathrm{min}\left(E_{\mathrm{total}}, E_{\mathrm{reheat}}\right)}{0.5V_\mathrm{vir}^{2}}\right].
\end{equation}
This mass is removed from the cold gas reservoir and redistributed in the hot gas component, $\Delta m_{\mathrm{cold}}= -\Delta m_{\mathrm{hot}}= -m_{\mathrm{reheat}}$. If there is still some energy left after reheating, the supernovae feedback will further remove hot gas from the galaxy, adding it to the ejected component
\begin{equation}\label{eq:m_eject}
\Delta m_{\mathrm{eject}}=
\mathrm{min}\left[m_\mathrm{hot}, \dfrac{\mathrm{max}\left(0, E_{\mathrm{total}}- 0.5m_\mathrm{reheat}V_\mathrm{vir}^{2}\right)}{0.5V_\mathrm{vir}^{2}}\right],
\end{equation}
which is removed from the hot gas reservoir, $\Delta m_{\mathrm{hot}}=-\Delta m_{\mathrm{eject}}$. 

In addition, the metals produced by supernovae enrich the environment, which then enhances the cooling rate (see equation \ref{eq:t_cool}). Moreover, mergers drive strong turbulence and hence increase the possibility of star formation. Major mergers generally introduce more energetic bursts than minor mergers since they induce strong inflows and easily trigger bar-like instabilities in the cloud \citep{Somerville2001}. Therefore, following mergers, {\sc Meraxes} also includes a starburst mechanism. Reionization feedback will be introduced in Section \ref{sec:reionization model}. There are more details of the semi-analytic model in \citet{Mutch2016}. This work extends the current model with a detailed black hole growth prescription based on \citet{Croton2016} and is described in the following subsection.

\subsection{Black hole growth}

In the updated model, every newly formed galaxy is seeded with a central black hole\footnote{Dual or multiple AGN are not considered.} of mass $1000\ h^{-1} \mathrm{M}_\odot$. The two gas reservoirs in the galaxy (i.e. hot and cold) lead to two different black hole growth scenarios, termed radio and quasar modes \citep{Croton2016}. In the normal quiescent state, black holes only accrete mass from the hot gas reservoir, resulting in radio emission in the centre of galaxy. However, mergers trigger rapid accretion on to the black hole from the cold gas disc, causing them to radiate as quasars. In this work, we do not distinguish AGN with different types and refer to them all as quasars unless specified otherwise.

\subsubsection{Accretion of hot gas}\label{sec:radio mode}

Whenever there is a static hot gas reservoir, $m_\mathrm{hot}$, around the galaxy, some of it will cool, $m_\mathrm{cool}$, and form a cold gas disc, $m_\mathrm{cold}$ (see the previous section), while some will be directly accreted by the central black hole. We adopt the Bondi-Hoyle accretion model proposed in \citet{Croton2016} to describe this hot gas accretion. The Bondi-Hoyle accretion rate is given by 
\begin{equation}\label{eq:m_bondi}
\dot{m}_\mathrm{Bondi} = \dfrac{2.5\pi G^2 m_\mathrm{bh}^2 \rho_\mathrm{hot}}{c_\mathrm{s}^3},
\end{equation}
where $G, m_\mathrm{bh}, c_\mathrm{s}$ and $\rho_\mathrm{hot}$ are the gravitational constant, the black hole mass, the speed of sound and the density of the hot gas reservoir, respectively. We define the Bondi radius as $r_\mathrm{Bondi} = \dfrac{2Gm_\mathrm{bh}}{c_\mathrm{s}^2}$, and by equating the sound crossing travel time from the centre to the Bondi radius with the local cooling time (see equation \ref{eq:t_cool}), we obtain
\begin{equation}\label{eq:rho}
\dfrac{\rho_\mathrm{hot}}{{c_\mathrm{s}^3}} = \dfrac{3\mu m_\mathrm{p} kT}{4G\Lambda m_\mathrm{bh}}.
\end{equation}
Then, assuming the accretion rate does not change during one time step,\footnote{It is generally true for the radio mode. However, the quasar mode introduced in Section \ref{sec:quasar mode} can induce significant increases in black hole mass during a single time step, leading to a non-negligible change of the accretion rate.} $\Delta t$, the accretion mass is
\begin{equation}\label{eq:bh,hot}
\Delta {m}_\mathrm{bh,hot} = \min\left(m_\mathrm{hot}, m_\mathrm{Edd}, k_\mathrm{h} \dot{m}_\mathrm{Bondi} \Delta t\right),
\end{equation}
where $k_\mathrm{h}$ is a free parameter, used to adjust the efficiency of black hole growth in the radio mode since black holes may not be accreting at the full Bondi-Hoyle accretion rate. $m_\mathrm{Edd}$ is the Eddington limit, which will be introduced later. This mass is removed from the hot gas reservoir, $\Delta m_\mathrm{hot} = - \Delta {m}_\mathrm{bh,hot}$. Assuming a fraction, $\eta$, of the accreted mass is radiated, the black hole only grows by
\begin{equation}\label{eq:eta}
\Delta {m}_\mathrm{bh} = (1-\eta)\Delta {m}_\mathrm{bh,hot}.
\end{equation}
Moreover, the radiation acting outward is limited by the Eddington luminosity
\begin{equation}\label{eq:eddington_limit}
\eta \dot{m}_\mathrm{bh}(t) c^2 = \epsilon\dfrac{4\pi G m_\mathrm{bh}(t) m_\mathrm{p} c}{\sigma_\mathrm{T}},
\end{equation}
where $\epsilon$ is the Eddington ratio ($\epsilon{=}1$ in this work), $m_\mathrm{bh}(t)$ and $\dot{m}_\mathrm{bh}(t)$ are the black hole mass and growth rate at time $t$, respectively, $\sigma_\mathrm{T}$ is the Thomson cross-section area and $c$ is the speed of light. By integrating equation \ref{eq:eddington_limit} through one time step, this provides the second limitation in equation (\ref{eq:bh,hot})
\begin{equation}\label{eq:medd}
m_\mathrm{Edd} = m_\mathrm{bh}\left[\exp\left(\dfrac{\epsilon\Delta t}{\eta t_\mathrm{Edd}}\right)-1\right],
\end{equation}
where $t_\mathrm{Edd}\equiv \dfrac{\sigma_\mathrm{T}c}{4\pi G m_\mathrm{p}} \approx450 \mathrm{Myr}$ is the Eddington accretion time-scale and $m_\mathrm{bh}$ is the black hole mass at the beginning of this time step.

Assuming adiabatic heating and that a fraction of the radiated energy, $\kappa_\mathrm{r}$, is coupled to the surrounding gas and therefore can contribute to feedback, then the heated mass due to black hole feedback can be calculated through
\begin{equation}\label{eq:reheat}
m_\mathrm{heat} =  \dfrac{\kappa_\mathrm{r} \eta \Delta{m}_\mathrm{bh,hot} c^2}{0.5V^2_\mathrm{vir}},
\end{equation}
which is subtracted from the cooling flow, $\Delta m_\mathrm{cool} {=} {-}m_\mathrm{heat}$. However, if the heating from black hole feedback is too strong, $m_\mathrm{heat}{>}m_\mathrm{cool}$ (see equation \ref{eq:mcool}), it will significantly suppress the cooling flow, which will consequently restrain the black hole accretion of hot gas. This suppression is referred to as radio mode feedback. In this case, following\footnote{We do not limit the heating radius (see \citealt{Croton2016}) to only move outwards.} \citet{Croton2016}, $\Delta m_\mathrm{bh,hot}$ is rescaled to be the amount of mass within the cooling radius, $\Delta m_\mathrm{bh,hot} = \Delta m_\mathrm{bh,hot} \times \dfrac{m_\mathrm{cool}}{m_\mathrm{heat}}$ and the heated mass consequently shrinks to be the cooling mass, $m_\mathrm{heat}{=}m_\mathrm{cool}$, resulting in a complete quenching of cooling, $\Delta m_\mathrm{cool} = -m_\mathrm{heat} \rightarrow m_\mathrm{cool} = 0$.  Radio mode feedback limits gas condensation in cluster cooling flows and regulates star formation in massive galaxies \citep{croton2006many}. We note that this does not have a significant impact on the results from \textit{Tiamat} at $z\ge2$ due to the limited number of massive objects in the box. However, black hole feedback is important in more massive objects at lower redshifts, which can be observed using the \textit{Tiamat-125-HR} halo merger trees. 

We note that at high redshift, accreted hot gas does not contribute significantly to black hole growth, with the accreted mass during the radio mode typically being only ${\sim}0.1$ per cent of the mass from the quasar mode (see Appendix \ref{app:Ngamma}) that is introduced in the next section. Additionally, because the radio mode accretion rate is approximately 3 orders of magnitude smaller than the Eddington accretion rate, we ignore the energy due to the radio mode for the calculation of the quasar luminosity. 

\subsubsection{Accretion of cold gas}\label{sec:quasar mode}
Many binary AGN have been detected in merging galaxies \citep{Shields2012,Comerford2013,Comerford2014,Comerford2015,Muller-Sanchez2015}. This suggests that galaxy mergers might trigger AGN activity, which is also supported by hydrodynamic simulations of galaxy mergers \citep{Capelo2015,Volonteri2015b,Volonteri2015a,Steinborn2016}. Following \citet{Croton2016}, when a merger occurs between two galaxies, the central black holes coalesce and their masses combine.\footnote{Ignoring the loss due to gravitational wave emission.} As mergers drive strong gas inflows towards the central region \citep{Capelo2015}, cold gas is funnelled to the central black hole of the resulting merged galaxy, significantly increasing its mass. The amount of accretion can be estimated by \citep{Bonoli2009,Croton2016}
\begin{equation}\label{eq:bh,max}
\Delta {m}_\mathrm{bh,max} = \min\left[m_\mathrm{cold}, \dfrac{k_\mathrm{c}\gamma m_\mathrm{cold}}{\left(1{+}\dfrac{280\mathrm{km\ s^{-1}}}{V_\mathrm{vir}}\right)^2}\right]+\Delta {m}_\mathrm{bh,max}^{\prime},
\end{equation}
where $m_\mathrm{cold}$ is the amount of mass available in the cold gas disc, $k_\mathrm{c}$ is a free parameter used to modulate the strength of black hole accretion and $\gamma{\leq}1$ is the mass ratio between the two merging galaxies, respectively. The term $\Delta {m}_\mathrm{bh,max}^{\prime}$ corresponds to the accretion mass left from the quasar mode in the previous snapshot, which will be introduced later. 

\begin{table*}
	\caption{A list of relevant parameters in the model with description and adopted value. The redshift varying $f_\mathrm{esc,*}$ model \citep[][hereafter \citetalias{Mutch2016}]{Mutch2016} is shown for comparison. The M16BH model includes black hole feedback compared to the \citetalias{Mutch2016} model but without reionization from quasars. We refer the interested reader to \citet{Mutch2016}.}
	\begin{threeparttable}
		\label{tab:parameters}
		\begin{tabular}{ccclccc}
			\hline \hline
			Parameter & Section & Equation & Description & Fiducial &M16BH&\citetalias{Mutch2016}\\
			\hline
			$\alpha_\mathrm{sf}$ &\ref{sec:galaxy_evolution}& \ref{eq:sfr}& Star formation efficiency& 0.08&0.03& 0.03\\
			$\alpha_\mathrm{energy}$&\ref{sec:galaxy_evolution} & \ref{eq:esn} & Energy coupling efficiency normalization& 1.0&0.5& 0.5\\
			$\alpha_\mathrm{mass}$&\ref{sec:galaxy_evolution} & \ref{eq:E_reheat} & Mass loading normalization& 15.0&9.0& 9.0\\
			\hdashline
			$\eta$ & \ref{sec:radio mode} & \ref{eq:eta} & Black hole efficiency of converting mass to energy & 0.06& 0.06& -\\
			$\epsilon$ & \ref{sec:radio mode} & \ref{eq:eddington_limit} & Eddington ratio &  1.0 & 1.0& -\\
			$k_\mathrm{h}$ & \ref{sec:radio mode} & \ref{eq:bh,hot} & Black hole growth efficiency for the radio mode & 0.3 &0.1& - \\ 
			$\kappa_\mathrm{r}$ & \ref{sec:quasar mode} & \ref{eq:eddington_limit} & Black hole feedback efficiency for the radio mode & 1.0&1.0&- \\ 
			$k_\mathrm{c}$ & \ref{sec:quasar mode} & \ref{eq:bh,max} & Black hole growth efficiency for the quasar mode &0.05&0.05&- \\ 
			$\kappa_\mathrm{q}$ & \ref{sec:quasar mode} & - & Black hole feedback efficiency for the quasar mode & 0.0005&0.0005&- \\ 
			$N_{\gamma,*}$& \ref{sec:reionization model} & \ref{eq:reionization_condition} & Mean number of ionizing photons produced per stellar baryon& 4,000 &4,000&4,000\\		
			$f_\mathrm{esc,q}$ & \ref{sec:reionization model} & \ref{eq:reionization_condition} & Ionizing photon escape fraction for quasars & 1.0\tnote{a}&0.0&-\\
			$f_\mathrm{esc,*}$ & \ref{sec:reionization model} & \ref{eq:reionization_condition} & Ionizing photon escape fraction for the stellar component & $f_\mathrm{esc,*,z}$\tnote{b,c} & $f_\mathrm{esc,*,M16}$\tnote{d}&$f_\mathrm{esc,*,M16}$\\
			\hline 
			\hline
		\end{tabular}
		\begin{tablenotes}
			\item[a] $f_\mathrm{esc,q}=0$ is adopted for the StellarReion model in which stars are the only reionization source.
			\item[b] $f_\mathrm{esc,*}=0$ is adopted for the QuasarReion model in which quasars are the only reionization source.			
			\item[c] $f_\mathrm{esc,*,z} =\min\left[0.06\times\left(\dfrac{1+z}{6}\right)^{0.5},1.0\right]$ for the fiducial and StellarReion models.
			\item[d] $f_\mathrm{esc,*,M16} = \min\left[0.04\times\left(\dfrac{1+z}{6}\right)^{2.5},1.0\right]$.
		\end{tablenotes}
	\end{threeparttable}
\end{table*}

Unlike the radio mode, black holes grow dramatically during the quasar mode. The AGN activity lifetime (a few $10^7$ yr up to a Gyr \citealt{Fiore2012}) is much longer than the 11Myr time step at $z>5$. In some cases, this leaves the central black hole insufficient time to consume all of the newly accreted gas, $\Delta {m}_\mathrm{bh,max}$, at the Eddington limit. Therefore, the mass actually accreted by the black hole is
\begin{equation}\label{eq:bh,cold}
\Delta {m}_\mathrm{bh,cold} = \min\left(m_\mathrm{Edd}, \Delta {m}_\mathrm{bh,max} \right),
\end{equation}
which is removed from the cold gas reservoir, $\Delta m_\mathrm{cold} = -\Delta {m}_\mathrm{bh,cold}$. In our model, during the quasar mode black holes are assumed to either accrete and radiate at the Eddington rate or stay quiescent if the accretion mass is not sufficient. Therefore, depending on the total available mass brought in, $\Delta {m}_\mathrm{bh,max}$, and the Eddington limit, $m_\mathrm{Edd}$, there are two possible scenarios when the central black hole is undergoing a merger:
\begin{enumerate}
	\item $\Delta {m}_\mathrm{bh,max} < m_\mathrm{Edd}$. In this case, there is inadequate mass to feed the central black hole at the Eddington rate for the entire time step. The duration of accretion can be calculated through
	\begin{equation}
	t_\mathrm{acc} = \ln\left(\dfrac{\Delta {m}_\mathrm{bh,cold}}{m_\mathrm{bh}} +1\right)\times\dfrac{\eta t_\mathrm{Edd}}{\epsilon}.
	\end{equation}
	When the quasar is observed at a random time $t_\mathrm{obs}$, the bolometric luminosity is
	\begin{equation}\label{eq:luminosity}
	L_\mathrm{bol} \equiv \epsilon m_\mathrm{bh}|_{t=t_\mathrm{obs}} \dfrac{c^2}{t_\mathrm{Edd}} = \epsilon m_\mathrm{bh}\exp\left(\dfrac{\epsilon t_\mathrm{obs}}{\eta t_\mathrm{Edd}}\right)\times \dfrac{c^2}{t_\mathrm{Edd}}, 
	\end{equation}
	when $t_\mathrm{obs}\leqslant t_\mathrm{acc}$, otherwise $L_\mathrm{bol} = 0$.
	\item  $\Delta {m}_\mathrm{bh,max} \geqslant m_\mathrm{Edd}$. In this case, the merger event delivers sufficient cold gas into the accretion disc. In the case of $\Delta {m}_\mathrm{bh,max} > m_\mathrm{Edd}$, instead of consuming this instantaneously, causing a super-Eddington accretion event, some of the mass is accreted by the central black hole, limited by the Eddington rate, while the rest, $\Delta {m}_\mathrm{bh,max}^\prime = \Delta {m}_\mathrm{bh,max} - m_\mathrm{Edd}$, is stored in the accretion disc to be consumed in the next time step. Similarly, when this quasar is observed at $t_\mathrm{obs}$, the bolometric luminosity can be calculated using equation (\ref{eq:luminosity}).
\end{enumerate} 

It is suggested that during mergers, black holes undergo rapid accretion for a certain time period, which is followed by a long quiescent phase \citep{Hopkins2005c,Hopkins2005a,Hopkins2005d,Hopkins2005b}. The assumption that black holes are either accreting at the Eddington rate ($\epsilon=1$, see equation \ref{eq:eddington_limit}) or stay quiescent has been shown to provide a good description of black hole growth for the majority of black holes at high redshift \citep{Bonoli2009}. 

The energy injected into galactic gas during the quasar mode is given by $\kappa_\mathrm{q} \eta\Delta m_\mathrm{bh,cold} c^2$, where $\kappa_\mathrm{q}$ represents the mass coupling factor in the quasar mode. Unlike the radio mode, this energy generates a wind, which heats the gas in the cold disc into the hot reservoir. Depending on the amount of energy provided by the quasar, the wind can further unbind and eject the hot gas in a manner similar to the stellar feedback prescription presented in Section \ref{sec:galaxy_evolution}.

\subsection{Quasar luminosity}\label{sec:QL}

In order to compare the predicted black hole population in our model with observations, the intrinsic \textit{B}-band and UV 1450 $\mathrm{\AA}$ band luminosities of quasars are calculated as follows:
\begin{enumerate}
	\item We calculate the bolometric magnitude through 
	\begin{equation}\label{eq:app:Lbol}
	M_\mathrm{bol} = 4.74 - 2.5\log_{10}\left(\dfrac{L_\mathrm{bol}}{\mathrm{L_\odot}}\right).
	\end{equation}
	\item We calculate the \textit{B}-band magnitude in the Vega magnitude system using the bolometric correction proposed by \citet{Hopkins2007},
	\begin{equation}
	M_\mathrm{bol} - M_\mathrm{B} = -2.5\log_{10}k_\mathrm{B},
	\end{equation}
	where
	\begin{equation}\label{eq:k_b}
	k_\mathrm{B} \equiv \dfrac{L_\mathrm{bol}}{L_\mathrm{B}}= 6.25\left(\dfrac{L_\mathrm{bol}}{10^{10}\mathrm{L_\odot}}\right)^{-0.37}+9.00\left(\dfrac{L_\mathrm{bol}}{10^{10}\mathrm{L_\odot}}\right)^{-0.012}.
	\end{equation}
	\item We convert the \textit{B}-band magnitude from the Vega system to the AB system following \citet{Glikman2010}
	\begin{equation}\label{eq:MABB}
	M_\mathrm{AB,B} - M_\mathrm{B} = -0.09
	\end{equation}
	\item We extrapolate the \textit{B}-band magnitude of which the effective wavelength is 4344 $\mathrm{\AA}$ \citep{Blanton2007} to the 1450$\mathrm{\AA}$ magnitude, assuming the quasar continuum between 1450 and 4344$\mathrm{\AA}$ has a power-law slope of $\alpha_{q,\mathrm{optical}} = 0.44$ relative to wavelength \citep{Schirber2003}. Thus
	\begin{equation}\label{eq:app:M1450}
	M_{1450} {=} M_\mathrm{AB,B} - 2.5\log_{10}\left(\dfrac{1450 \mathrm{\AA}}{4344\mathrm{\AA}}\right)^{\alpha_{q,\mathrm{optical}}} {=} M_\mathrm{AB,B}+0.524.
	\end{equation} 
\end{enumerate}

We will further discuss the quasar luminosity function in Section \ref{sec:QLF}.

\section{Galaxy and black hole properties}\label{sec:properties}

We summarize the relevant model parameters in Table \ref{tab:parameters} compared to the original value adopted in \citet{Mutch2016}. All other parameters remain the same as \citet{Mutch2016}. In this work, we constrain\footnote{Calibration using the MCMC technique is ongoing. In this work, the calibration is performed by hand.} our model against:
\begin{enumerate}
	\item the observed evolution of the galaxy stellar mass function \citep{Pozzetti2007,Drory2009,Marchesini2009ApJ...701.1765M,Gonzalez2011,Mortlock2011MNRAS.413.2845M,Santini2012,Ilbert2013,Muzzin2013,Duncan2014,Tomczak2014,Grazian2015,Song2015,Huertas-Company2016,Stefanon2016arXiv161109354S,Davidzon2017} between $z{\sim}0.6$ and 7. We note that the observed stellar mass functions, based on a diet Salpeter IMF, a \citet{chabrier2003galactic} IMF or a \citet{Kroupa2001MNRAS.322..231K} IMF, were all converted into a standard Salpeter IMF by adding $-0.15$, 0.22 or 0.18 dex, respectively, to the logarithm of the stellar masses;
	\item the observed black hole mass function \citep{Graham2007,Shankar2009,Vika2009,Davis2014,Mutlu-Pakdil2016} and Magorrian relation \citep{Thornton2008,Jiang2011,Mathur2012,Jiang2013,Reines2013,Scott2013,Busch2014,Sanghvi2014,Yuan2014} at low redshift ($z\lesssim0.5$);
	\item the latest integrated free electron Thomson scattering optical depth measurement \citep{PlanckCollaboration2016};
	\item the predicted global ionizing emissivity from Ly$\alpha$ opacities \citep{Becker2013}.
\end{enumerate}

\subsection{Black hole properties}

\begin{figure*}
	\begin{minipage}{\textwidth}
		\centering
		\begin{tabular}{lcr}
			\begin{minipage}{0.333\textwidth}
				\subfigure{\includegraphics[width=\textwidth]{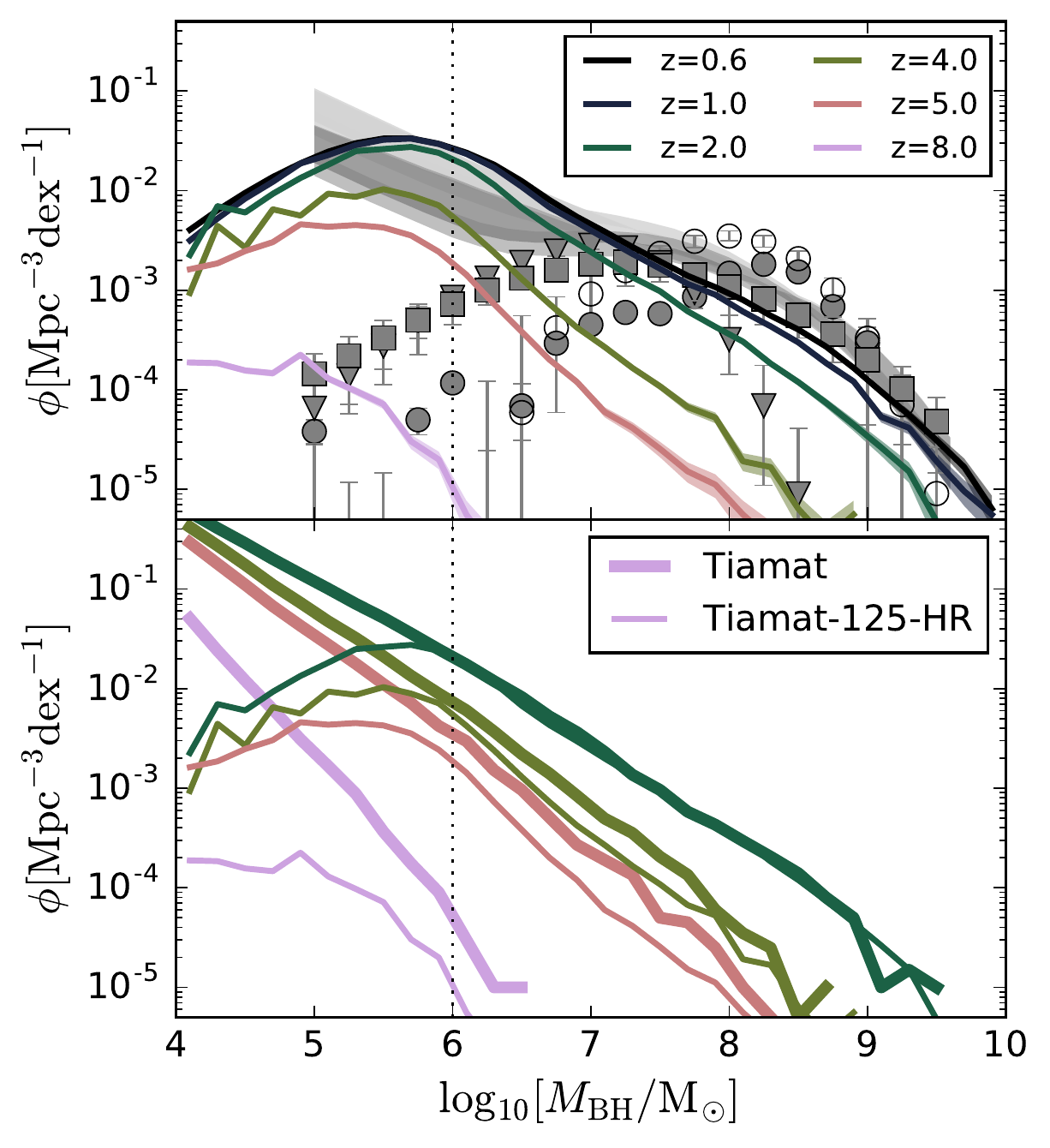}}
			\end{minipage}
			\begin{minipage}{0.333\textwidth}
				\subfigure{\includegraphics[width=\textwidth]{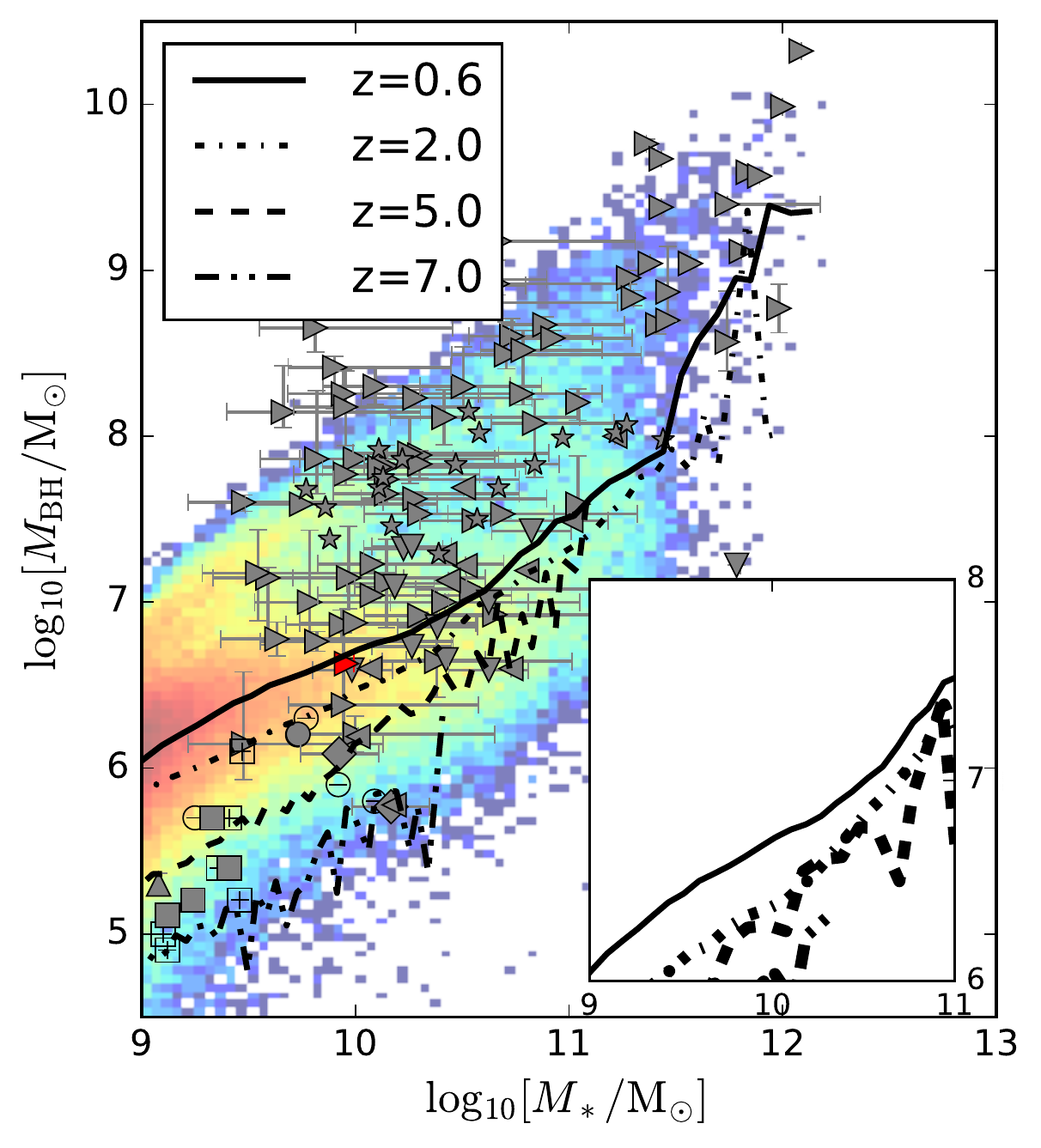}}
			\end{minipage}
			\begin{minipage}{0.333\textwidth}
				\subfigure{\includegraphics[width=\textwidth]{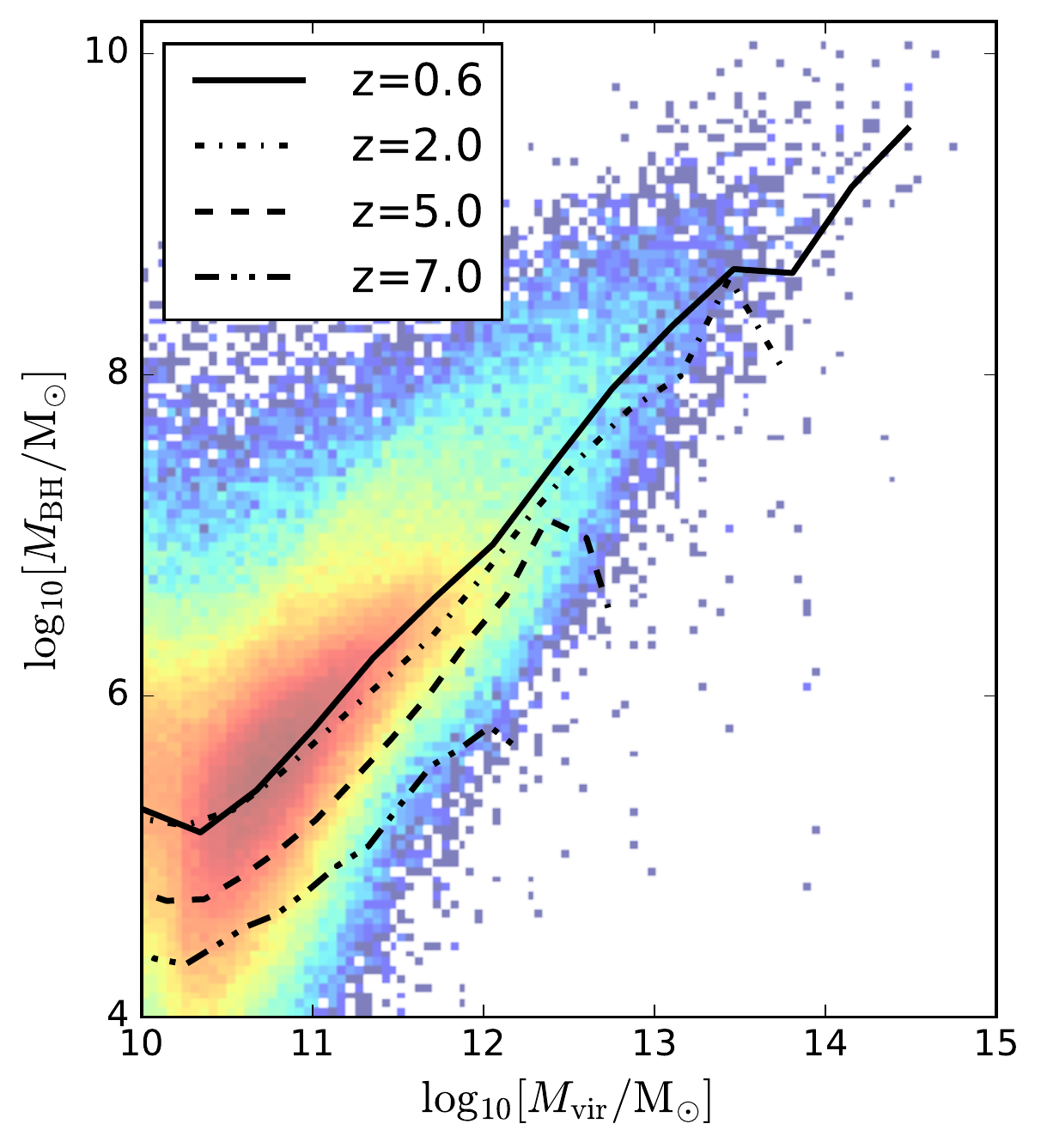}}
			\end{minipage}
		\end{tabular}\\ 
	\end{minipage}
	\caption{\label{fig:BHMFandMagorrian} \textit{Left-hand panel:} black hole mass functions from $z{\sim}8-0.6$ in the fiducial model. The results calculated using the \textit{Tiamat} and \textit{Tiamat-125-HR} trees are shown with thick and thin lines, respectively. The shaded regions represent the $1\sigma$ Poisson uncertainties for the \textit{Tiamat-125-HR} result. Observational data are shown with grey colours: {\color{gray}{\CIRCLE}}\citet{Graham2007}, \Circle\citet{Vika2009}, {\color{gray}{\DOWNarrow}}\citet{Davis2014}, {\color{gray}{$\blacksquare$}}\citep{Mutlu-Pakdil2016} and \citep{Shankar2009} at {\color{lightgray}{\hdashrule[0.6mm]{6mm}{5pt}{}}}$z\sim0$ and {\color{gray}{\hdashrule[0.6mm]{6mm}{5pt}{}}}$z\sim0.5$. The vertical dotted line represents the resolution limit in \textit{Tiamat-125-HR}. \textit{Middle panel:} correlation between black hole mass and stellar mass. The 2D histogram shows the distribution of galaxies in the fiducial model using the \textit{Tiamat-125-HR} halo merger trees at $z\sim0.6$ while the solid line represents the mean. The Magorrian relations from the model at $z=2$, 5 and 7 are shown with dash--dotted, dashed and dash-dot-doted lines, respectively. The results calculated using the \textit{Tiamat} trees are shown with thick lines in the bottom right subplot for comparison with the $z\sim0.6$ \textit{Tiamat-125-HR} Magorrian relation. Observations of the local Universe are indicated with different symbols: {\color{gray}{\UParrow}}\citet{Thornton2008}, \Circle\citet{Jiang2011}, {\color{gray}{\DOWNarrow}}\citet{Mathur2012}, {\color{gray}{\CIRCLE}}\citet{Jiang2013}, {\color{gray}{\RIGHTarrow}}\citet{Scott2013}, $\square$\citet{Reines2013} with BPT AGNs, {\color{gray}{$\blacksquare$}}\citet{Reines2013} with BPT composites, {\color{gray}{$\blacklozenge$}}\citet{Yuan2014}, {\color{gray}{\LEFTarrow}}\citet{Busch2014}, {\color{gray}{$\star$}}\citet{Sanghvi2014} at $z{\sim}0.5-1$, and {\color{red}{\RIGHTarrow}} representing the Milky Way \citet{Scott2013}. \textit{Right-hand panel:} correlation between the black hole mass and the virial mass from the fiducial model using the \textit{Tiamat-125-HR} trees. The 2D histogram shows the distribution of galaxies at $z\sim0.6$ while the solid line represents the mean. The scaling relations from the model at $z=2$, 5 and 7 are shown with dash--dotted, dashed and dash--dot--dotted lines, respectively.}
\end{figure*}

The black hole mass functions at $z\sim8.0-0.6$ are shown with different colours in the left-hand panel of Fig. \ref{fig:BHMFandMagorrian}. The results calculated using the \textit{Tiamat} and \textit{Tiamat-125-HR} trees are shown with thick and thin lines, respectively. The shaded regions represent the $1\sigma$ Poisson uncertainties.\footnote{In order to avoid crowded presentations, only the \textit{Tiamat-125-HR} uncertainty is shown when the \textit{Tiamat} and \textit{Tiamat-125-HR} results are both present in one plot.} Estimates of the local black hole mass function are shown with points and grey shaded regions. We see that the discrepancy between various observations is substantial. This is a result of inconsistent correlations between black hole mass and observable quantities, such as the S\'{e}rsic indices, bulge velocity dispersion or luminosity and galaxy geometry, and from the intrinsic scatter of these adopted scaling relations. Extrapolations of these observed scaling relations have impacts on the black hole mass function at the high-mass end, while different treatments of the spiral galaxy bulge can significantly change the low-mass end \citep{Shankar2009}. In this work, the model is therefore calibrated against the black hole mass function between $10^{7.5}\mathrm{M}_\odot$ and $10^{9}\mathrm{M}_\odot$. In the bottom left-hand panel of Fig. \ref{fig:BHMFandMagorrian}, we see that the mass function converges at lower redshifts above a black hole mass of $10^6\mathrm{M}_\odot$ (shown as the vertical dotted line). The different mass resolutions of \textit{Tiamat} and \textit{Tiamat-125-HR} result in different merger rates, especially when approaching the resolution limit. At high redshift, because the growth of black hole is dominated by the merger triggered quasar mode, the number density of small black holes is relatively lower in the \textit{Tiamat-125-HR} result (e.g. comparing the $z=8.0$ thick and thin lines). At low redshift, $z\sim0.6$, the model is in agreement with the observational estimations.

The middle panel of Fig. \ref{fig:BHMFandMagorrian} shows the relation between black hole mass and stellar mass (the Magorrian relation\footnote{A black hole mass -- galaxy property scaling relation usually accounts for the bulge property. However, the majority of systems with black holes are expected to be bulge dominated. Therefore, using the total stellar mass as a proxy does not lead to a significant bias in Fig. \ref{fig:BHMFandMagorrian}. We also note that recent studies suggest that while the $M_\mathrm{bh}-M_*$ relation is significantly biased (see Section \ref{sec:discussion}), it is the bulge velocity dispersion that connects supermassive black holes and host galaxies. However, interpreting the scaling relation is beyond the scope of this work. We leave a detailed analysis of the black hole - galaxy scaling relation to future work when bulge properties are included (e.g. \citealt{Tonini2016}).}, \citealt{Magorrian1998}). The 2D histogram indicates the distribution of galaxies\footnote{We exclude the recently identified haloes, in order to minimize the effect of black hole seeding. However, this does not have a significant impact in Fig \ref{fig:BHMFandMagorrian}.} in logarithm from the fiducial model using the \textit{Tiamat-125-HR} halo merger trees at $z\sim0.6$ while the solid line represents the mean. The Magorrian relations at $z=2$, 5 and 7 from \textit{Tiamat-125-HR} are also shown with dash--dotted, dashed and dash--dot--dotted lines, respectively. The right bottom subplot shows the $z=2$, 5 and 7 Magorrian relations of the \textit{Tiamat} result with thick lines compared with the $z\sim0.6$ \textit{Tiamat-125-HR} Magorrian relation. Observations from the local Universe are indicated with different symbols \citep{Thornton2008,Jiang2011,Mathur2012,Jiang2013,Reines2013,Scott2013,Busch2014,Sanghvi2014,Yuan2014}. We see that the model predicts a similar Magorrian relation at $z\sim0.6$ compared to the local observations and we find an increasing normalization towards lower redshifts in the mass range of $10^{10}\mathrm{M}_\odot<M_*<10^{12}\mathrm{M}_\odot$. 

The evolution of the Magorrian relation in our model is due to the black hole and stellar mass evolving with the underlying dark matter halo mass differently. In \citet{Mutch2016}, we have shown that the median relation between stellar mass and virial mass does not evolve in our model and can be described by $M_*\propto M_\mathrm{vir}^{7/5}$ in the range of $10^{8}\mathrm{M}_\odot<M_*<10^{11}\mathrm{M}_\odot$, which is supported by a simple analytic model of supernova energy conversation \citep{Wyithe2003}. On the other hand, haloes with a given virial mass host less massive black holes at earlier times in our model (see the right-hand panel of Fig. \ref{fig:BHMFandMagorrian}). These result in an increasing normalization of the Magorrian relation towards lower redshifts.

From the right-hand panel of Fig. \ref{fig:BHMFandMagorrian}, we see that the $M_\mathrm{bh}-M_\mathrm{vir}$ scaling relation does not get suppressed in massive haloes (at least to $M_\mathrm{vir}\sim10^{14}\mathrm{M}_\odot$ or $M_\mathrm{bh}\sim10^{9}\mathrm{M}_\odot$). Both black holes and stars grow from the cold gas disc.\footnote{Radio mode accretion and merger-driven starburst are relatively less important to the growth.} However, AGN feedback significantly suppresses the cooling flow in massive galaxies \citep{croton2006many}, preventing stellar mass from growing. On the other hand, black holes are able to continue accreting until there is enough energy in feedback to overcome the halo potential and unbind the gas \citep{Booth2010}. Because of these, the slope becomes steeper in the Magorrian relation at $M_\mathrm{bh}>10^{8}\mathrm{M}_\odot$.

\subsection{Galaxy properties}\label{sec:galaxy_properties}

\begin{figure*}
	\begin{minipage}{\textwidth}
		
		\begin{tabular}{lcr}
			\begin{minipage}{0.333\textwidth}
				\subfigure{\includegraphics[width=\textwidth]{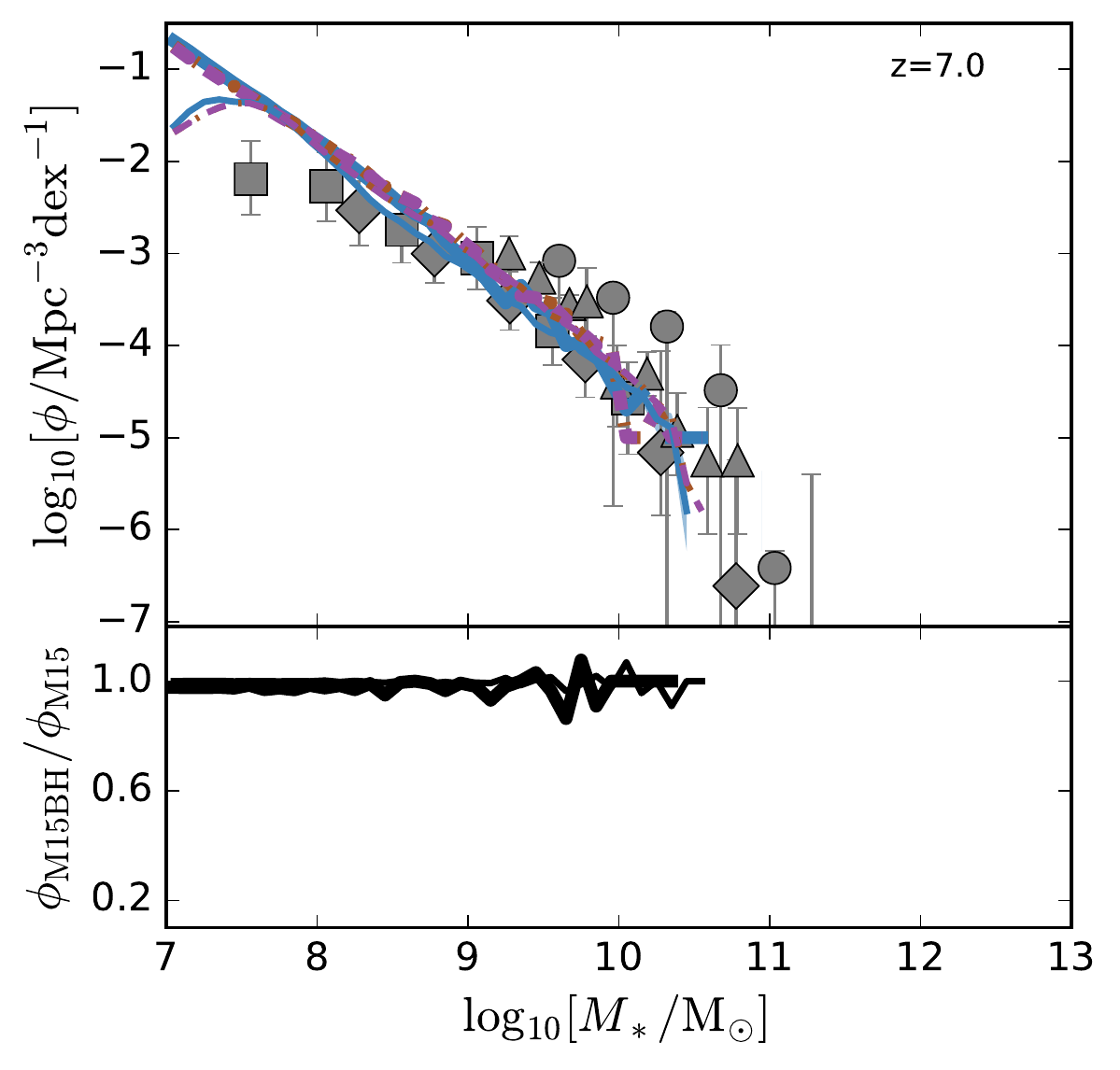}} 
			\end{minipage}
			\begin{minipage}{0.333\textwidth}
				\subfigure{\includegraphics[width=\textwidth]{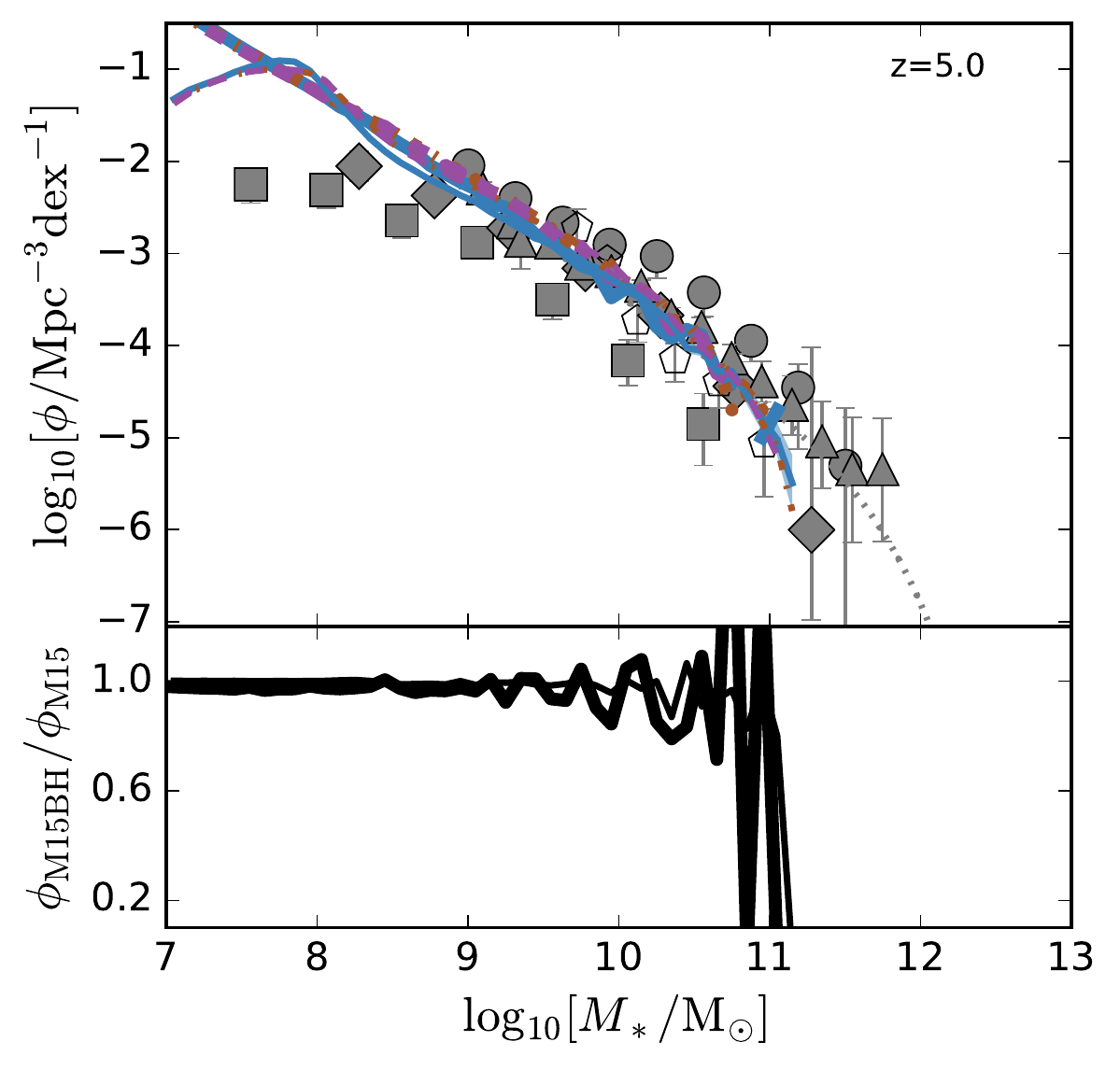}}
			\end{minipage}
			\begin{minipage}{0.333\textwidth}
				\subfigure{\includegraphics[width=\textwidth]{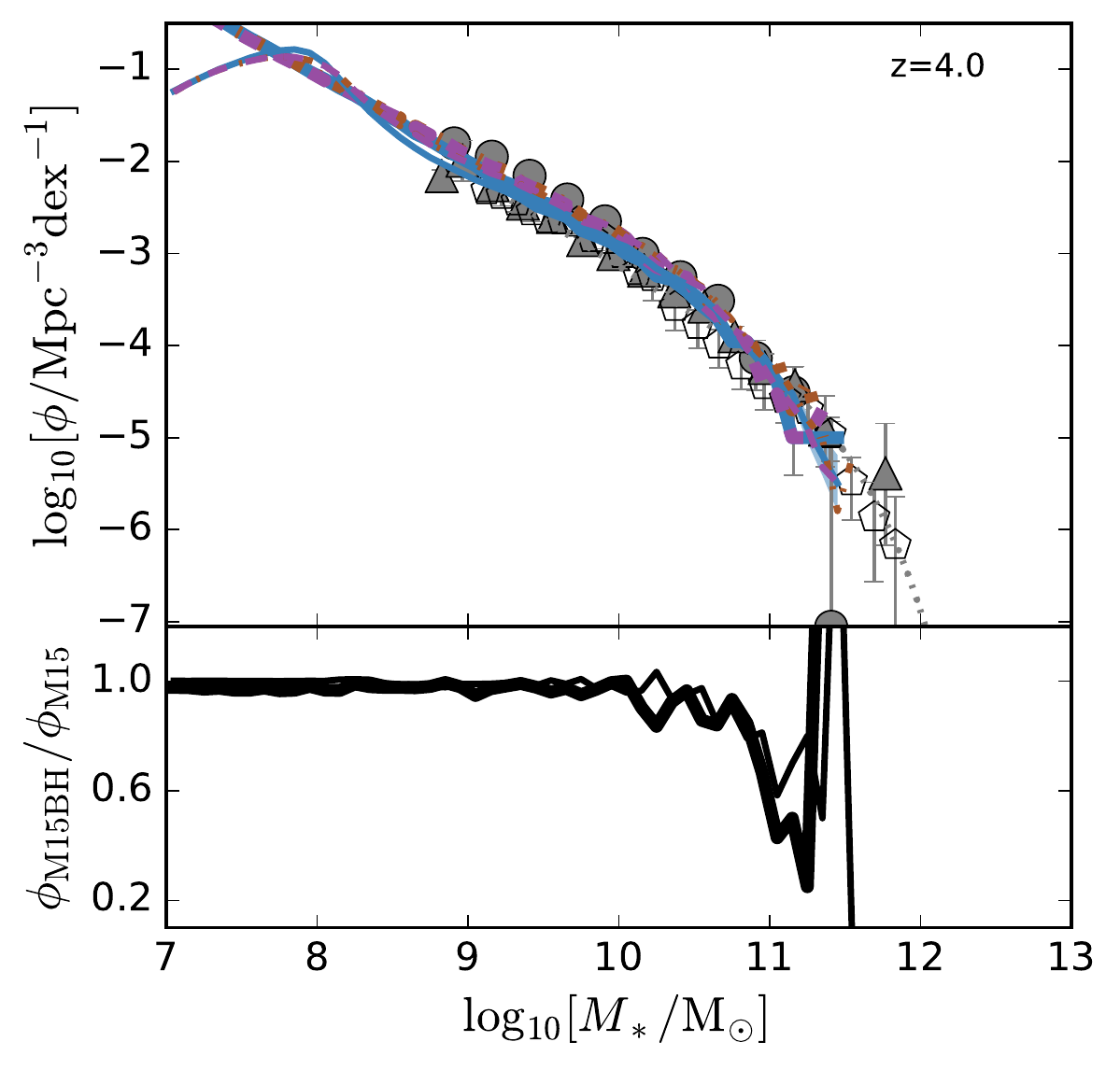}}
			\end{minipage}
		\end{tabular}\\	
		\begin{tabular}{lcr}
			\begin{minipage}{0.333\textwidth}
				\subfigure{\includegraphics[width=\textwidth]{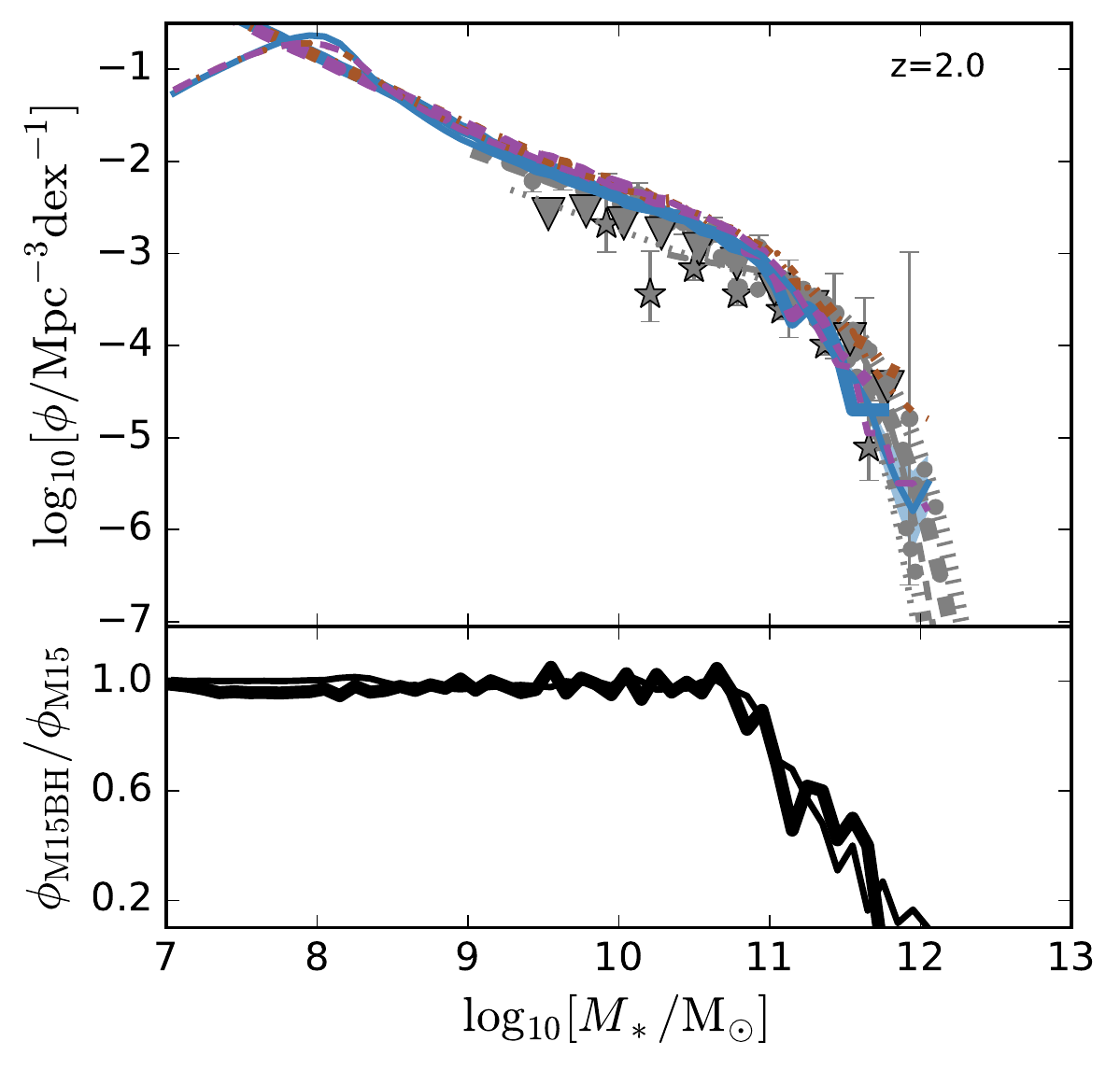}} 
			\end{minipage}
			\begin{minipage}{0.333\textwidth}
				\subfigure{\includegraphics[width=\textwidth]{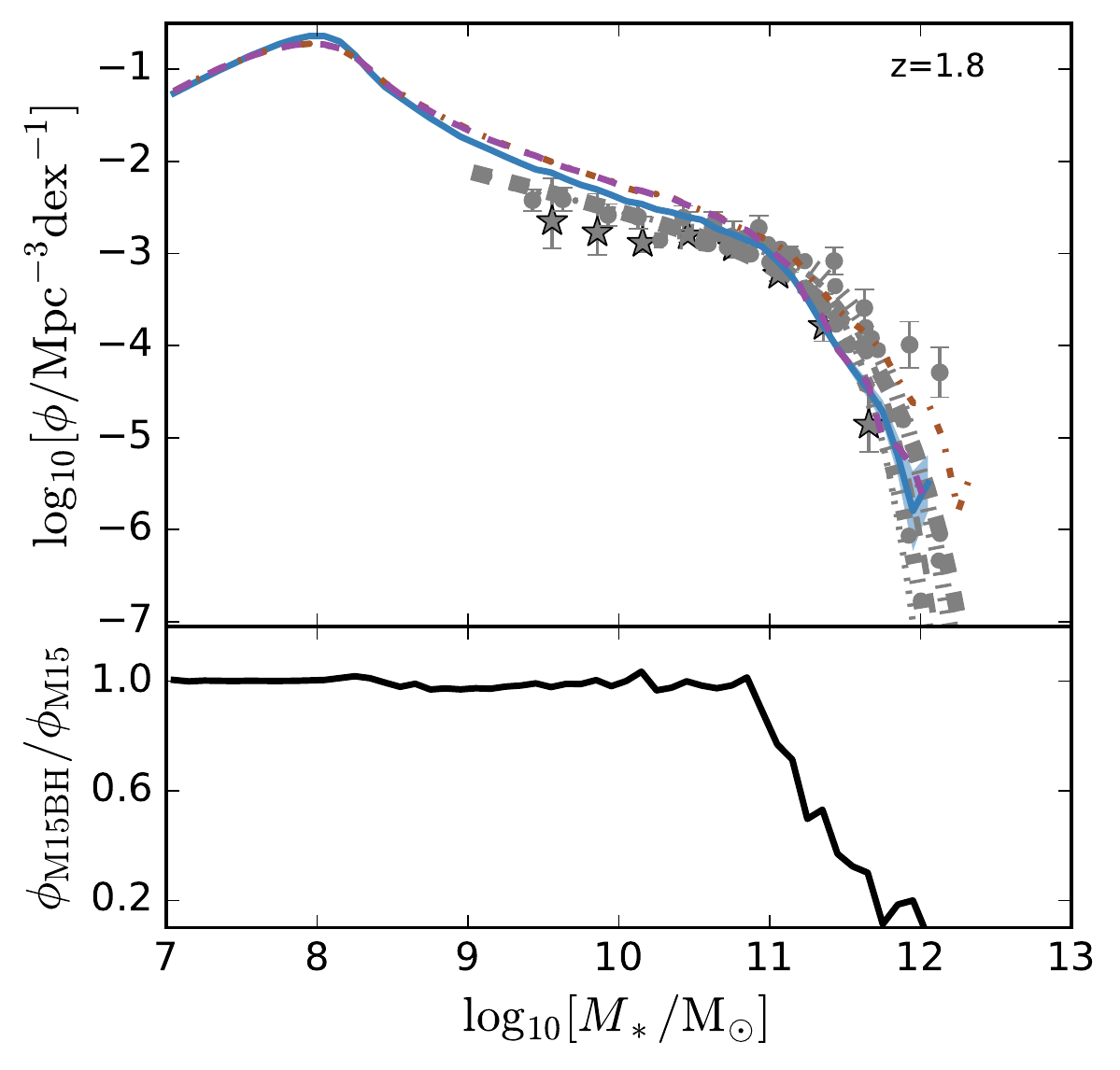}}
			\end{minipage}
			\begin{minipage}{0.333\textwidth}
				\subfigure{\includegraphics[width=\textwidth]{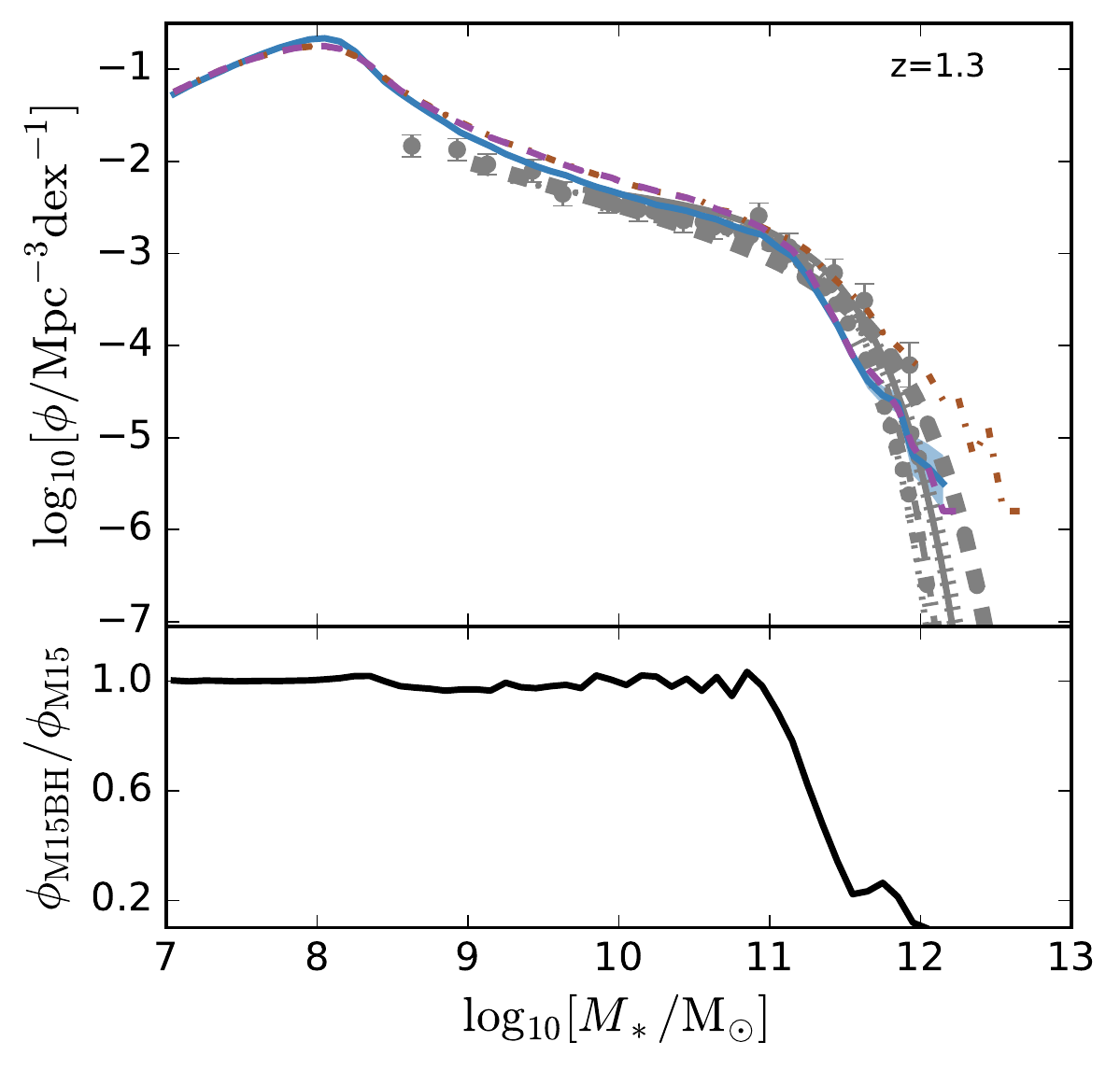}}
			\end{minipage}
		\end{tabular}\\	
		\begin{tabular}{lcl}
			\begin{minipage}{0.333\textwidth}
				\subfigure{\includegraphics[width=\textwidth]{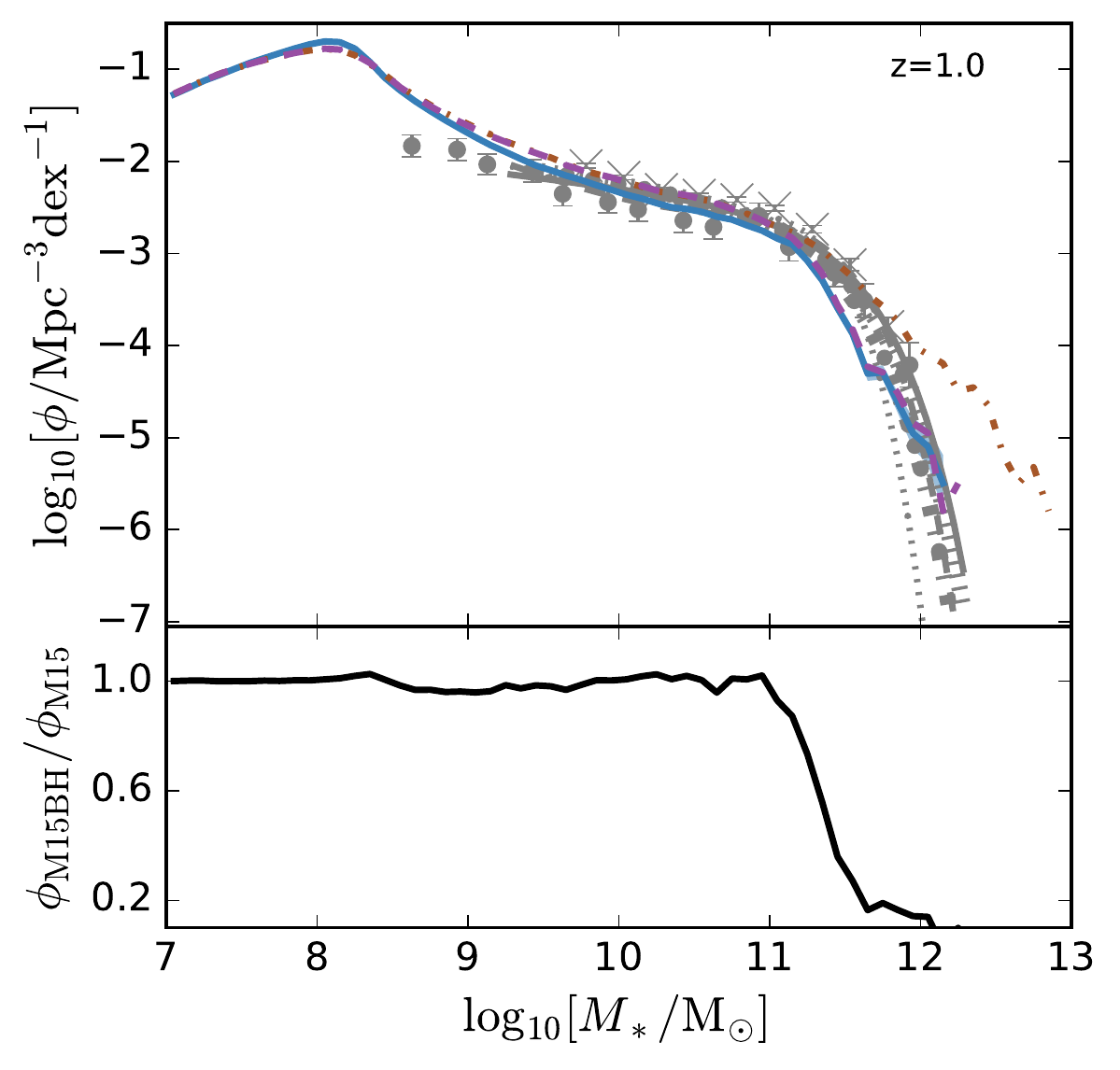}}
			\end{minipage}
			\begin{minipage}{0.333\textwidth}
				\subfigure{\includegraphics[width=\textwidth]{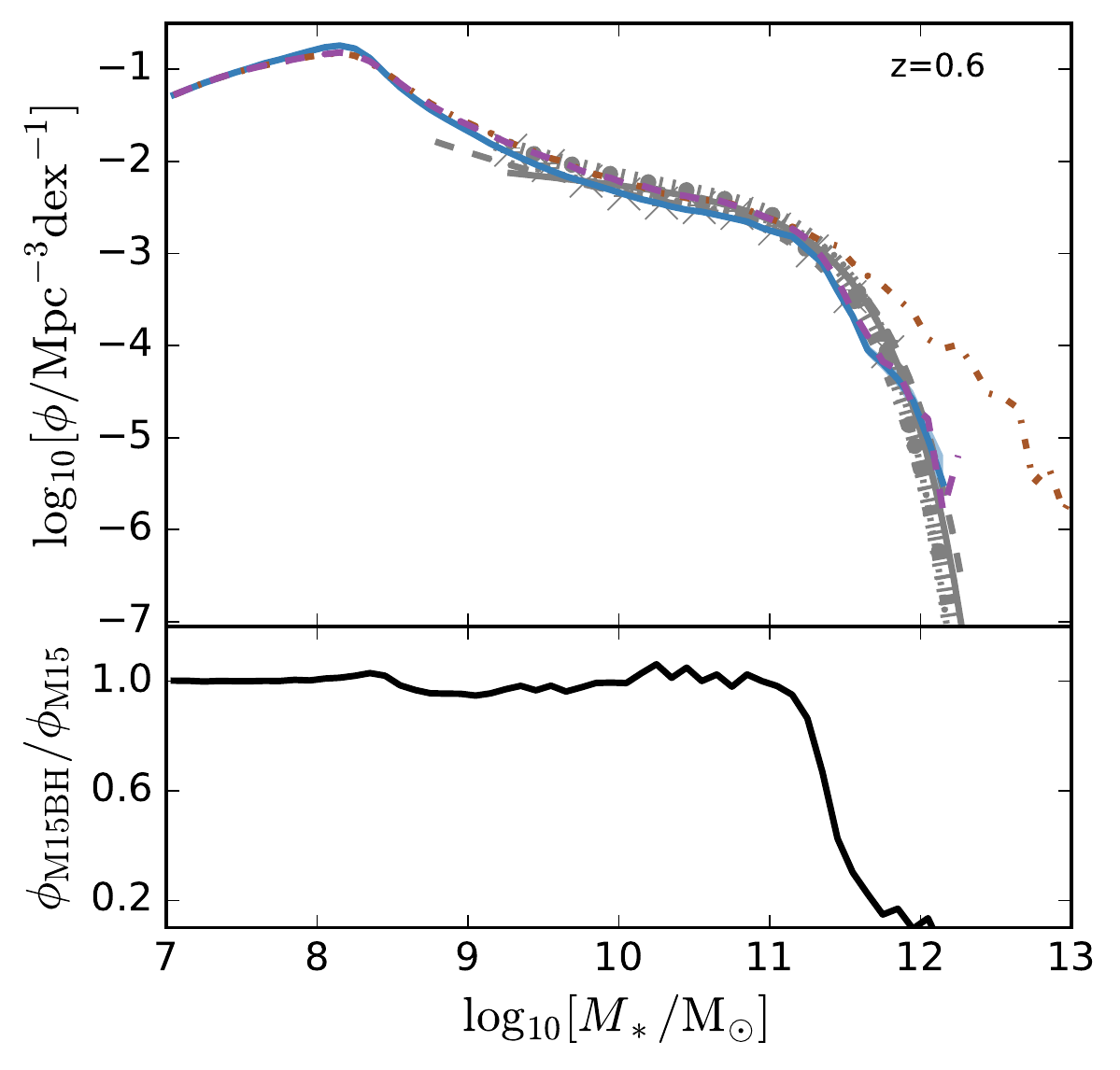}}
			\end{minipage}
			\begin{minipage}{0.333\textwidth}
				\subfigure{\includegraphics[width=\textwidth]{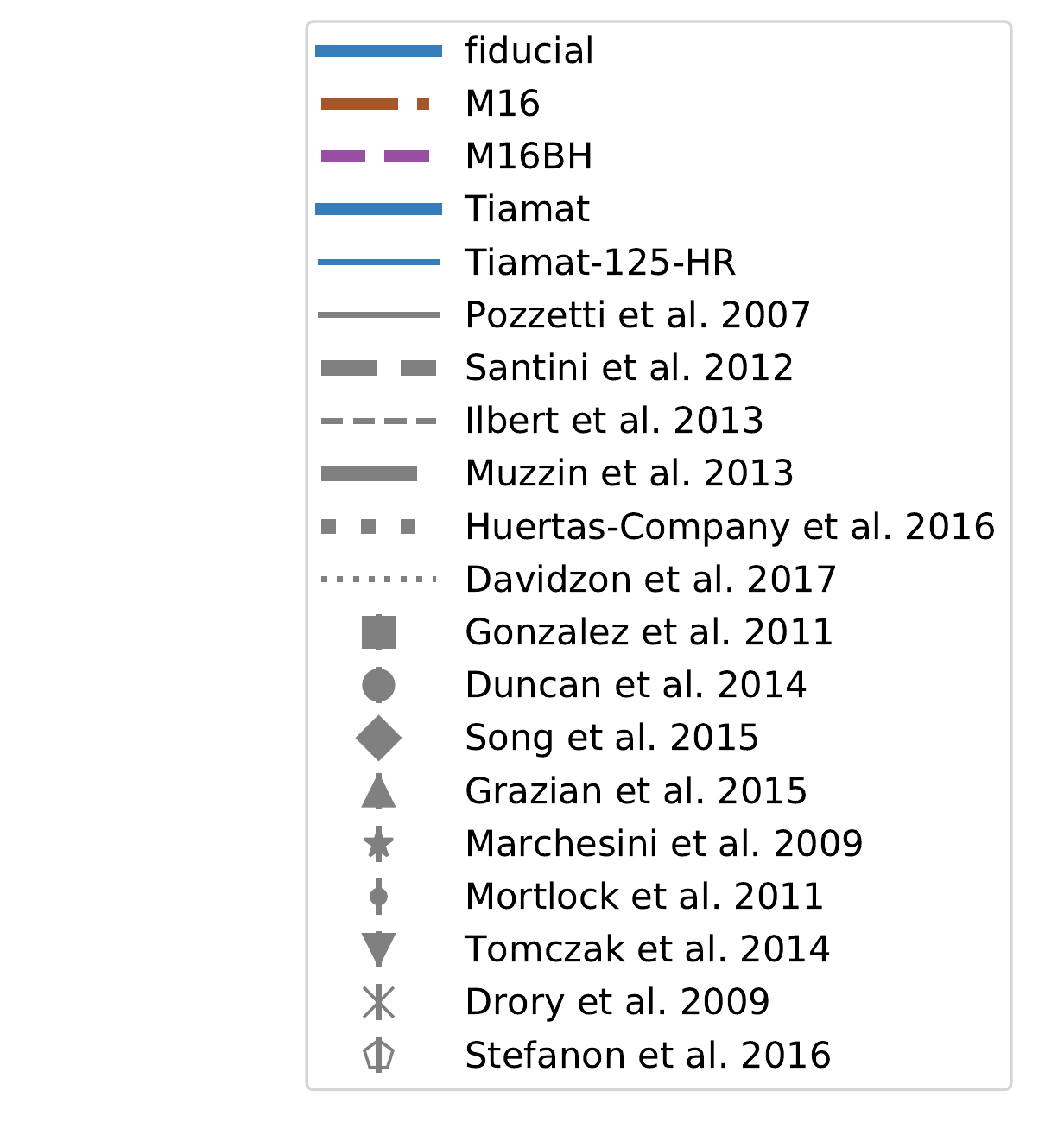}} 
			\end{minipage}
		\end{tabular}\\ 
	\end{minipage}
	\caption{\label{fig:GSMF} Galaxy stellar mass functions at $z\sim7-0.6$ from the fiducial ({\color{colorfiducial}{\hdashrule[0.6mm]{6mm}{2.5pt}{}}}), \citetalias{Mutch2016} ({\color{brown}{\hdashrule[0.6mm]{8mm}{2.5pt}{1.6mm 0.4mm 0.4mm 0.4mm}}}) and M16BH ({\color{Plum}{\hdashrule[0.6mm]{6mm}{2.5pt}{1.5mm 0.5mm}}}) models compared to the observational data ({\color{gray}{\hdashrule[0.6mm]{6mm}{1.5pt}{}}}\citealt{Pozzetti2007}, {\color{gray}{$\times$}}\citealt{Drory2009}, {\color{gray}{$\bigstar$}}\citealt{Marchesini2009ApJ...701.1765M}, {\color{gray}{$\bullet$}}\citealt{Mortlock2011MNRAS.413.2845M},  {\color{gray}{$\blacksquare$}}\citealt{Gonzalez2011}, {\color{gray}{{\hdashrule[0.6mm]{6mm}{3pt}{1.5mm 0.5mm}}}}\citealt{Santini2012}, {\color{gray}{\hdashrule[0.6mm]{6mm}{1.5pt}{1.5mm 0.5mm}}}\citealt{Ilbert2013}, {\color{gray}{\hdashrule[0.6mm]{6mm}{3pt}{1.5mm 0.4mm 0.4mm 0.4mm}}}\citealt{Muzzin2013}, {\color{gray}{$\blacktriangledown$}}\citealt{Tomczak2014}, {\color{gray}{\CIRCLE}}\citealt{Duncan2014}, {\color{gray}{$\blacklozenge$}}\citealt{Song2015}, {\color{gray}{$\blacktriangle$}}\citealt{Grazian2015}, {\color{gray}{\hdashrule[0.6mm]{6mm}{3pt}{0.4mm 0.4mm 0.4mm 0.4mm}}}\citealt{Huertas-Company2016}, {\color{gray}{$\pentagon$}}\citealt{Stefanon2016arXiv161109354S},{\color{gray}{\hdashrule[0.6mm]{6mm}{1.5pt}{0.4mm 0.4mm 0.4mm 0.4mm}}}\citealt{Davidzon2017}). The ratios of M16BH to the \citetalias{Mutch2016} result are shown in the bottom subpanels. The results calculated using the \textit{Tiamat} and \textit{Tiamat-125-HR} trees are shown with thick and thin lines, respectively. The shaded regions represent the $1\sigma$ Poisson uncertainties for the \textit{Tiamat-125-HR} result.}
\end{figure*}

Fig. \ref{fig:GSMF} presents the galaxy stellar mass functions from our fiducial model for comparison with the available observational data \citep{Pozzetti2007,Drory2009,Marchesini2009ApJ...701.1765M,Gonzalez2011,Mortlock2011MNRAS.413.2845M,Santini2012,Ilbert2013,Muzzin2013,Duncan2014,Tomczak2014,Grazian2015,Song2015,Huertas-Company2016,Stefanon2016arXiv161109354S,Davidzon2017} at redshifts 7 to ${\sim}0.6$. The results calculated using the \textit{Tiamat} and \textit{Tiamat-125-HR} trees are shown with thick and thin lines, respectively. The shaded regions represent the $1\sigma$ Poisson uncertainties. We see that the fiducial model is able to reproduce the observed galaxy stellar mass function across the redshift range of $z=7-0.6$. 

\begin{figure*}	
	\begin{minipage}{1.\textwidth}
		\begin{tabular}{lcr}
			\begin{minipage}{0.333\textwidth}
				\subfigure{\includegraphics[width=\textwidth]{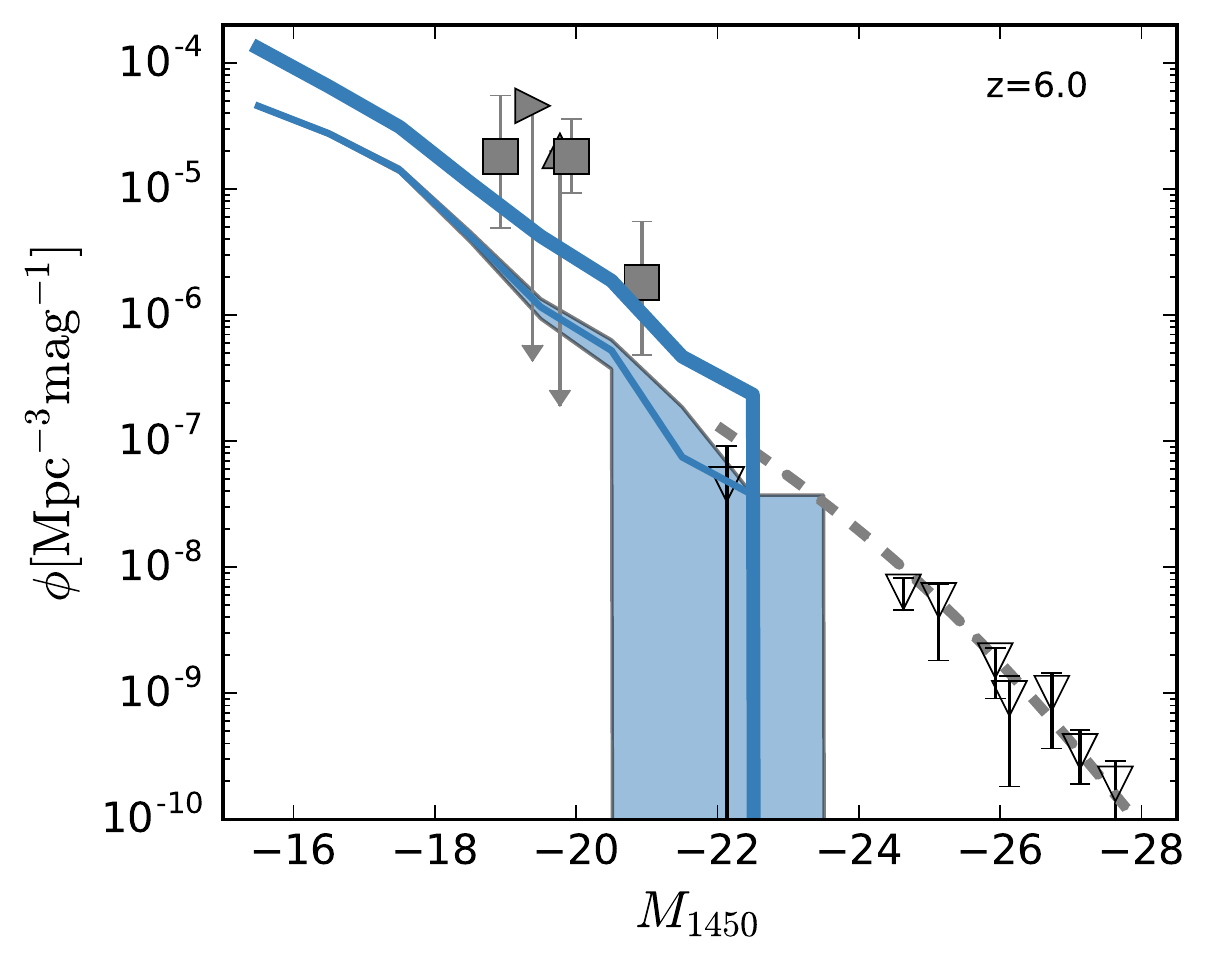}} 
			\end{minipage}
			\begin{minipage}{0.333\textwidth}
				\subfigure{\includegraphics[width=\textwidth]{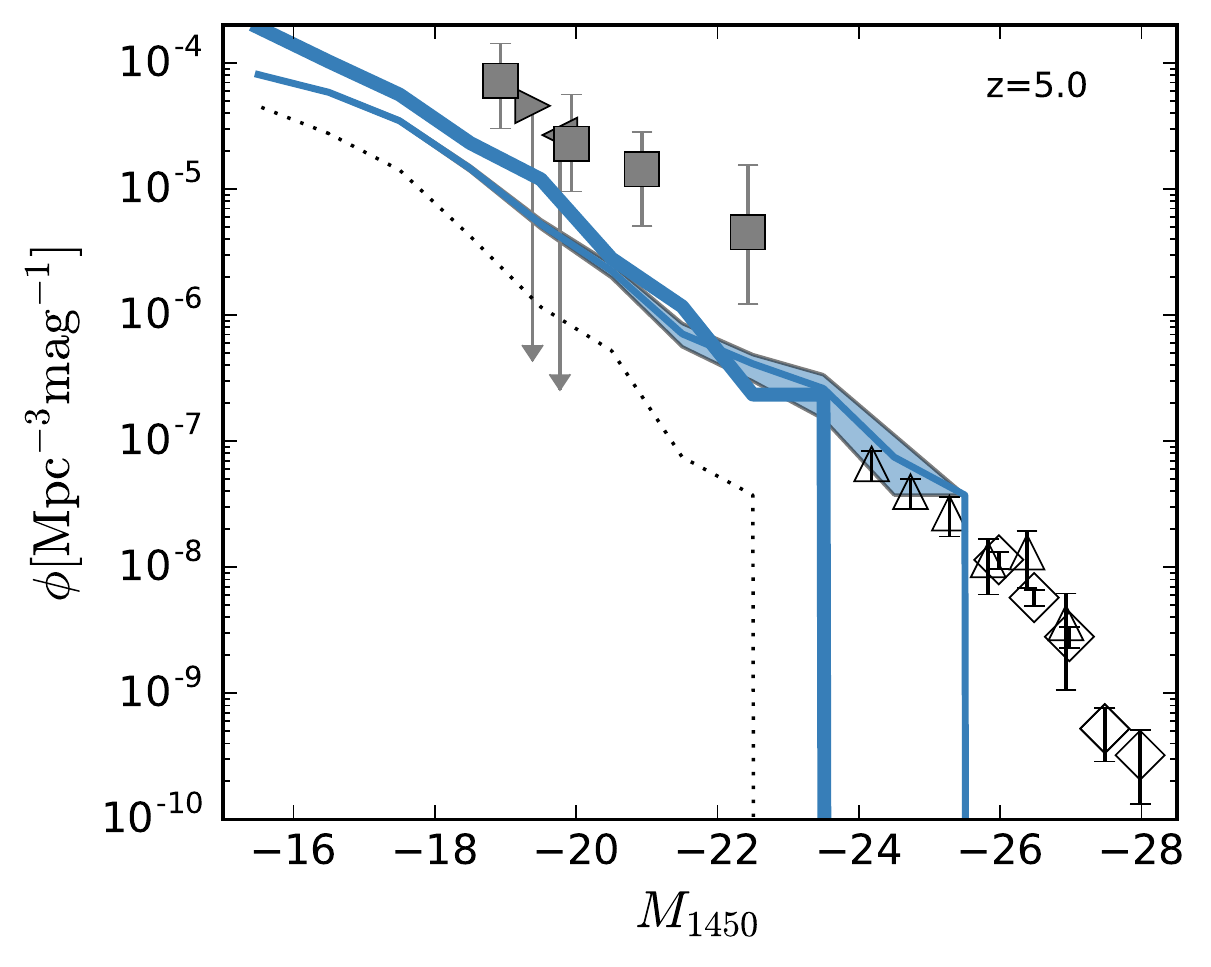}}
			\end{minipage}
			\begin{minipage}{0.333\textwidth}
				\subfigure{\includegraphics[width=\textwidth]{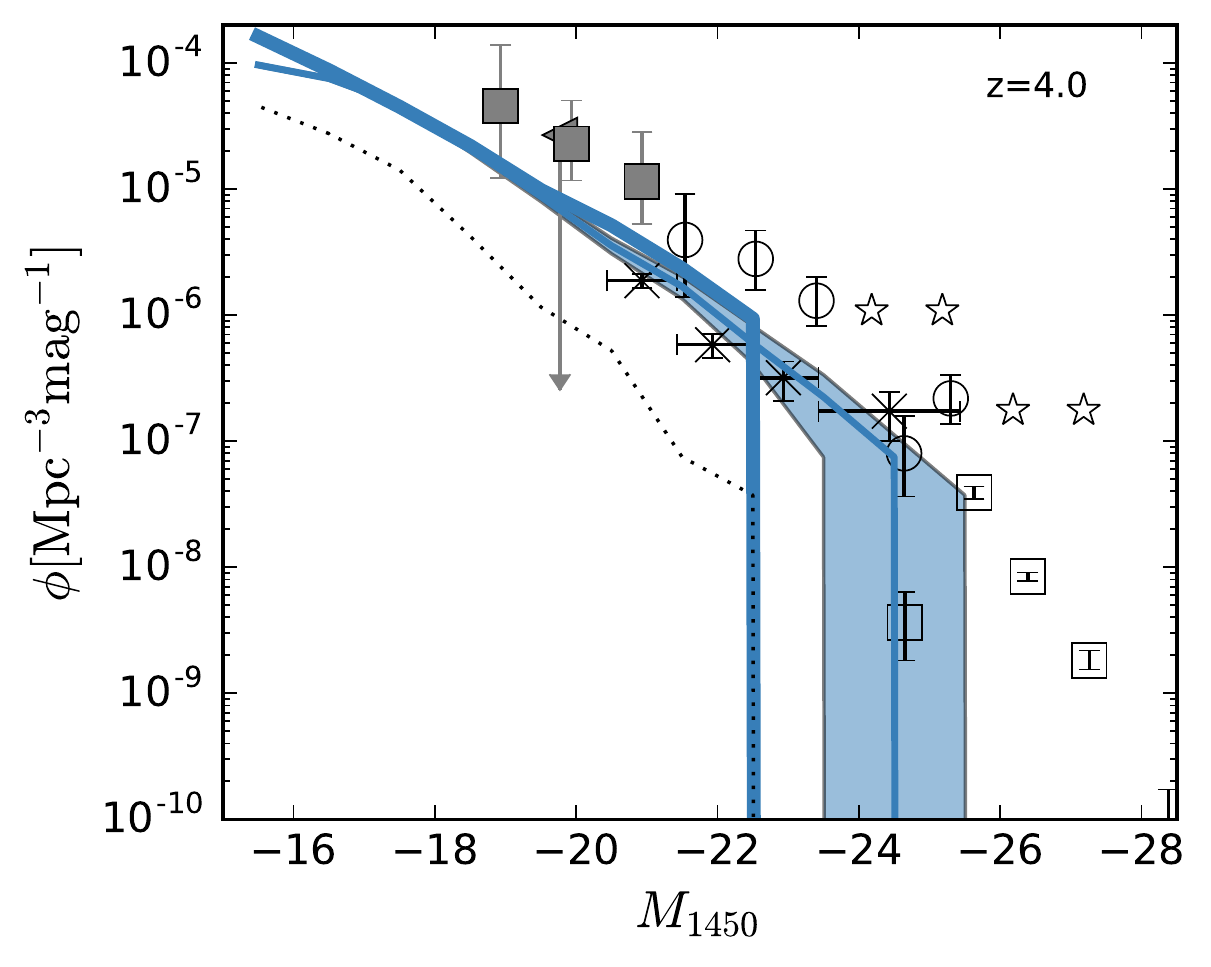}}
			\end{minipage}
		\end{tabular}\\	
		\begin{tabular}{lcr}
			\begin{minipage}{0.333\textwidth}
				\subfigure{\includegraphics[width=\textwidth]{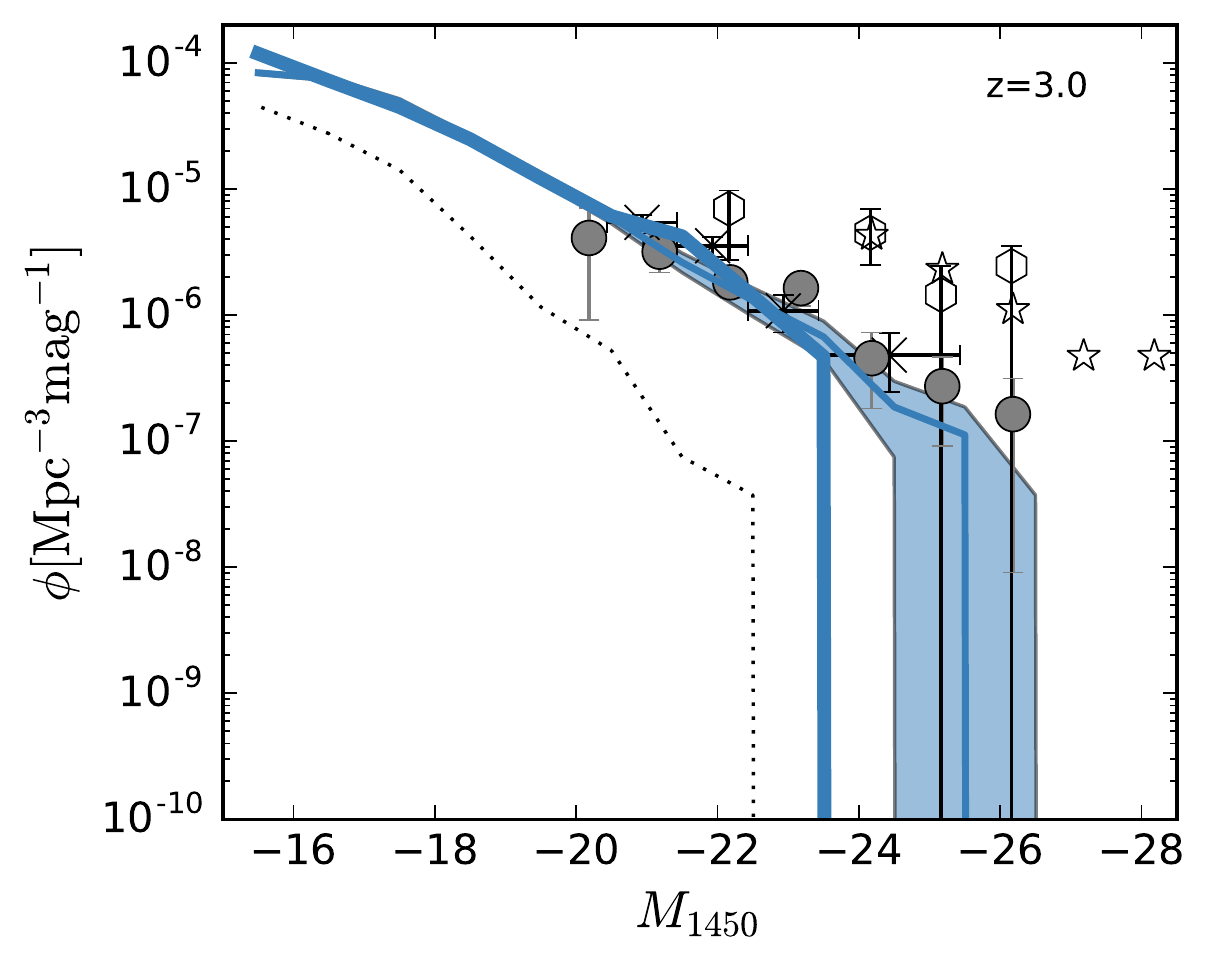}} 
			\end{minipage}
			\begin{minipage}{0.333\textwidth}
				\subfigure{\includegraphics[width=\textwidth]{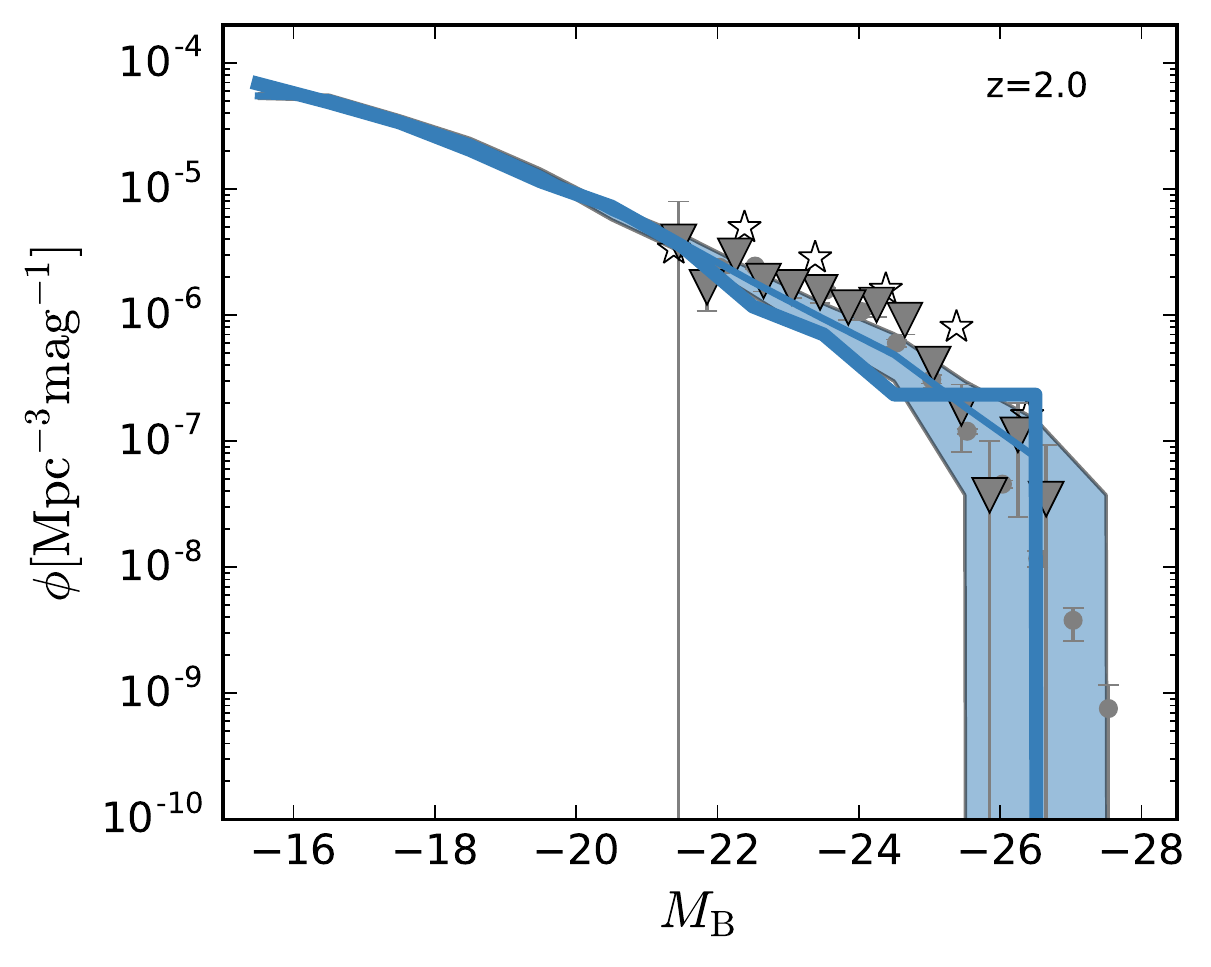}}
			\end{minipage}
			\begin{minipage}{0.333\textwidth}
				\subfigure{\includegraphics[width=\textwidth]{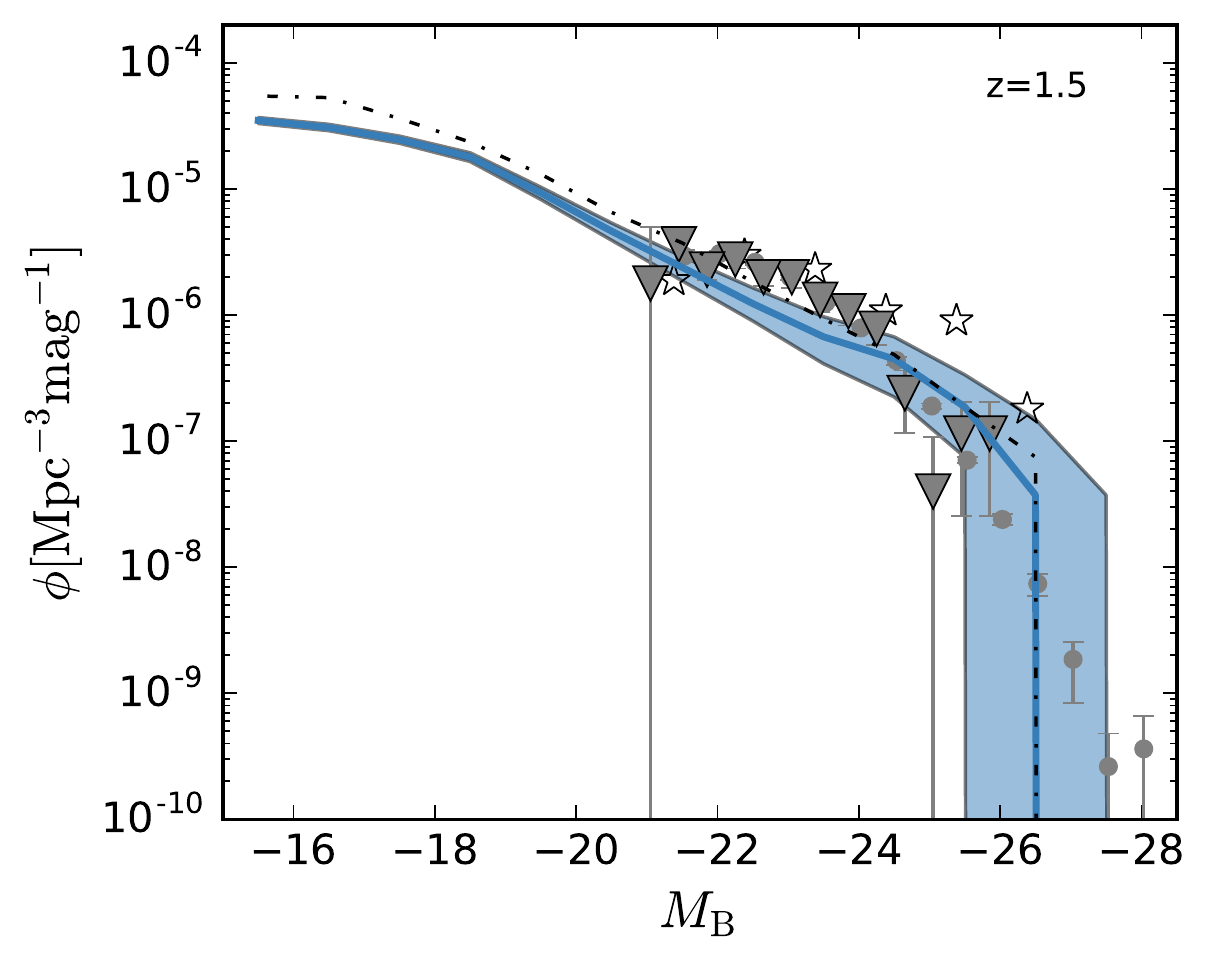}}
			\end{minipage}
		\end{tabular}\\
		\begin{tabular}{lcr}
			\begin{minipage}{0.333\textwidth}
				\subfigure{\includegraphics[width=\textwidth]{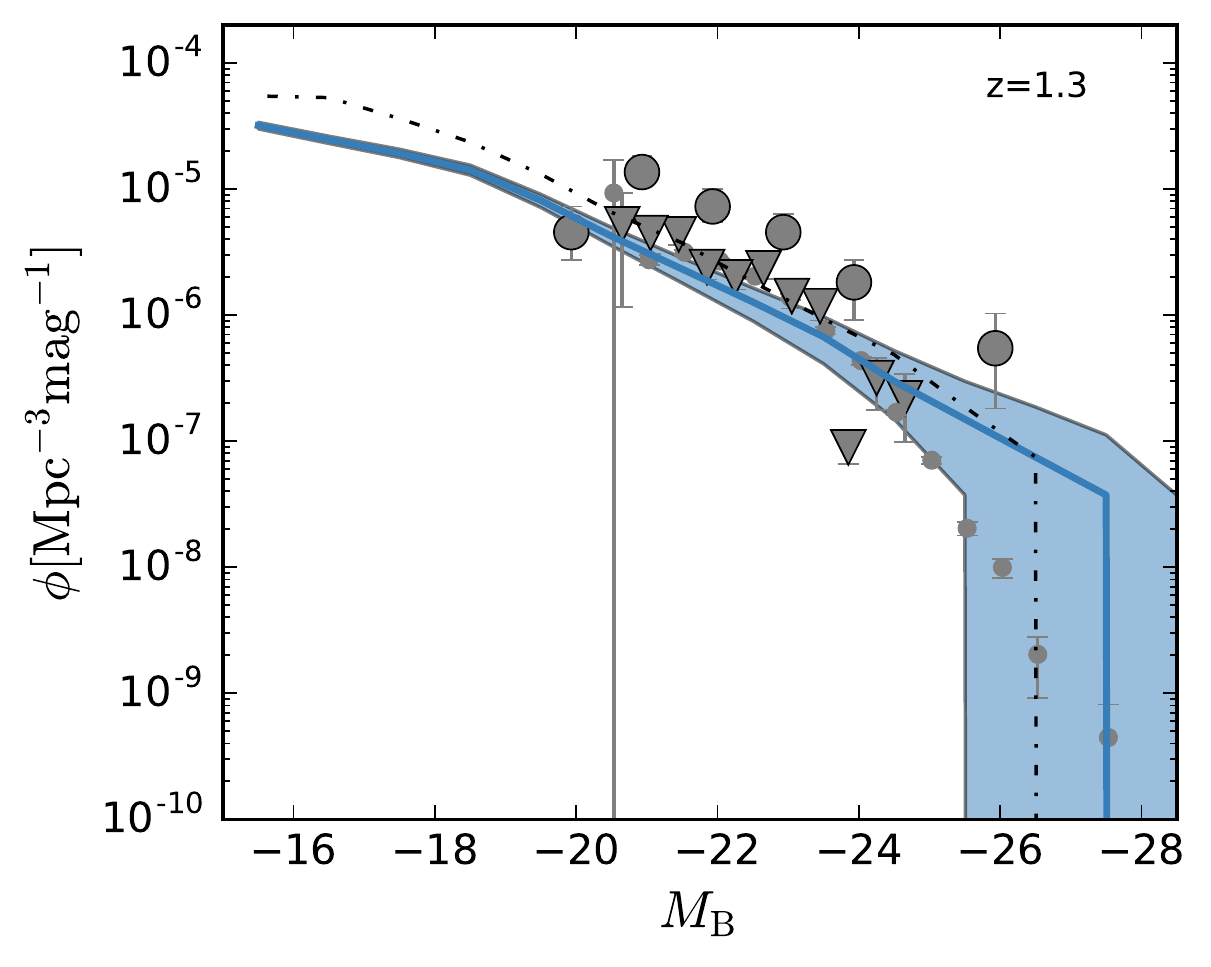}}
			\end{minipage}
			\begin{minipage}{0.333\textwidth}
				\subfigure{\includegraphics[width=\textwidth]{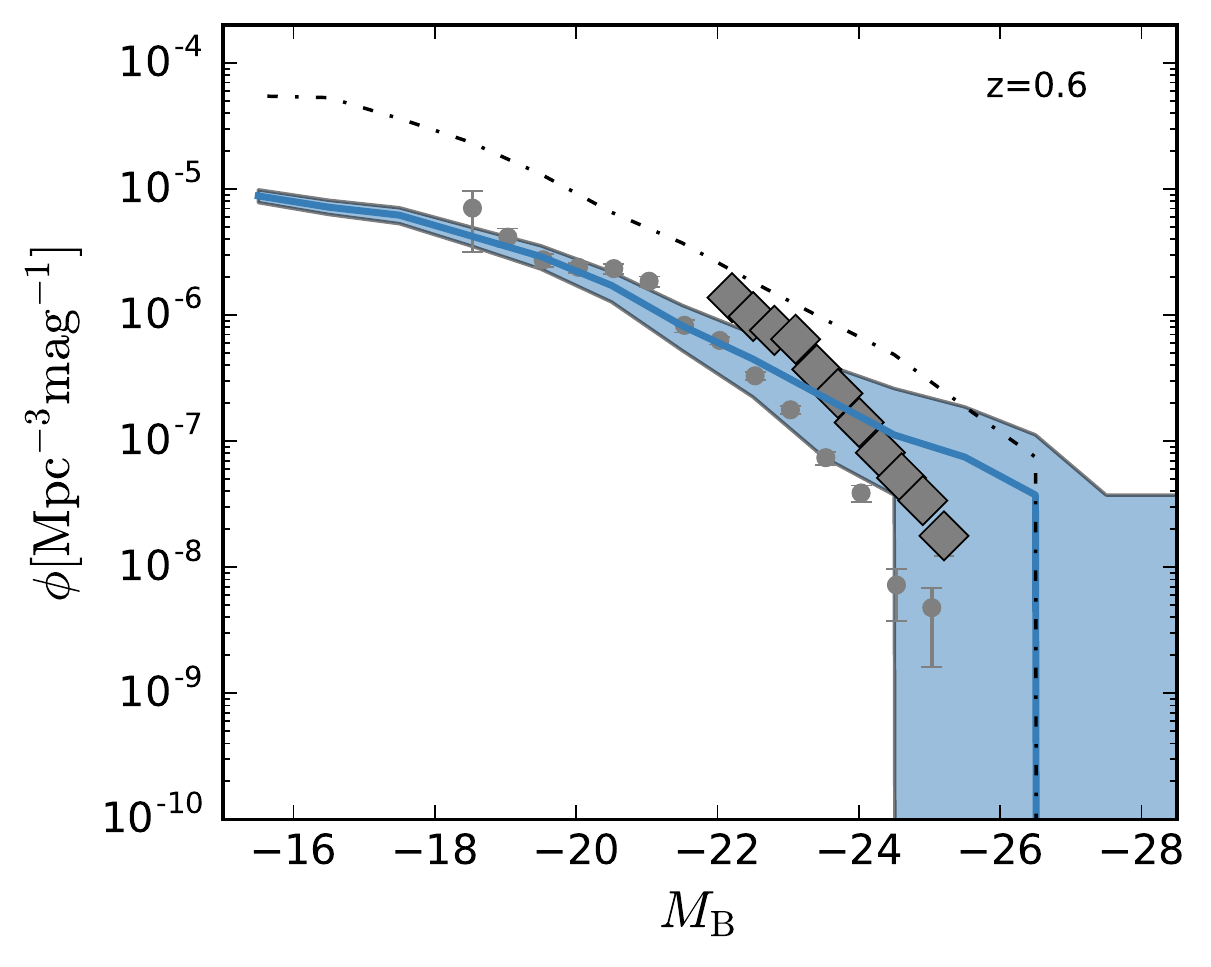}}
			\end{minipage}
			\begin{minipage}{0.333\textwidth}
				\subfigure{\includegraphics[width=\textwidth]{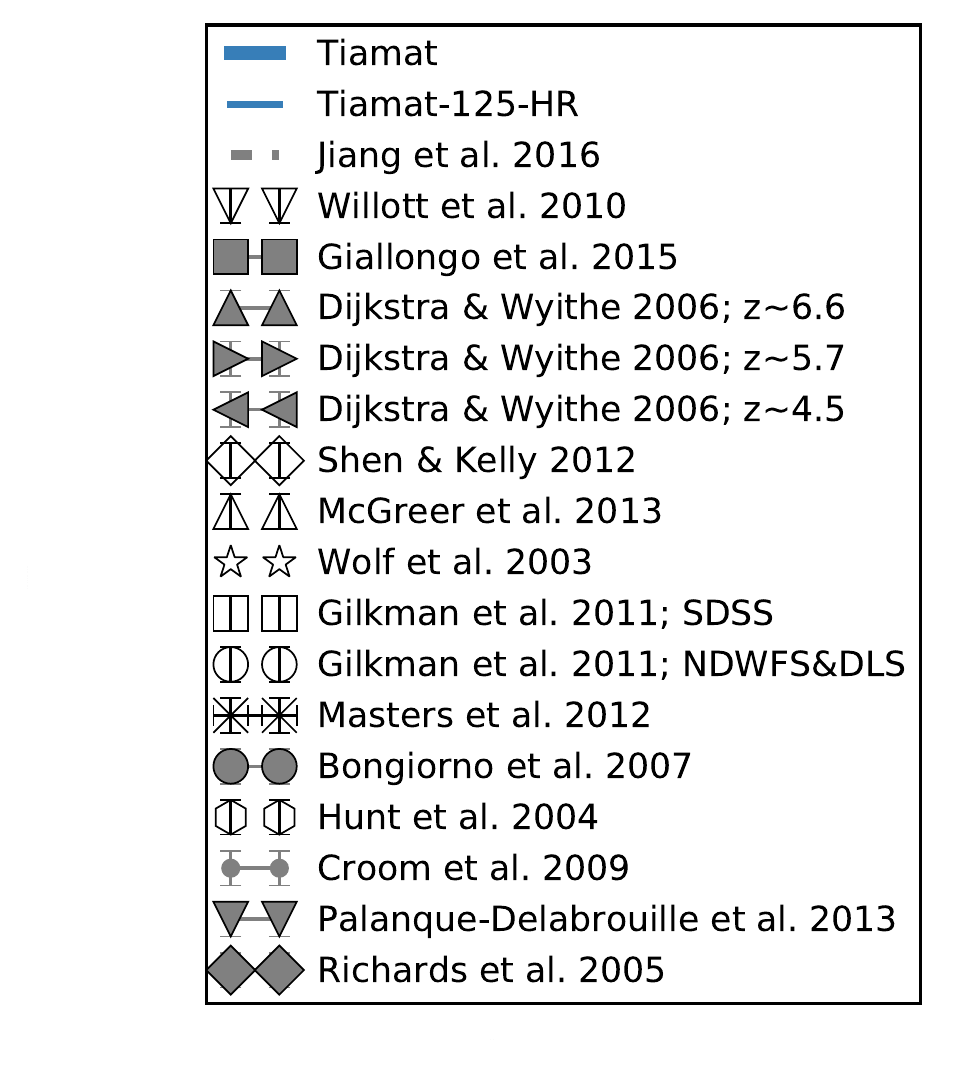}}
			\end{minipage}
		\end{tabular}
	\end{minipage}
	\vspace*{-0.5cm}
	\caption{\label{fig:BHMFandLF} UV 1450 $\mathrm{\AA}$ luminosity functions of quasars at $z\sim6$, 5, 4 and 3, and \textit{B}-band luminosity functions at $z\sim2.0$, 1.5, 1.3 and 0.6. The results using the \textit{Tiamat} and \textit{Tiamat-125-HR} halo merger trees are shown with blue thick and thin lines, respectively. The shaded regions represent the 95 per cent confidence intervals around the mean using 100000 bootstrap re-samples for the \textit{Tiamat-125-HR} result. In the panels showing UV luminosity functions at $z\sim5.0$, 4.0 and 3.0, the $z=6.0$ UV luminosity function of \textit{Tiamat-125-HR} is indicated with black thin dotted lines for comparison. The $z=2.0$ \textit{B}-band luminosity function of \textit{Tiamat-125-HR} is shown with black thin dash--dotted at $z\sim1.5$, 1.3 and 0.6. The observational data are shown with different symbols: \citet{wolf2003} ({\color{gray}{$\bigstar$}}); \citet{Hunt2004faint} ({\color{gray}{$\varhexagon$}}); \citet{Richards2005} using the 2dF-SDSS LRG and QSO survey ({\color{gray}{$\blacklozenge$}}) \citet{Dijkstra2006} using Ly$\alpha$ constraints and providing upper limits at $z{\sim}4.5$ ({\color{gray}{\LEFTarrow}}), $z\sim5.7$ ({\color{gray}{\RIGHTarrow}}) and $z\sim6.6$ ({\color{gray}{\UParrow}}); \citet{Bongiorno2007vvds} using VVDS ({\color{gray}\CIRCLE}); \citet{Croom2009} using the 2dF-SDSS LRG and QSO survey ({\color{gray}{$\bullet$}}); \citet{willott10} using the CFHQS data ($\bigtriangledown$); \citet{Glikman2011} at $z\sim4$ using the SDSS data ($\square$) and using the NOAO Deep Wide-Field Survey and the Deep Lens Survey ($\Circle$); \citet{Shen2012} using SDSS DR7 data at $z\sim4.75$ ($\Diamond$); \citet{Masters2012} using COSMOS ({\color{black}{$\times$}}); \citet{McGreer2013} using SDSS, UKIDSS and MMT at $z\sim4.7-5.1$ ($\vartriangle$); \citet{Palanque-Delabrouille13} with SDSS-\Rom{3} and MMT data; \citet{Giallongo2015} using \textit{Chandra, HST, Spitzer} and various ground-based telescopes ({\color{gray}{$\blacksquare$}}); \citet{Jiang2016} with SDSS ({\color{gray}{\hdashrule[0.6mm]{6mm}{2.5pt}{1.5mm 0.5mm}}}).}
\end{figure*}

We also present two additional models in Fig. \ref{fig:GSMF}, \citetalias{Mutch2016} and M16BH. \citetalias{Mutch2016} adopts identical parameters as the redshift varying $f_\mathrm{esc,*}$ model ($f_\mathrm{esc,*,M16} = \min\left[0.04\times\left(\dfrac{1+z}{6}\right)^{2.5},1.0\right]$, see Table \ref{tab:parameters}) presented in \citet{Mutch2016}. This model is able to reproduce the evolution of the stellar mass function at high redshift ($z>5$). The redshift dependence of $f_\mathrm{esc,*,M16}$ was chosen to simultaneously reproduce the normalization and flat slope of the \citet{McQuinn2011} emissivity measurement at $z\sim5$ and the Planck 2015 optical depth measurement (\citealt[][hereafter Planck15]{PlanckCollaboration2015}). Details of the \citetalias{Mutch2016} model at high redshift ($z>5$) can be found in \citet{Mutch2016}. In this work, we extend the model\footnote{We note that the \textit{Tiamat} halo merger trees have been improved (Poole et al. in preparation) since \citet{Mutch2016}. However, the impact on galaxy formation is trivial.} to lower redshifts and find that, without AGN feedback regulating galaxy formation, the model fails to reproduce the observed stellar mass functions at $z<2$, especially in larger mass ranges ($M_*>10^{11}\mathrm{M}_\odot$). However, when AGN feedback is implemented, shown as the M16BH model, the model shows better agreement with observations \citep{croton2006many}. These two models also suggest that radio-mode feedback does not play a significant role in galaxy formation during the EoR, and because reionization is dominated by low-mass galaxies (\citealt{Liu2016MNRAS.462..235L}, see also Section \ref{sec:discussion}), AGN feedback is expected to have no significant impact on reionization.\footnote{Ignoring the impact on the ionizing photon escape fraction from AGN feedback.}

With respect to \citetalias{Mutch2016}, the fiducial model presented in this work employs a stronger star formation efficiency ($\alpha_\mathrm{sf}$) with maximized supernova feedback ($\alpha_\mathrm{energy}$ and $\alpha_\mathrm{mass}$) and more intense radio mode feedback ($k_\mathrm{h}$), in order to gain better agreement with the observed stellar mass function in the intermediate mass range ($10^{9}\mathrm{M}_\odot<M_*<10^{11}\mathrm{M}_\odot$) at $1<z<2$.

\subsection{Quasar luminosity function}\label{sec:QLF}

In the previous two subsections, we presented the predicted properties of the central massive black holes and host galaxies, together with their correlations. These show that our model is able to produce the evolution of the galaxy stellar mass function over a large time-scale ($z\sim7-0.6$), as well as the observed black hole properties at low redshift. Before we start exploring the contribution of quasars to reionization, we discuss the quasar luminosity function in this section.

Semi-analytic models can predict quasar bolometric luminosity through the modelled central black hole mass. In order to compare with observations, three corrections are usually required: bolometric corrections, duty cycles and obscured fractions. In our model, we calculate the duty cycle self-consistently by assuming Eddington accretion of available gas and a random observation time between snapshots (see Section \ref{sec:quasar mode}). Black holes that are inactive when observed are not considered in the census. We convert the total luminosity into a particular band using the bolometric correction of \citet{Hopkins2007} (see Section \ref{sec:QL}). Finally, we need to take the obscuration due to the presence of a dusty torus surrounding the AGN into account. In practice, each quasar is weighted by $1-\cos\dfrac{\theta}{2}$, where $\theta$ represents the opening angle of AGN radiation, during the calculation of the quasar luminosity function. We note that the dependence of $\theta$ or the obscured fraction (the ratio of obscured to unobscured AGN) is very complicated. It has been suggested that $\theta$ may depend on wavelength, redshift and luminosity \citep{Elvis1994,Hopkins2007}. For simplicity, we only consider constant $\theta$. The opening angle is a free parameter in our model, chosen to reproduce the quasar luminosity function amplitude. In this work, $\theta=80$ deg is adopted. This corresponds\footnote{An opening angle of $\theta=80$ deg corresponds to a solid angle of $\Omega = 2\pi(1-\cos\dfrac{\theta}{2})\sim1.47$. Considering a symmetric radiation from both sides of the accretion disc, the un-obscured fraction is $f_\mathrm{obs}\equiv\dfrac{2\Omega}{4\pi}\sim0.234$.} to a fraction of visible objects,\footnote{The opening angle is used to illustrate the observable fraction assuming an orientation model. We note that the observable fraction can also be interpreted by the line-of-sight absorption column density when the evolutionary model is assumed.} $f_\mathrm{obs}$, to be $\sim23.4$ per cent, in agreement with \citet{Hopkins2007}, who suggests a luminosity-dependent observable fraction, $f_\mathrm{obs}=0.26\left(L_\mathrm{bol}/10^{12.4}\mathrm{L}_\odot \right)^{0.082}$ for the \textit{B} band.

We present the UV 1450 $\mathrm{\AA}$ luminosity functions of quasars at redshift 6, 5, 4 and 3, and the \textit{B}-band luminosity functions at $z\sim2$, 1.5, 1.3 and 0.6 in Fig. \ref{fig:BHMFandLF}, compared to the observational data from \citet{wolf2003}, \citet{Hunt2004faint}, \citet{Richards2005}, \citet{Dijkstra2006}, \citet{Bongiorno2007vvds}, \citet{Croom2009}, \citet{willott10}, \citet{Glikman2011}, \citet{Shen2012}, \citet{Masters2012}, \citet{McGreer2013}, \citet{Palanque-Delabrouille13}, \citet{Giallongo2015} and \citet{Jiang2016}. The results calculated using the \textit{Tiamat} and \textit{Tiamat-125-HR} trees are shown with thick and thin lines, respectively. The shaded regions represent the 95 per cent confidence intervals around the mean using 100000 bootstrap re-samples (due to the random number $t_\mathrm{obs}$ in equation \ref{eq:luminosity}). Similarly to the black hole mass functions, the \textit{Tiamat-125-HR} results show a lower number density at high redshift compared to the \textit{Tiamat} results, and they converge at lower redshifts ($z\lesssim4$). We see that the model shows good agreement with observations across a large redshift range ($z\sim6-0.6$). At high redshift the model is consistent with the samples of bright quasars \citep{willott10,Shen2012,McGreer2013,Jiang2016} while it predicts a lower number density of faint quasars compared to the \citet[][hereafter the \citetalias{Giallongo2015}]{Giallongo2015} sample. On the other hand, the model produces a significant number of faint quasars. The turnover (not shown here) of the predicted UV luminosity function is around $M_{1450}{\sim}{-}11$, which is much fainter than the observed faintest quasars, $M_{1450}{\sim}{-}18$ (\citetalias{Giallongo2015}). In addition, the observed population of bright quasars ($M_{1450}{<}-23$) at high redshift is not present in our model due to our limited simulation volume. All of these factors may have an impact on the contribution of AGN to the reionization history, which is discussed in Section \ref{sec:reionization}.

\subsection{The luminosity function of galaxies with AGN}

The large number of faint AGN identified by \citetalias{Giallongo2015} has prompted renewed discussion of the contribution of quasars to reionization \citep{Madau2015,Mitra2015,Kulkarni2017}. We therefore further discuss the low number density in the faint end predicted by the model compared with the G15 data. 

When constructing an observed AGN luminosity function, the sample is typically identified spectroscopically or via colour--colour selections if spectra are not available, following which contamination from host galaxies is removed. For example, some observers model the surface brightness distributions of host galaxies and fit galaxies (e.g. S\'{e}rsic profiles) and point sources (using point-spread functions, \citealt{Dunlop2003MNRAS.340.1095D,Peng2006ApJ...649..616P,Du2014,Martinez-Paredes2017}) to the images.\footnote{Also with a constant to model the sky background.} Others examined the SED using a combination of AGN and galaxy emission with possible extinction of the AGN flux when spectroscopic data are available \citep{Bongiorno2007vvds,Croom2009,Masters2012,Mechtley2012ApJ...756L..38M,Lyu2016ApJ...816...85L}. In most cases, ignoring the contribution from host galaxies does not make a significant difference to bright quasars \citep{Hopkins2007}. Therefore, some AGN samples do not exclude stellar light (\citet{wolf2003,Hunt2004faint,Richards2005,willott10,Shen2012,McGreer2013,Palanque-Delabrouille13}; \citetalias{Giallongo2015}). However, this may not be the case for faint AGN. In particular, the G15 AGN sample is selected using X-ray activity and no AGN-galaxy separation is possible. Thus, the total UV luminosity may have a large fraction of stellar light, suggesting that the G15 sample may be potentially impacted by stellar light contamination. Despite the recent claim that only 12 of the 22 reported X-ray detections in G15 are high-redshift AGN \citep{Parsa2017}, this conjecture is supported by \citet{Ricci2017}, who used X-ray observations as a proxy and derived the quasar UV luminosity function down to much fainter ranges. They showed that the luminosity function is in agreement with UV/optical observations (e.g. \citealt{Croom2009,Glikman2010,Masters2012,Palanque-Delabrouille13}), and have much lower amplitudes than the G15 results, and that the high number density of faint AGN in G15 can be explained by the contribution from the luminosity of host galaxy with $M_{1450}\sim-20$.

\begin{figure}
	\begin{minipage}{0.95\columnwidth}
		\centering
		\includegraphics[width=\textwidth]{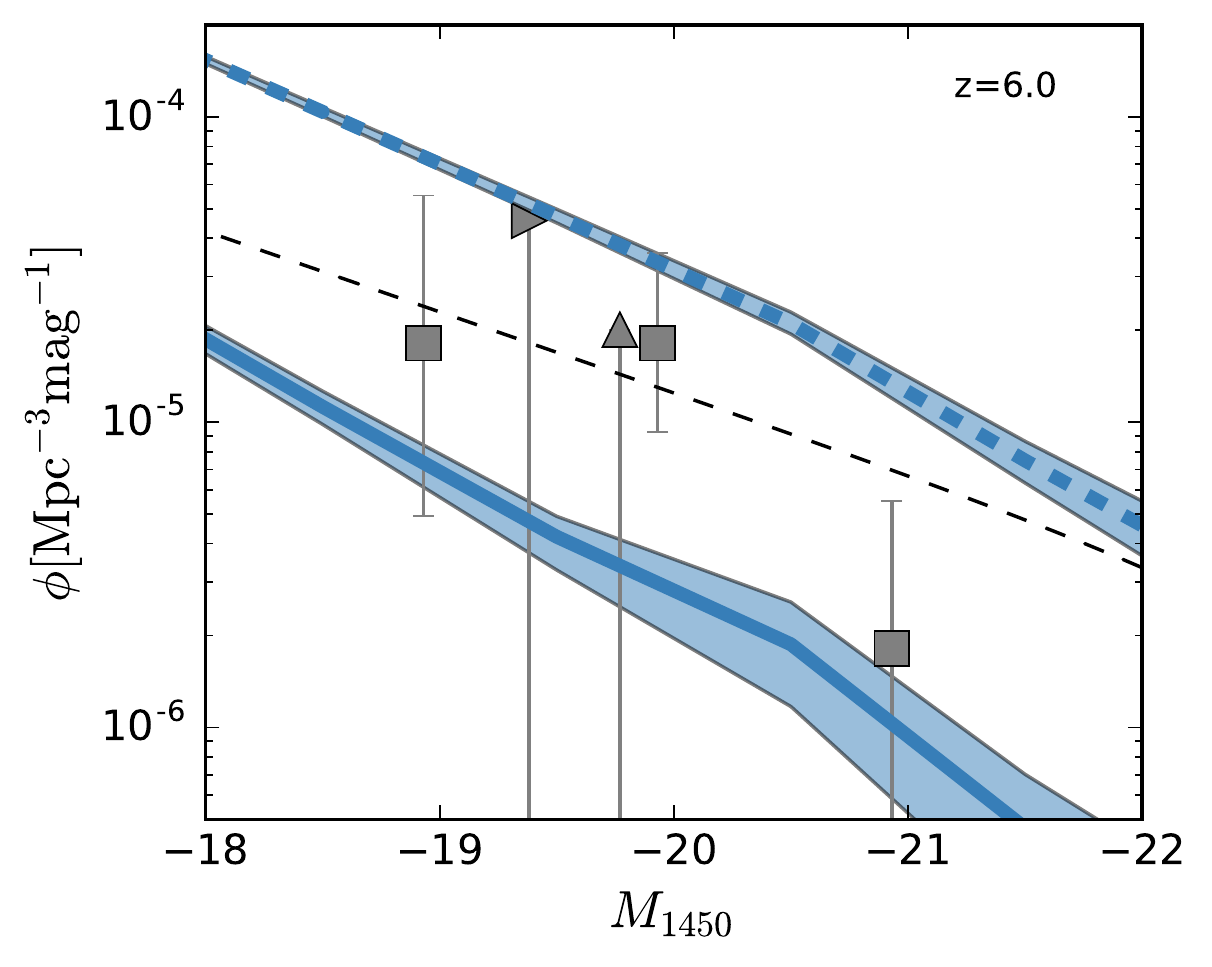}\\	\includegraphics[width=\textwidth]{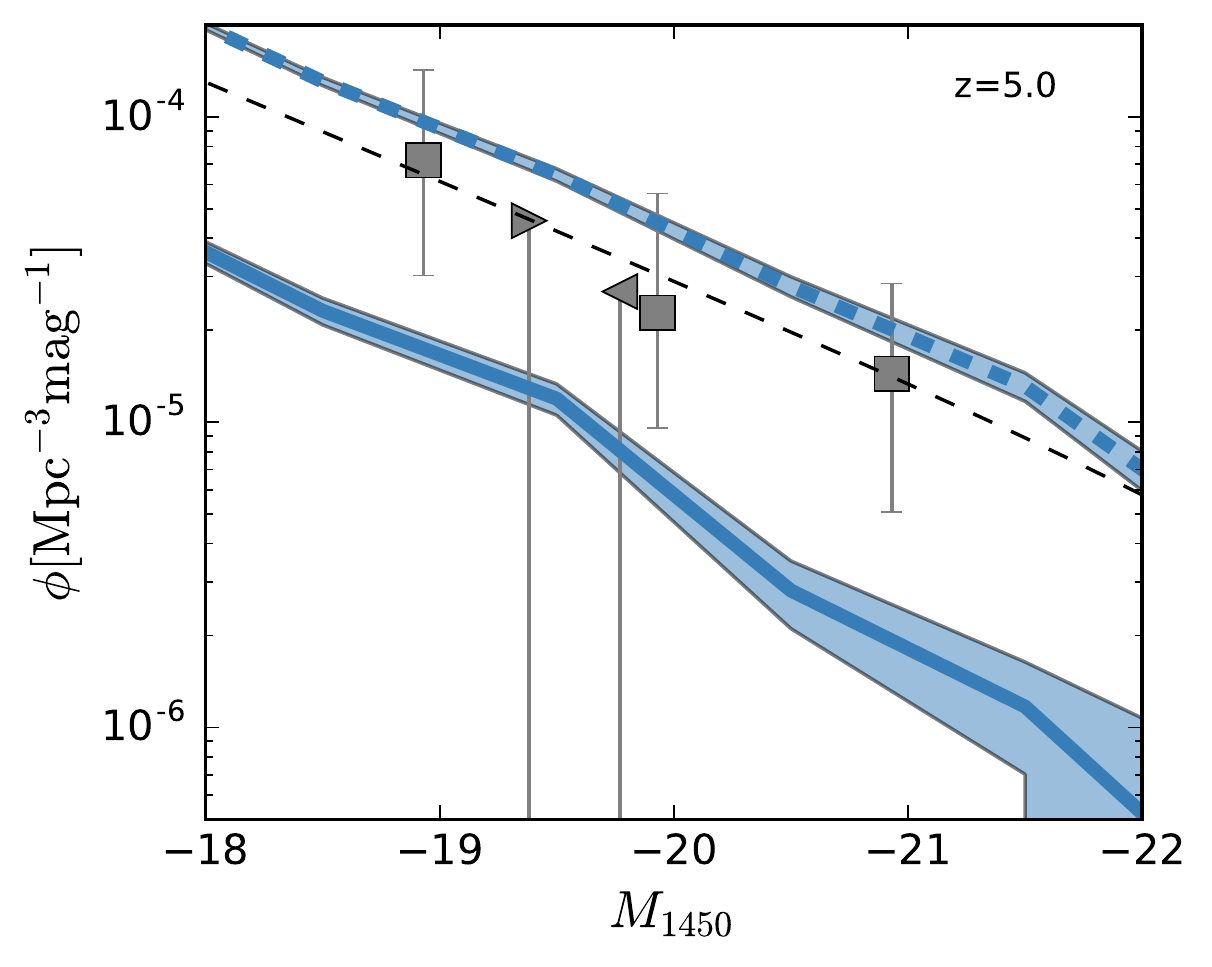}\\	\includegraphics[width=\textwidth]{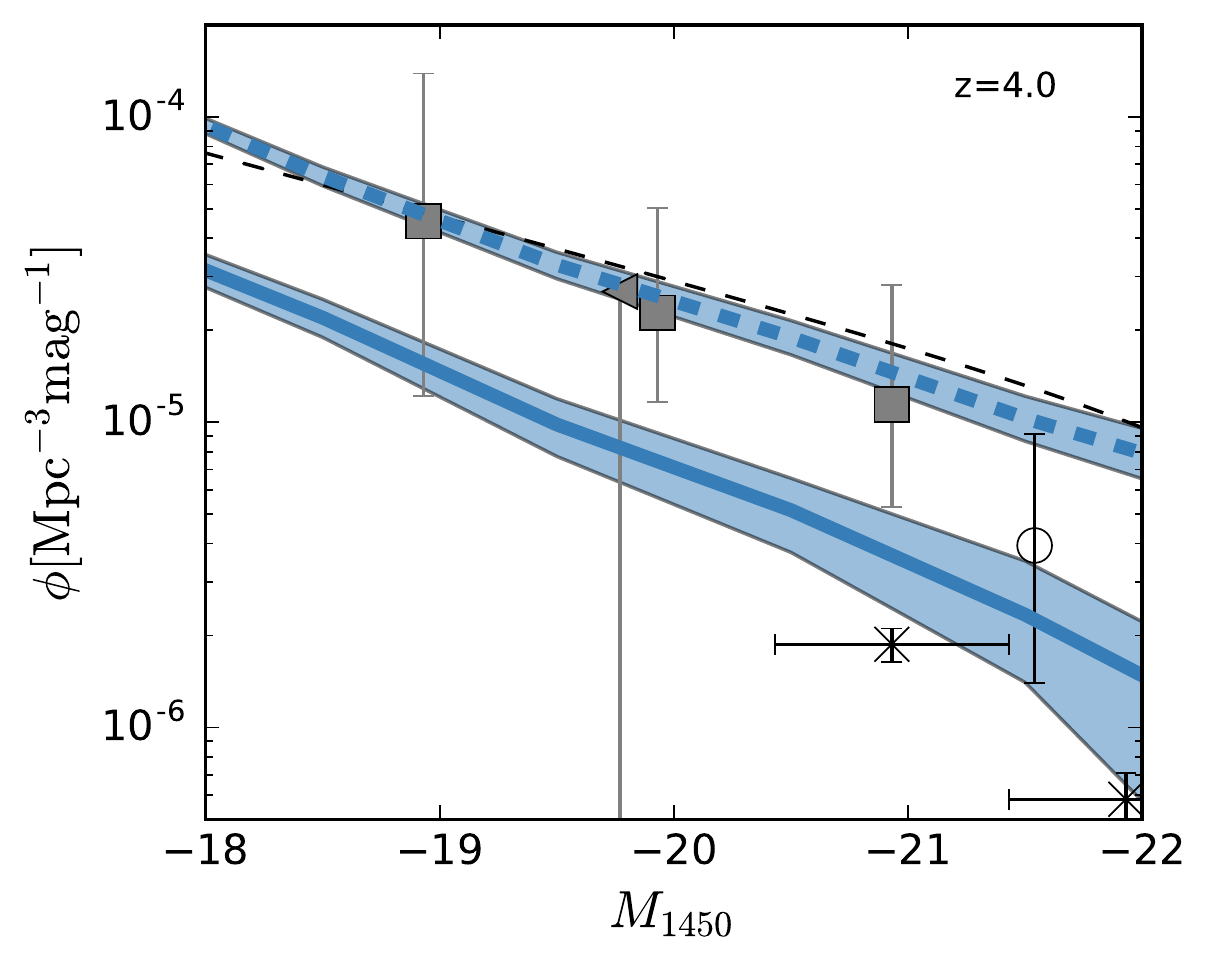}
		\caption{\label{fig:qglf} UV 1450 $\mathrm{\AA}$ luminosity functions of faint quasars at $z\sim6-4$ using the \textit{Tiamat} halo merger trees. Solid lines represent the luminosity functions calculated through the light only from AGN (same as in Fig. \ref{fig:BHMFandLF}) while dashed lines correspond to the calculation accounting for contributions from both AGN and their host galaxies. The shaded regions represent the 95 per cent confidence intervals around the mean using 100000 bootstrap re-samples. Observations are shown with different symbols (see the caption of Fig. \ref{fig:BHMFandLF}). In particular, the \citetalias{Giallongo2015} sample is shown with squares and the fitting function provided by \citetalias{Giallongo2015} is shown with thin black dashed lines.}
	\end{minipage}
\end{figure}

In order to account for this, we calculate the galaxy UV luminosity by integrating model SEDs based on the modelled star formation history \citep{Liu2016MNRAS.462..235L} and add the stellar light to mimic the observed total UV luminosity\footnote{Dust attenuation is not considered because these relatively faint galaxies have little dust in our models \citep{Liu2016MNRAS.462..235L}.} (AGN+galaxy). We present the resulting luminosity function of faint objects ($M_{1450}{\sim}{-}18$ to $-22$) in Fig. \ref{fig:qglf} with dashed lines. The result is calculated using the \textit{Tiamat} halo merger trees and is shown with shaded regions representing the 95 per cent confidence intervals around the mean using 100000 bootstrap re-samples. The luminosity function calculated using only the AGN light (AGN only, as shown in Fig. \ref{fig:BHMFandLF}) is shown as solid lines for comparison. We see that including stellar light can significantly increase the number density of faint AGN inferred at high redshift, by up to ${\sim}1$ order of magnitude. The fitting functions provided by \citetalias{Giallongo2015} as shown with thin black dashed lines in Fig. \ref{fig:qglf} are more consistent with the AGN+galaxy luminosity function. If this is the case, the estimated emissivity at high redshift based on \citetalias{Giallongo2015} is likely overestimated. 

\section{Reionization from quasars}\label{sec:reionization}

Motivated by the G15 sample, which suggests a numerous population of faint AGN at $z=4-6$ (\citetalias{Giallongo2015}), \citet{Madau2015} extrapolated the emissivity calculated from \citetalias{Giallongo2015} to higher redshifts and assessed a model of reionization, in which quasars are the dominant ionizing sources. They found that due to the high escape fraction of ionizing photons produced by those luminous objects, quasars are able to ionize the neutral hydrogen by $z{\sim}5.7$ if the high emissivity of quasars derived at $z{\sim}5$ continues to higher redshifts. Later, \citet{Mitra2015} revisited the model with a revised extrapolation and also found that quasars have a significant role during the EoR. However, their analysis still prefers models with a non-zero escape fraction of ${\sim}12$ per cent from galaxies. It is worth noting that with different formalism, \citet{Manti2017} fit the observed quasar UV luminosity function at $z=0.5-6.5$, including the G15 sample. They recalculated the emissivity by integrating the luminosity function and confirmed a large number of ionizing photons from quasars at high redshift using the Schechter luminosity function. However, the fitting result using a double power law presents a rapidly decreasing emissivity at $z>6$, suggesting that the extrapolated high-redshift quasar emissivity is strongly dependent on the assumed shape of the quasar luminosity function. Taking advantage of the {\sc Meraxes} semi-analytic model with 21{\sc cm}FAST \citep{Mesinger2011}, we investigate the contribution of quasars to reionization, within a frame work that accounts for black hole growth and feedback on star formation.\footnote{The \textit{Tiamat-125-HR} halo merger trees cannot resolve small objects; therefore we cannot consider ionizing photons from faint galaxies or quasars. We only show the result calculated using the \textit{Tiamat} trees in the following sections.}

\subsection{Reionization model}\label{sec:reionization model}

The semi-numerical reionization code 21{\sc cm}FAST \citep{Mesinger2011} uses an excursion set formalism to identify {\HII} bubbles in which the cumulative number of ionizing photons is more than the number of absorbing atoms:
\begin{equation}\label{eq:reionization_condition}
N_{*}N_{\gamma,*} f_\mathrm{esc,*} + N_\mathrm{q} N_{\gamma,q} f_\mathrm{obs} f_\mathrm{esc,q} \geq \left(1+\bar{N}_\mathrm{rec}\right) N_\mathrm{H_I},
\end{equation}
where $N_*$ and $N_\mathrm{q}$ are the numbers of baryons in stars and quasars,\footnote{The mass of black hole seed is subtracted because they do not produce any ionizing photons. Reionization from the progenitor of black hole seeds will be considered in the future when Population III stars are implemented.} $N_{\gamma,*}\sim4000$ \citep{Loeb2001} and $N_{\gamma,q}$ (see the calculation in Appendix \ref{app:Ngamma}) are the mean numbers of ionizing photons produced per baryon incorporated into the stellar or quasar components. The parameters $f_\mathrm{esc,*}$ and $f_\mathrm{esc,q}$ (see Table \ref{tab:parameters}) are the escape fractions of ionizing photons produced by stars and quasars. $f_\mathrm{obs}{\sim}0.234$ represents the observable fraction due to obscuration\footnote{One may also define $f_\mathrm{esc,q} f_\mathrm{obs}$ as the quasars escape fraction.} (see Section \ref{sec:QLF}). $N_\mathrm{H_I}$ is the cumulative number of atoms being ionized and $\bar{N}_\mathrm{rec}$ is the mean number of recombinations per baryon. Inhomogeneous recombinations are ignored, which can have a large impact \citep{Sobacchi2014}. In this work, $\bar{N}_\mathrm{rec}$ is set to be 0 as suggested by the high-redshift Ly$\alpha$ forest in the IGM \citep{Bolton2007,McQuinn2011}. Expanding equation (\ref{eq:reionization_condition}) gives
\begin{equation}
\dfrac{\xi_* m_* + \xi_\mathrm{q} m_\mathrm{bh}}{\dfrac{4}{3}\pi R^3 \Omega_\mathrm{m}\rho_\mathrm{c}\left(z\right)}  \geq 1,
\end{equation}
with
\begin{equation}
\xi_* = \dfrac{N_{\gamma,*} f_\mathrm{esc,*}}{f_\mathrm{b} (1-0.75Y_\mathrm{He})}\mathrm{\ and\ }
\xi_\mathrm{q} = \dfrac{N_{\gamma,q} f_\mathrm{obs} f_\mathrm{esc,q}}{f_\mathrm{b} (1-0.75Y_\mathrm{He})},
\end{equation}
where $m_*$ is cumulative stellar mass that excludes the loss due to supernova\footnote{The stellar mass recycled to the ISM through supernova also contributes ionizing photons in the {\HII} bubble.} and $R$ is the radius of the {\HII} bubble. $\xi_*$ and $\xi_\mathrm{q}$ are the {\HII} ionizing efficiencies for stars and quasars, $f_\mathrm{b}{=}\dfrac{\Omega_\mathrm{b}}{\Omega_\mathrm{m}}$ and $Y_\mathrm{He}{=}0.24$ are the fraction of baryons in the Universe and the fraction of helium, and $\rho_\mathrm{c}\left(z\right)$ is the comoving critical density of the universe, respectively. 

When the local volume around a galaxy is ionized, the UV background provides an extra heating mechanism, which modifies the baryonic fraction of the host halo (see $\chi_\mathrm{r}$ in equation \ref{eq:minfall}). Following \citet{Sobacchi2013}, when the virial mass, $M_\mathrm{vir}$ is smaller than a filtering mass, which can be calculated through
\begin{equation}
M_\mathrm{filt} = 2.8\times10^9\mathrm{M}_\odot J_{21}^{0.17}\left(\dfrac{1+z}{10}\right)^{-2.1}\left[1-\left(\dfrac{1+z}{1+z_\mathrm{ion}}\right)^2\right]^{2.5},
\end{equation}
the suppression of gas becomes significant ($\chi_\mathrm{r}\equiv2^{-{M_\mathrm{filt}}/{M_\mathrm{vir}}}$). Here, $z_\mathrm{ion}$ is the redshift when the local volume is first ionized, which is determined by the criteria given by equation (\ref{eq:reionization_condition}). $J_{21}$ represents the intensity of the local UV background. This can be calculated through
\begin{equation}\label{eq:J21}
J_{21} {=} \dfrac{3(1{+}z)^2}{8\pi^2 R^3 m_\mathrm{p}}\lambda_\mathrm{mfp}h \left(\alpha_*f_\mathrm{esc,*} N_{\gamma,*}\dot{m}_*{+} \alpha_\mathrm{q} f_\mathrm{obs} f_\mathrm{esc,q} N_{\gamma,q}\dot{m}_\mathrm{q}\right),
\end{equation}
where $h$ is the Planck constant. $\lambda_\mathrm{mfp}$ is the mean-free path of ionizing photons, which is approximated by the {\HII} bubble radius, $R$. The parameters $\alpha_*=5.0$ \citep{Loeb2001} and $\alpha_\mathrm{q}=1.57$ (\citetalias{Giallongo2015}) are the spectral indexes for a stellar-driven and a quasar-driven spectra in the UV band. $\dot{m}_*$ and $\dot{m}_\mathrm{q}$ are the growth rates of stellar mass and black hole mass, respectively. 

\subsection{Reionization history}

\begin{figure}
	\includegraphics[width=\columnwidth]{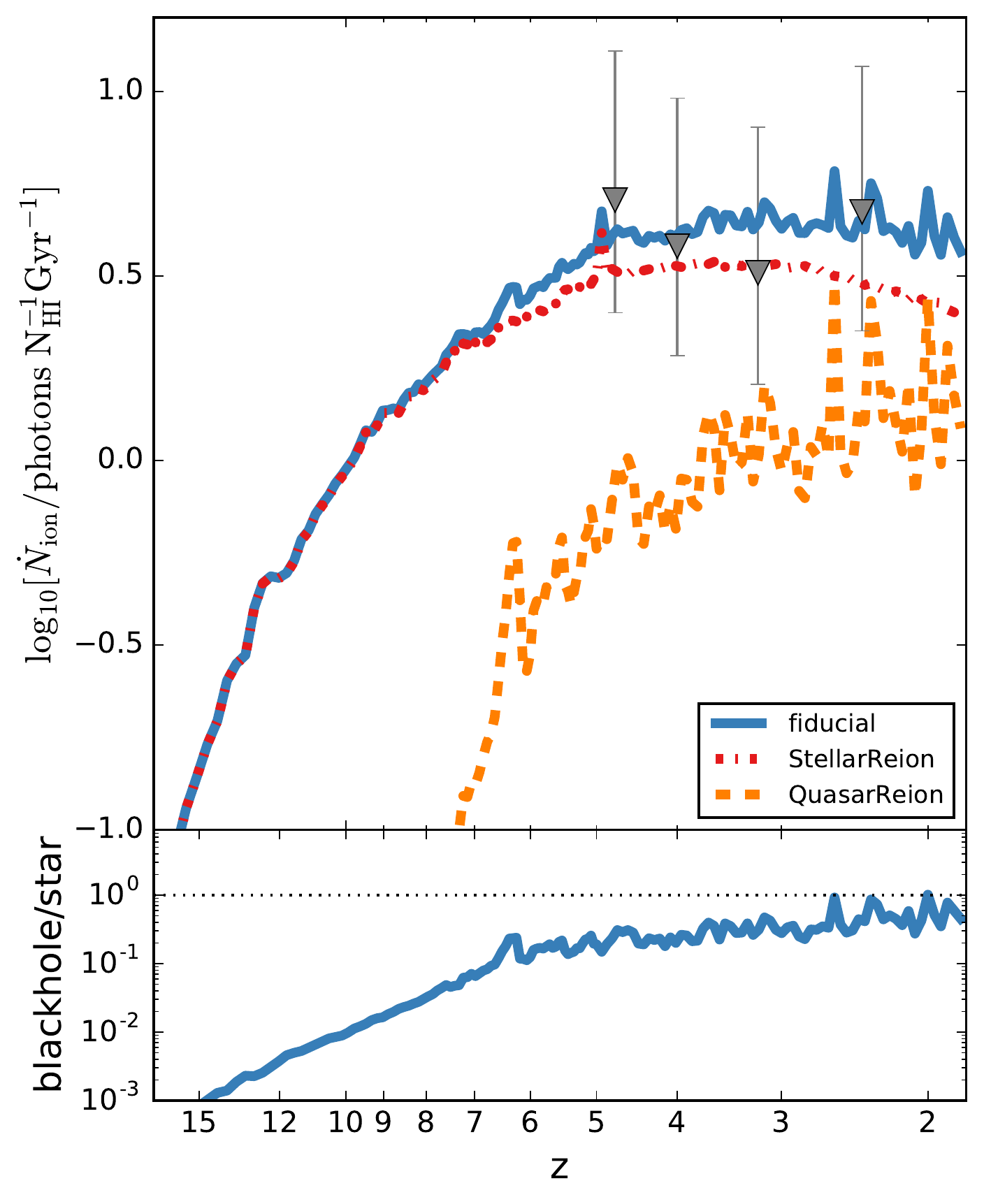}
	\caption{\label{fig:Nion} {\textit{Top panel:}} the evolution of the total 
		ionizing emissivity in units of number of ionizing photons per Hydrogen atom per Gyr for the fiducial model ({\color{colorfiducial}{\hdashrule[0.6mm]{6mm}{2.5pt}{}}}), the StellarReion model ({\color{colorM16}{\hdashrule[0.6mm]{6mm}{2.5pt}{1.5mm 0.4mm 0.4mm 0.4mm}}}) in which stars are the only ionizing photon source, and the QuasarReion model ({\color{colorQuasarReion}{\hdashrule[0.6mm]{6mm}{2.5pt}{1.5mm 0.5mm}}}) where quasars are the only reionization contributor. The total emissivity from \citetalias{Becker2013} are indicated with grey triangles ({\color{gray}{$\blacktriangledown$}}). \textit{Bottom panel:} the ratio of emissivities from black holes to the stellar component in the fiducial model.}
\end{figure}

\begin{figure}
	\includegraphics[width=\columnwidth]{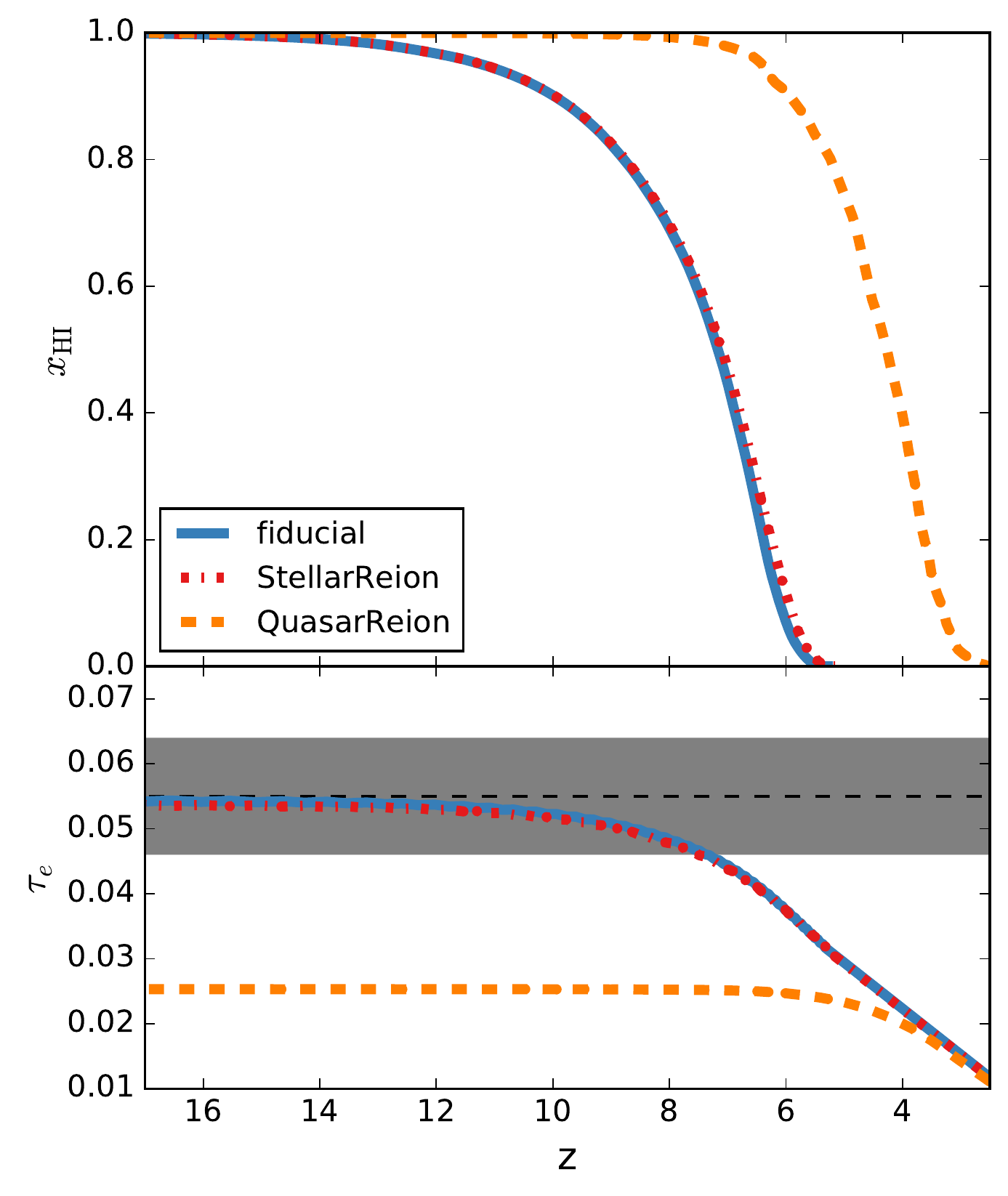}
	\caption{\label{fig:xH} {\textit{Top panel:}} the evolution of the mass-weighted global neutral hydrogen fraction for the fiducial model ({\color{colorfiducial}{\hdashrule[0.6mm]{6mm}{2.5pt}{}}}), the StellarReion model ({\color{colorM16}{\hdashrule[0.6mm]{6mm}{2.5pt}{1.5mm 0.4mm 0.4mm 0.4mm}}}) in which stars are the only ionizing photon source, and the QuasarReion model where quasars are the only reionization contributor ({\color{colorQuasarReion}{\hdashrule[0.6mm]{6mm}{2.5pt}{1.5mm 0.5mm}}}). {\textit{Bottom panel:}} the Thomson scattering optical depth as a function of redshift. The dashed line with shaded regions indicates the Planck 2016 measurement with $1\sigma$ uncertainties (\citetalias{PlanckCollaboration2016}).}
\end{figure}

Fig. \ref{fig:Nion} shows the instantaneous ionizing emissivity averaged over the entire \textit{Tiamat} simulation volume from different models as a function of redshift
\begin{equation}
\dot{N}_\mathrm{ion,*} = \dfrac{N_{\gamma,*} f_\mathrm{esc,*}\dot{m}_\mathrm{*,tot}}{f_\mathrm{b} (1-0.75Y_\mathrm{He})\rho_\mathrm{c}V_\mathrm{tot}},
\end{equation}
and 
\begin{equation}
\dot{N}_\mathrm{ion,q} = \dfrac{N_{\gamma,q} f_\mathrm{obs} f_\mathrm{esc,q}\dot{m}_\mathrm{bh,tot}}{f_\mathrm{b} (1-0.75Y_\mathrm{He})\rho_\mathrm{c}V_\mathrm{tot}},
\end{equation}
where $\dot{N}_\mathrm{ion,*}$, $\dot{N}_\mathrm{ion,q}$, $\dot{m}_\mathrm{*,tot}$ and $\dot{m}_\mathrm{q,tot}$ are the ionizing emissivities for the stellar component and quasars, and the total growth rates of stellar mass excluding the loss from supernovae and black holes in the model, respectively. From the Ly$\alpha$ opacity, several measurements of the total emissivity (AGN and stars) at high redshift have been estimated \citep{Bolton2007,McQuinn2011,Becker2013}. There are relatively large uncertainties in these measurements. In this work, we compare our models using the most recent data\footnote{The systematic error due to recombination radiation is ignored.} from \citet[][hereafter BB13]{Becker2013}. The top panel of Fig. \ref{fig:Nion} shows that the fiducial model agrees with the \citetalias{Becker2013} data and the bottom shows the ratio between ionizing photons from black holes and stars in the fiducial model, suggesting that during the EoR, quasars are subdominant in our model. The evolutions of the mass-weighted global neutral hydrogen fraction and the integrated Thomson scattering optical depth $\tau_e$ (see \citetalias{Mutch2016} for the calculation of $\tau_e$) are shown in Fig. \ref{fig:xH}. It shows that the fiducial model has a reasonable reionization history, with the mean global neutral hydrogen fraction decreasing from 90 per cent at $z\sim10$ to 0 by $z\sim6$ and a Thomson scattering optical depth in agreement with the latest Planck limits (\citealt[][hereafter \citetalias{PlanckCollaboration2016}]{PlanckCollaboration2016}).

There are two additional models in the top panel of Fig. \ref{fig:Nion} as well as in Fig. \ref{fig:xH}: StellarReion and QuasarReion. The ionizing source in the StellarReion model is only stars, with $f_\mathrm{esc,*} = \min\left[0.06\times\left(\dfrac{1+z}{6}\right)^{0.5},1.0\right]$ and $f_\mathrm{esc,q} = 0$, while quasars are the only reionization contributor in the QuasarReion model, with $f_\mathrm{esc,q} = 1$ and $f_\mathrm{esc,*} = 0$ (see Table \ref{tab:parameters}). Note that only changing the feedback from reionization has little impact on the stellar mass function (see \citetalias{Mutch2016}), the quasar luminosity function or the Magorrian relation. By preventing gas infall, reionization only affects less massive objects in our model.

In Fig. \ref{fig:Nion}, we see that the emissivity of quasars grows rapidly in the QuasarReion model (also in the fiducial model) by a factor of 10 from $z\sim7-5$. However, if quasars are the only reionization contributor, even with $f_\mathrm{esc,q}=1$ the number of ionizing photons cannot reach the \citetalias{Becker2013} data. Moreover, due to the deficiency in the photon budget at high redshift, quasars can only start reionization at $z\sim6$ resulting in an end at $z\sim3$. Together with the predicted optical depth, our model rules out the quasar-only reionization scenario. We note that one may recalibrate the model with a more efficient black hole growth rate at high redshift (e.g. by incorporating a redshift dependence in the equation \ref{eq:bh,max}, see \citealt{Bonoli2009}), in order to match the G15 luminosity function and the estimated emissivity. \citet{Mitra2015} also suggest that if G15 emissivity is correct, quasar-only reionization is possible and it results in a small value of $\tau_e$ due to the rapid evolution of the Lyman-limit systems. However, simultaneously matching the model with the G15 faint AGN luminosity function and the other observations of bright systems at high redshift is difficult. For example, comparing to observations at $z\sim6$, our models produce a flatter luminosity function, which is more consistent with the bright quasar sample. This suggests that a mass-dependent black hole growth efficiency (e.g. $k_\mathrm{c}$, see Section \ref{sec:quasar mode}) would be required, in order to steepen the luminosity function and produce more small quasars. In the following sections, we explore the relative contribution of quasars to reionization based only on the presented black hole growth model.

Comparing the neutral hydrogen fraction and the optical depth between the fiducial and StellarReion models also suggests that quasars do not have a significant role in reionization in this model. Their contribution helps finish reionization earlier by $\Delta z{\lesssim}0.1$ and decreases the optical depth by less than 10 per cent.

\section{Discussion}\label{sec:discussion}
Our models are calibrated against the black hole -- galaxy scaling relation and quasar luminosity function, in order to reproduce a realistic AGN catalogue for the study of the contribution of quasars to hydrogen reionization. However, it has recently been suggested that the black hole sample used to derive scaling relations is likely different from the entire population, leading to a selection bias\footnote{If it is not a selection bias, it is possible that not every galaxy hosts a central massive black hole.} \citep{Bernardi2007}. For instance, using Monte Carlo simulations \citet{Shankar2016} recovered the intrinsic scaling relation assuming the selection bias comes from unresolved black holes (e.g. for galaxies with a given velocity dispersion, $\sigma_*$, it is more difficult to resolve smaller black holes; \citealt{Batcheldor2010}) and showed that such a bias can lead to factors of $\ge 3$ and ${\sim}50-100$ higher normalizations of the $M_\mathrm{bh}-\sigma_*$ and $M_\mathrm{bh}-M_*$ relations, respectively. We note that a Magorrian relation with a smaller normalization can be achieved in our model using a smaller black hole growth efficiency (i.e. $k_\mathrm{c}$ in equation \ref{eq:bh,max}). In this case, the reconstructed black hole population becomes less massive, more efficient AGN feedback (e.g. $k_\mathrm{h}$ in equation \ref{eq:bh,hot}) and a larger fraction of observable AGN in the UV band are required to simultaneously reproduce the observed stellar mass function and quasar luminosity function. With the model calibrated against the quasar luminosity function, the black hole -- galaxy scaling relation is coupled with the observable AGN fraction -- a lower Magorrian relation requires a larger fraction of observable AGN. We note that the total emissivity of quasars is integrated from the luminosity function. Therefore, scaling relations do not have a significant impact on reionization in this work. Noting the difficulty of observationally determining the fraction of obscured AGN and the large uncertainties in the black hole -- galaxy scaling relation, in this section we use a range of models which predict similar Magorrian relations as shown in Fig. \ref{fig:BHMFandMagorrian} and explore the contribution of quasars to reionization.

\subsection{A larger opening angle}

We have shown that with an opening angle of 80 deg, the model is able to reproduce the observed quasar luminosity function from $z{\sim}6-0.6$ (see Fig. \ref{fig:BHMFandLF}). However, at high redshift, the model predicts significant stellar contribution to UV flux in the G15 sample (see Fig. \ref{fig:qglf}), and consequently less ionizing photons from AGN. Based on this, we find that quasars do not have a significant role during EoR. 

\begin{figure}
	\includegraphics[width=\columnwidth]{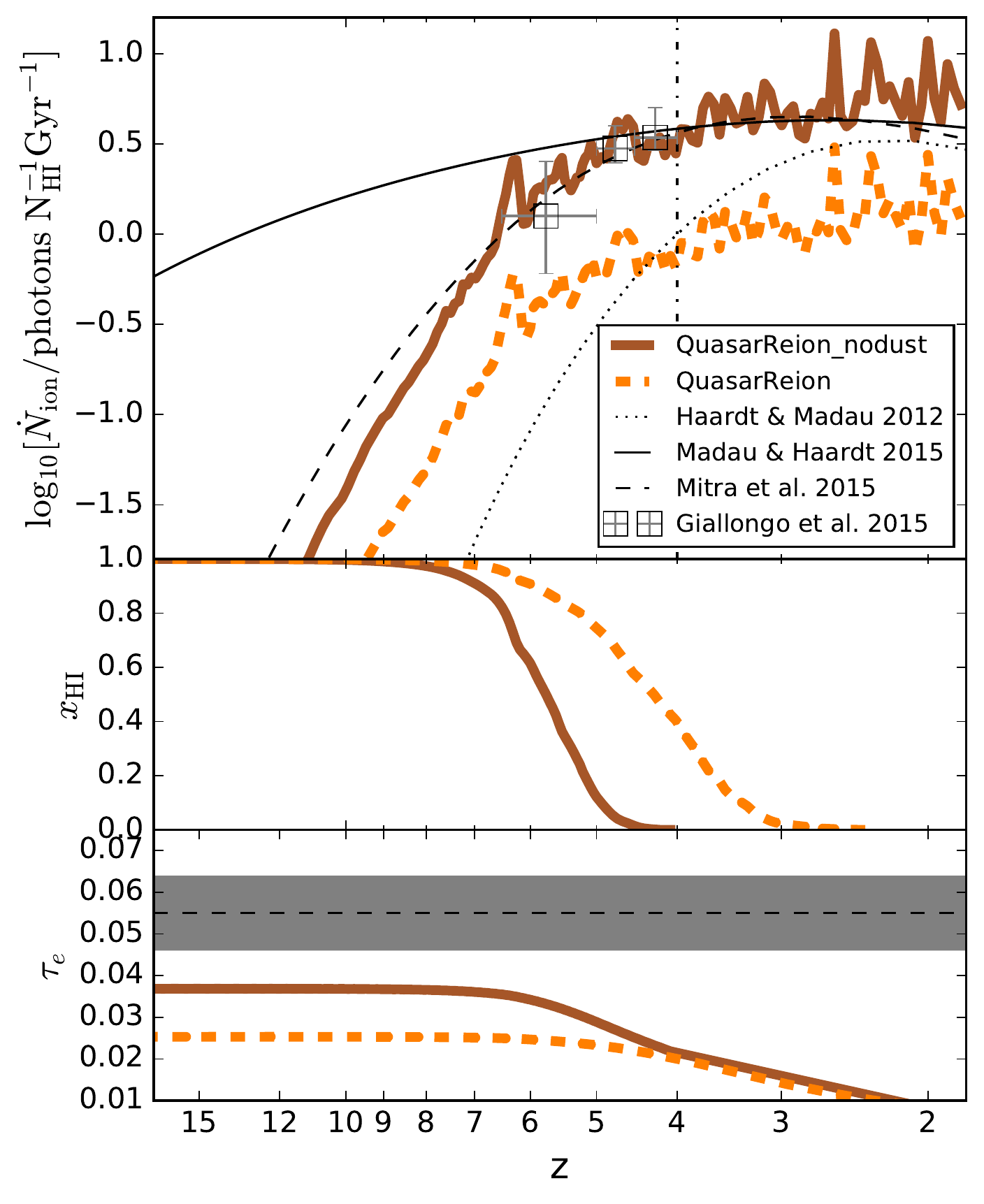}
	\caption{\label{fig:dot_Nion_G15} {\textit{Top panel:}} the evolution of ionizing emissivity for the QuasarReion ({\color{colorQuasarReion}{\hdashrule[0.6mm]{6mm}{2.5pt}{1.5mm 0.5mm}}}) and QuasarReion\_nodust models ({\color{colorQuasarReionnodust}{\hdashrule[0.6mm]{6mm}{2.5pt}{}}}). The estimation from \citetalias{Giallongo2015} is shown as squares while models from \citet{Haardt2012,Madau2015} and \citet{Mitra2015} are indicated using black dotted, solid and dash--dotted lines, respectively. The vertical dash--dotted line represents the redshift when the model starts underestimating the quasar emissivity. {\textit{Middle panel:}} the evolution of the mass-weighted global neutral hydrogen fraction. {\textit{Bottom panel:}} the Thomson scattering optical depth as a function of redshift. The dashed line with shaded region indicates the Planck 2016 measurement with $1\sigma$ uncertainties (\citetalias{PlanckCollaboration2016}).}
\end{figure}

In the top panel of Fig. \ref{fig:dot_Nion_G15}, we show the estimated total emissivity from \citetalias{Giallongo2015} with the model proposed by \citet{Haardt2012,Madau2015} and \citet{Mitra2015}. The modelled emissivity from QuasarReion is shown for comparison. Our quasar reionization-only model (QuasarReion) predicts lower emissivities compared to these two estimations, with only a third of the G15 value at $z{\sim}6$. Consequently, in disagreement with \citet{Madau2015} and \citet{Mitra2015} we conclude that quasars cannot be the dominate sources during reionization. We could increase the emissivity by excluding the obscuration from dust (setting $\theta=180$, shown as QuasarReion\_nodust in Fig. \ref{fig:dot_Nion_G15}), which gives a closer emissivity compared to the G15 estimation and in agreement with the model proposed by \citet{Mitra2015}. In this model, quasars have a more significant role during the EoR and can reionize the IGM alone by $z\sim4.5$. However, this model overestimates the number density of bright and low-redshift AGN, leading to an incorrect evolution of the quasar luminosity function. A lower fraction of observable AGN, $f_\mathrm{obs}$, towards brighter luminosities and lower redshifts is required to solve this conflict. However, observations suggest the opposite trend in optical, infrared and X-ray bands \citep{Hopkins2007} and more constraints are required to clearly establish a $f_\mathrm{obs}-z$ relation.

\begin{figure*}	
	\begin{tabular}{lcr}
		\begin{minipage}{0.29\textwidth}
			\subfigure{\includegraphics[width=\textwidth]{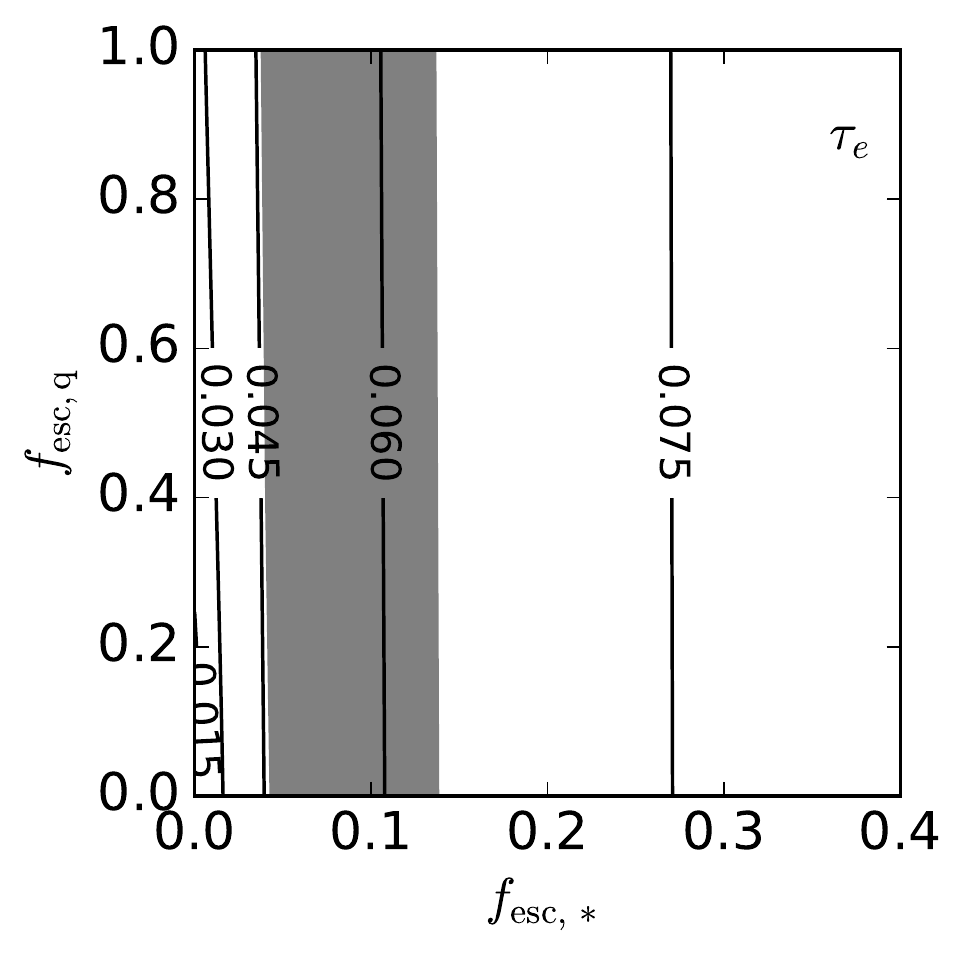}} 
		\end{minipage}
		\begin{minipage}{0.29\textwidth}
			\subfigure{\includegraphics[width=\textwidth]{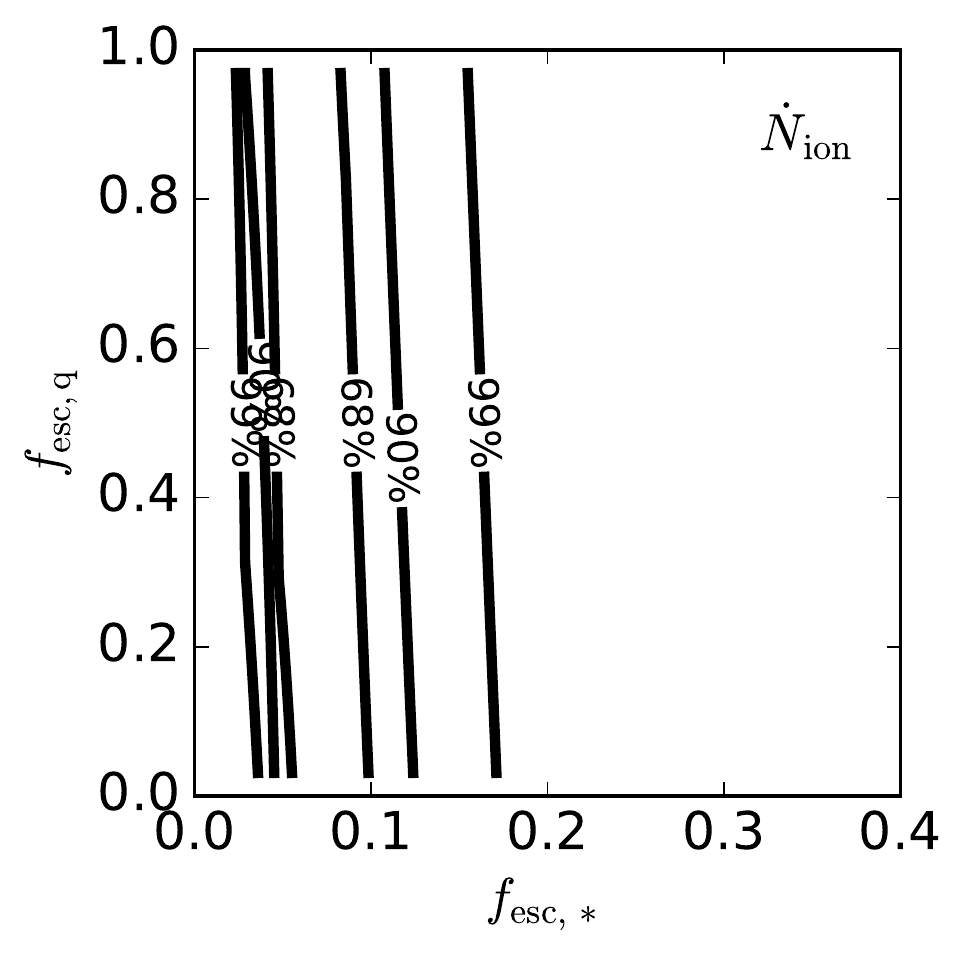}}
		\end{minipage}
		\begin{minipage}{0.42\textwidth}
			\subfigure{\includegraphics[width=\textwidth]{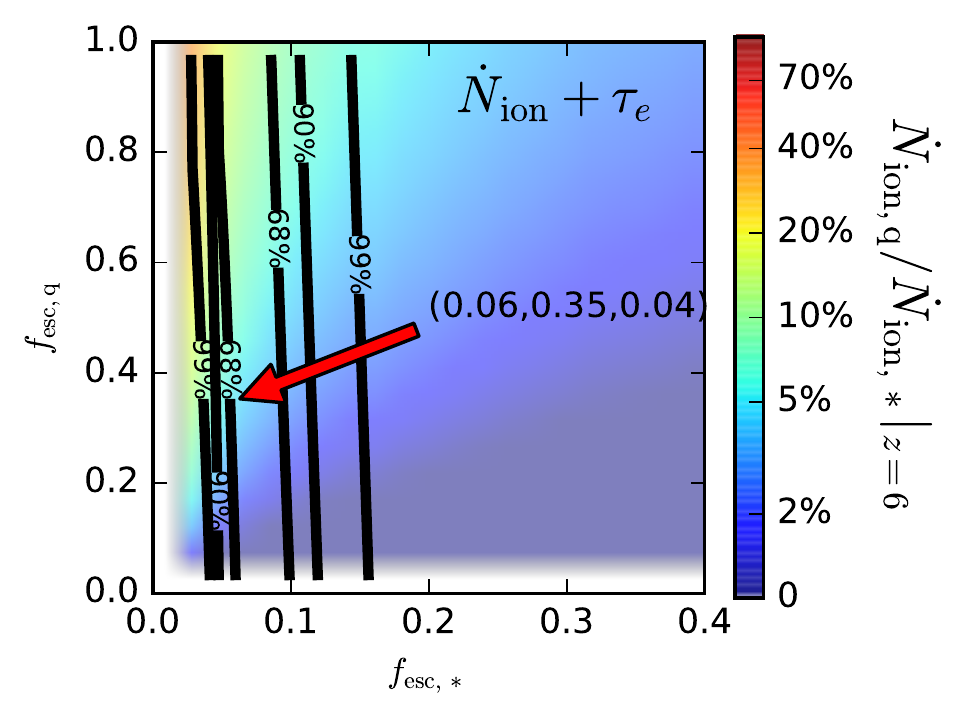}}
		\end{minipage}
	\end{tabular}\\	\vspace*{-5mm}
	\begin{tabular}{lcr}
		\begin{minipage}{0.29\textwidth}
			\subfigure{\includegraphics[width=\textwidth]{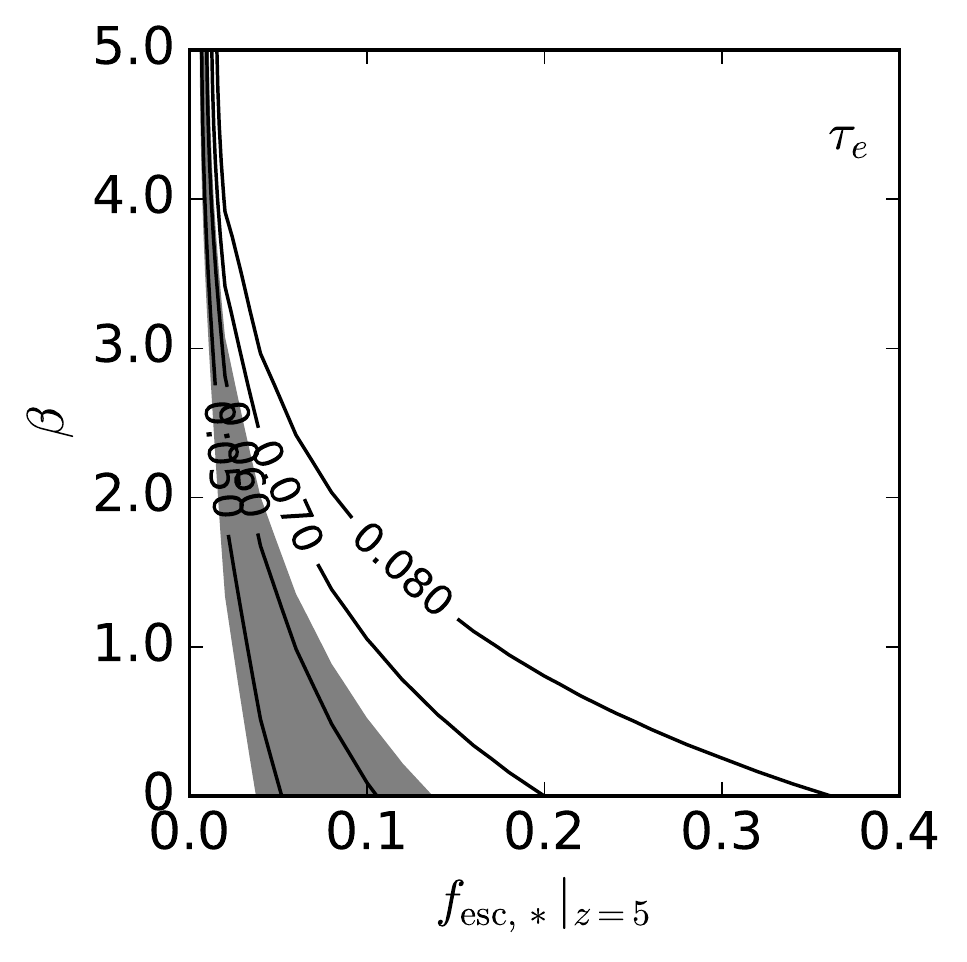}} 
		\end{minipage}
		\begin{minipage}{0.29\textwidth}
			\subfigure{\includegraphics[width=\textwidth]{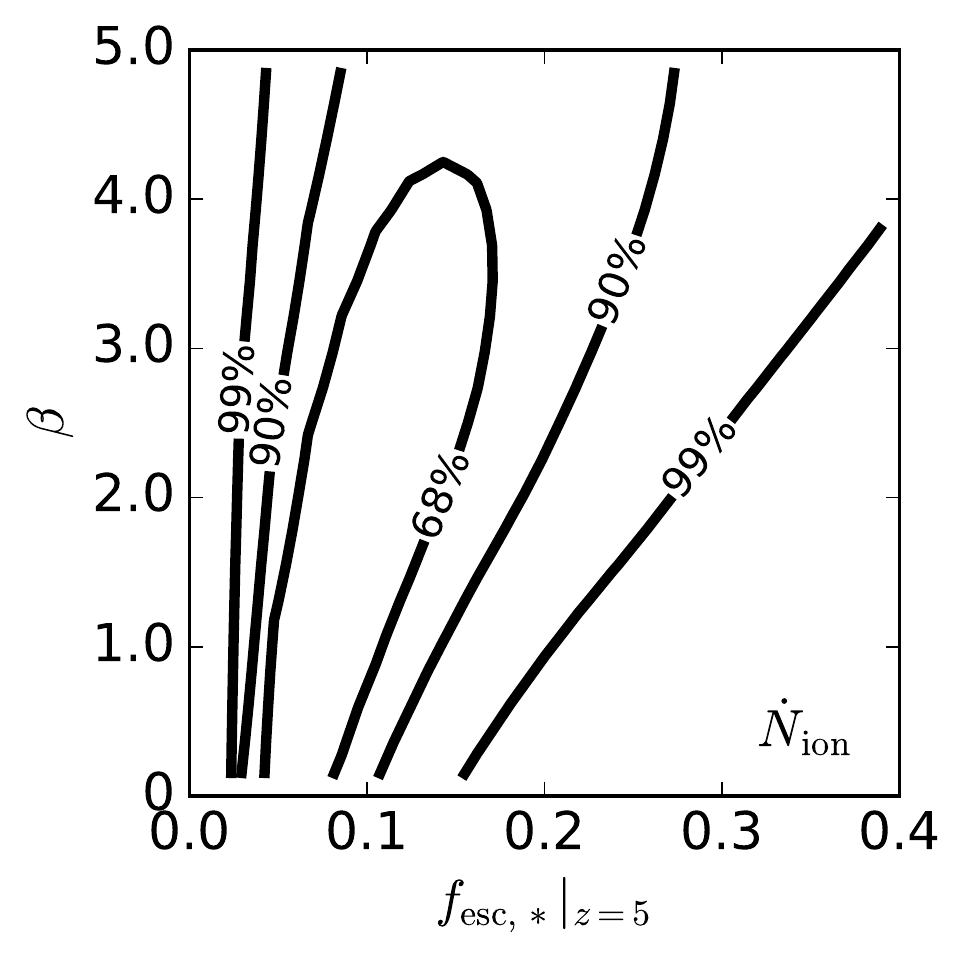}}
		\end{minipage}
		\begin{minipage}{0.4\textwidth}
			\subfigure{\includegraphics[width=\textwidth]{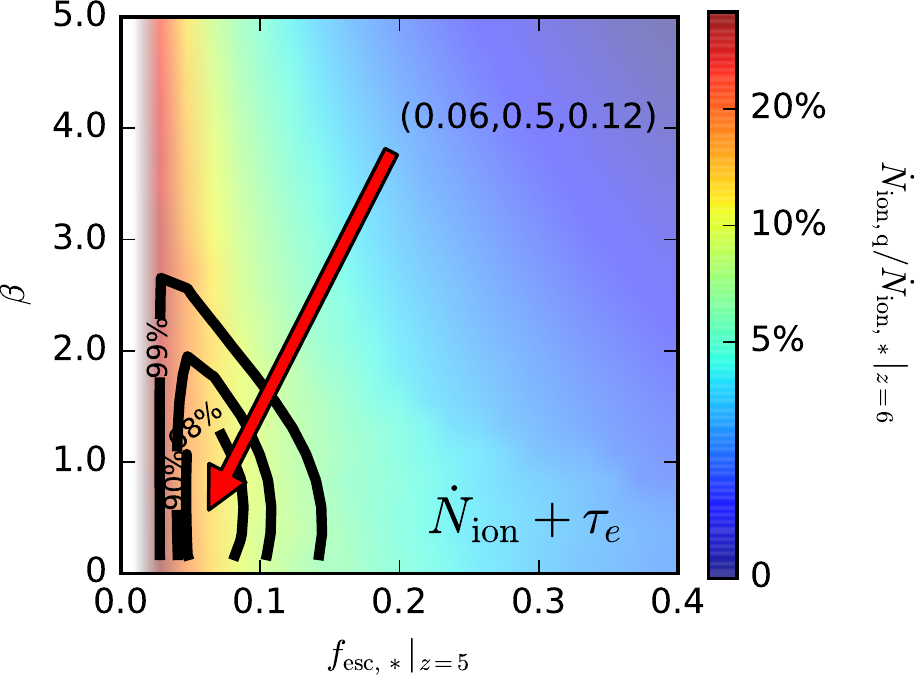}}
		\end{minipage}
	\end{tabular}\\
	\caption{\label{fig:varying} \textit{Left panel:} the Thomson scattering optical depth $\tau_e$. The measurements with $1\sigma$ uncertainties by \citetalias{PlanckCollaboration2016} are shown in the shaded region. \textit{Middle panel:} the 68, 90 and 99 per cent confidence limits on each parameter of the best fit via the minimum-$\chi^2$ technique based on the estimated total emissivities and their errors from \citet{Becker2013} at $z\sim2-5$. \textit{Right panel:} the 68, 90 and 99 per cent confidence limits based on \citetalias{Becker2013} and \citetalias{PlanckCollaboration2016}. The 2D histogram shows the distribution of the ratios of quasar emissivity to stellar emissivity at $z\sim6$. The position of the minimum $\chi^2$ value and the corresponding value of the 2D histogram are indicated with arrows. \textit{Top panel:} the results as functions of the escape fractions of stars, $f_\mathrm{esc,*}$ and quasars, $f_\mathrm{esc,q}$. \textit{Bottom panel:} the results as functions of $f_\mathrm{esc,*}|_{z=5}$ and $\beta$, which are the normalization and scaling of the redshift-dependent stellar escape fraction, $f_\mathrm{esc,*}(z) = \min\left[f_\mathrm{esc,*}|_{z=5}\times\left(\dfrac{1+z}{6}\right)^{\beta},1\right]$. In these models, the quasar escape fraction is assumed to be $f_\mathrm{esc,q}= 1$.}
\end{figure*}

\subsection{Escape fraction of ionizing photons from quasars}\label{sec:discuss:fesaq}

In this section, we explore possible combinations of stars and quasars that could result in an overall photon budget at $z>5$ consistent with the observed optical depth and ionizing flux at $z\sim2-5$. In the Section \ref{sec:reionization}, as well as in \citetalias{Mutch2016} where only galaxies are considered, we demonstrated the requirement of an evolving escape fraction for stars to explain the observed emissivity at $z\sim5$. Noting this requirement, in this section we assume constant escape fractions both for simplicity and to ease interpretation.

Motivated by the recent claim that the escape fraction of low-luminosity AGN is possibly less than unity at high redshift \citep{Micheva2016}, we run {\sc Meraxes} with different combinations of $f_\mathrm{esc,*}$ and $f_\mathrm{esc,q}$, without any changes to the other parameters.  In the top panel of Fig. \ref{fig:varying}, the left-hand panel shows the Thomson scattering optical depth, $\tau_e$. For comparison, shaded regions are shown corresponding to the best fit and 1$\sigma$ range of the \citetalias{PlanckCollaboration2016} measurements. Based on the \citetalias{Becker2013} data at $z\sim2-5$ and the \citetalias{PlanckCollaboration2016} measurement, the top right two panels show the 68, 90 and 99 per cent confidence contours on each parameter of the best fit via the standard minimum-$\chi^2$ technique. The 2D histogram shows the distribution of the ratios of quasar emissivity to stellar emissivity at $z\sim6$. We see that a lower escape fraction of ionizing photons from stars, $f_\mathrm{esc,*}$ requires a higher contribution from quasars, in order to reach the observational constraint. This also returns a higher ratio of quasar emissivity to stellar emissivity. However, because there is not a significant number of quasars at high redshift, changing $f_\mathrm{esc,q}$ has little impact to the optical depth. Through the best fitting contours, we see that if the escape fraction of ionizing photons from stars is only a few percent ($<5$ per cent; \citealt{Ciardullo2014,Matthee2016b}), the model requires $f_\mathrm{esc,q}\sim1.0$. 

\subsection{An evolving escape fraction}

Although the escape fraction of ionizing photons depends on the local environment, many theoretical and observational works suggest an evolving or mass-dependent escape fraction with a decreasing average value at lower redshifts or in more massive galaxies \citep{Kuhlen2012,Haardt2012,Paardekooper2013,Kimm2014,Wise2014,Bauer2015,Price2016}. Therefore, as discussed in \citetalias{Mutch2016}, in order to simultaneously match the normalization and flat slope of the observed ionizing emissivity at $z\lesssim6$, and the Planck $\tau_e$ measurements,\footnote{\citetalias{Mutch2016} constrained the model using the observed emissivities from \citet{McQuinn2011}, which are lower than the \citetalias{Becker2013} data, and the \citetalias{PlanckCollaboration2015} result, which has a larger optical depth than the \citetalias{PlanckCollaboration2016} data.} a redshift-dependent escape fraction for galaxies was proposed:
\begin{equation}
f_\mathrm{esc,*}(z) = \min\left[f_\mathrm{esc,*}|_{z=5}\times\left(\dfrac{1+z}{6}\right)^{\beta},1\right].
\end{equation}
We have shown two models with evolving escape fractions (fiducial and StellarReion) in Section \ref{sec:reionization}. In this section, we further explore the possible evolution of the escape fraction by running the semi-analytic model with different combinations of $f_\mathrm{esc,*}|_{z=5}$ and $\beta$. Note that all of the ionizing photons from quasars are included in this section ($f_\mathrm{esc,q}=1$) in order to investigate the evolution of the stellar escape fraction with the contribution of quasars to reionization. 

The bottom panels of Fig. \ref{fig:varying} show the optical depth and the best-fitting confidence limits as functions of the normalization of $f_\mathrm{esc,*}(z)$, $f_\mathrm{esc,*}|_{z=5}$ and the scaling, $\beta$, when $f_\mathrm{esc,q}= 1$. We see that because a larger scaling suppresses the escape fraction at lower redshifts, which results in a lower emissivity at $z\lesssim 5$, a larger normalization is required. In addition, a larger $\beta$ gives less ionizing photons at $z\sim 8$, which slows the process of reionization and consequently increases the optical depth. When $f_\mathrm{esc,*}|_{z=5}$ reaches 0, the model becomes quasar-dominated, returning a low $\tau_e$ of around a half of the \citetalias{PlanckCollaboration2016} measurement (see Fig. \ref{fig:xH}). We see that when including the contribution from quasars, the model prefers a combination of $f_\mathrm{esc,*}|_{z=5}\sim6$ per cent and $\beta\sim0.5$. This corresponds to $f_\mathrm{esc,*}\sim6.5$ per cent at $z=6$ with a ratio between the emissivities of quasars and stars of $\sim0.12$.

\subsection{Faintest and brightest quasar contributors}\label{sec:discuss:Mlim}

\begin{figure}
	\includegraphics[width=\columnwidth]{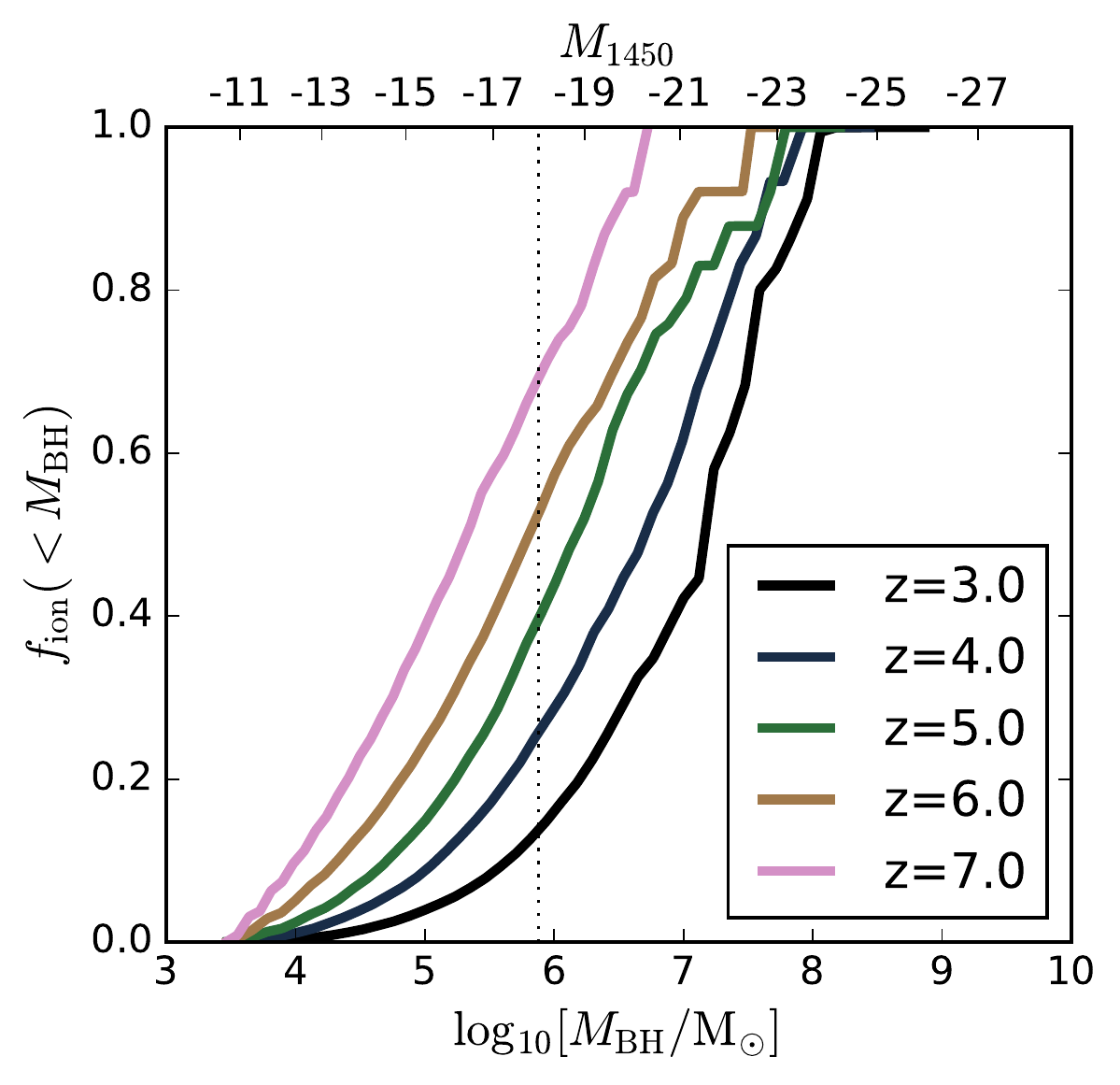}
	\caption{\label{fig:fion} The cumulative fraction of ionizing photons as a function of black hole mass (or the corresponding UV magnitude $M_{1450}$ during the Eddington state as shown in the top axis) at $z{\sim}7-3$ in the \textit{Tiamat} volume. The vertical dotted line represents the faintest AGN detected by \citetalias{Giallongo2015}.}
\end{figure}

In addition to the possibility that quasars do not have a very high escape fraction \citep{Barkana2000}, we note that the very faintest quasars predicted in the model are not observed. The recent detection by \citetalias{Giallongo2015} only reaches to $M_{1450}{\sim}{-}18$, while our fiducial models predict a significant population of faint quasars down to $M_\mathrm{1450}{\sim}{-}11$. Whether those undetected quasars are able to contribute a significant amount of ionizing photons is still unknown. For instance, they might be buried in the dust with a large obscuration fraction \citep{Hopkins2007}. The critical mass above which quasars can contribute ionizing photons, coupled with the previously discussed escape fraction, represents limiting cases of a mass-dependent escape fraction for quasars. In the top panel of Fig. \ref{fig:fion}, we present the cumulative fraction of ionizing photons as a function of black hole mass (or the corresponding UV magnitude $M_{1450}$ during the Eddington state as shown in the top axis) assuming $f_\mathrm{esc,q}=1$ from the fiducial model using the \textit{Tiamat} simulation. We see that quasars fainter than $M_{1450}=-18$ contribute approximately 80 per cent of the total emissivity at $z{\sim}7$ with a decreasing contribution towards lower redshifts (10 per cent at $z{\sim}3$, the end of the EoR in the QuasarReion model). This suggests that the number of fainter quasars becomes relatively smaller at later times, which can also be observed from the slope of the predicted quasar luminosity function becoming flatter from $z{\sim}7-3$ (see Fig. \ref{fig:BHMFandLF}). However, at redshifts higher than $z{\sim}6$, the total emissivity from quasars is low. For example, the total emissivity at $z{\sim}7$ is five times lower than $z{\sim}5$, suggesting that faint quasars below current observational limits provide only a small contribution to reionization. 

In addition, the AGN light curve adopted in this work, which assumes that black holes are either accreting with $\epsilon=1$ or stay quiescent ($\epsilon=0$) depending on the amount of accretion mass (see Section \ref{sec:quasar mode}), has been shown to underestimate the number density of faint AGN at low redshift \citep{Bonoli2009}. For instance, allowing $\epsilon$ to decrease progressively when the accretion disc has been mostly consumed predicts more faint AGN with $M_\mathrm{B}\sim-16$ by a factor of 2 at $z\sim0.1$. However, the impact becomes insignificant at brighter ranges and higher redshifts. Due to the small contribution of ionizing photons from faint quasars, AGN light curves are therefore not expected to have a significant impact on our conclusions regarding reionization.

On the other hand, due to the limited simulation volume the brightest quasars at high redshift ($z\ge4$) in our model only reach $M_\mathrm{1450}{\sim}{-}23$, above which the contribution of ionizing photons is not considered. However, the G15 emissivity accounts for bright quasars up to $M_{1450}{=}{-}28$, 100 times brighter than the brightest quasar in our model. In order to estimate the emissivity of the missing bright quasars, we integrate the fitting functions provided by \citetalias{Giallongo2015} with a magnitude interval of $-28<M_{1450}<-23$. We find that the total emissivity at high redshift increases by less than 1 per cent with the inclusion of the ionizing photons from these quasars. Therefore, the conclusion that quasars do not have a significant role during the EoR is not affected by the volume size. However, with the flattening luminosity function at lower redshifts ($z<4$, indicated with the vertical dash--dotted line in Fig. \ref{fig:dot_Nion_G15}), bright quasars become more important and their contribution to reionization is not ignorable.\footnote{We fit the predicted quasar luminosity function at $z=2$ using a single power law at $M_{1450}>-24$ and estimate the number of missing photons with a magnitude interval of $-24<M_{1450}<-22.5$. We find the emissivity of quasars can be increased by a factor of 2 with the inclusion of the missing bright quasars.} This results in a lower emissivity of quasars in our model compared to the \citet{Haardt2012} model at low redshift (see Fig. \ref{fig:dot_Nion_G15}). We note that including these objects will bring forward reionization in the QuasarReion model, but have no impact to the fiducial model.

\subsection{Black hole seed mass}

\begin{figure}
	\includegraphics[width=\columnwidth]{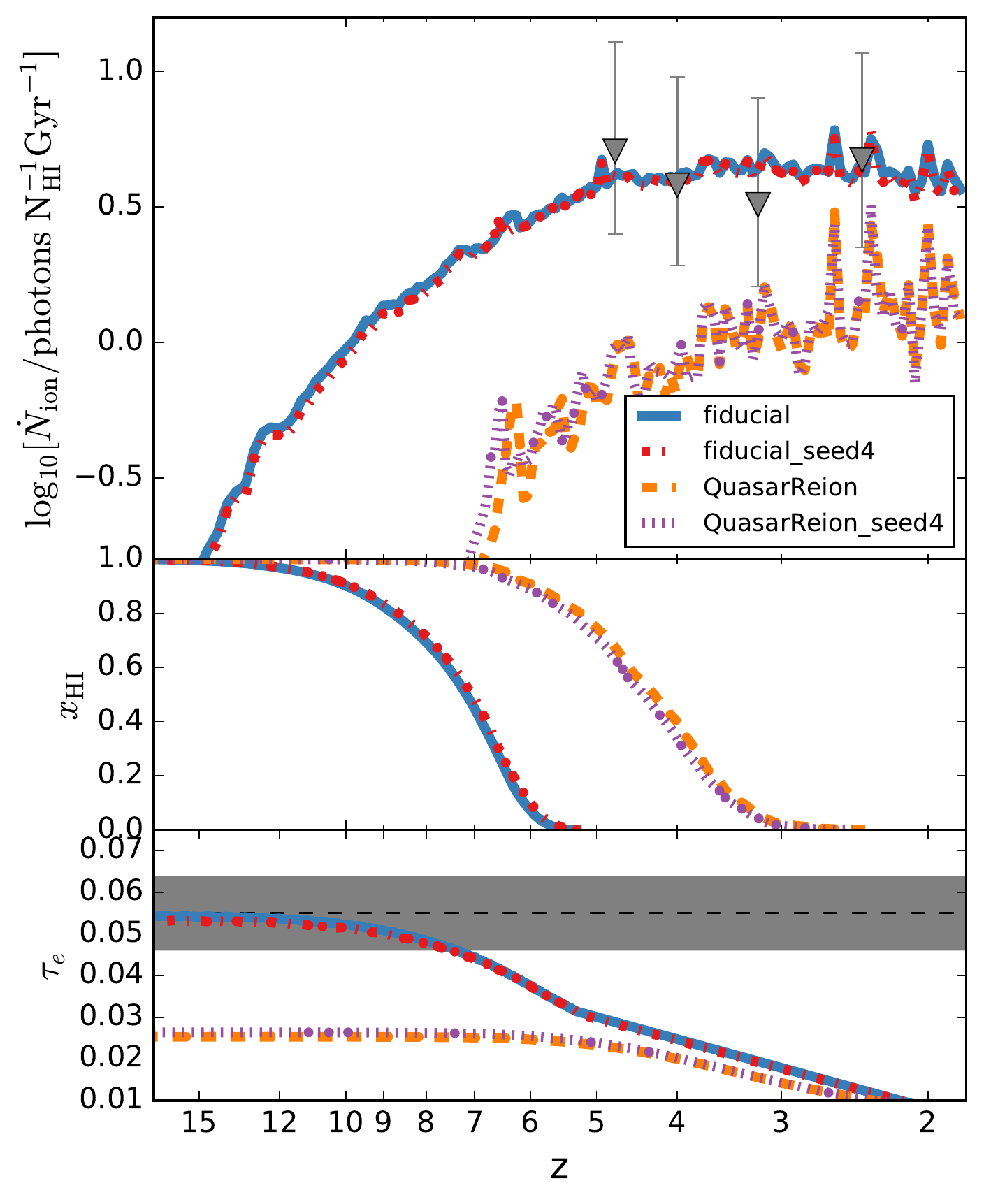}
	\caption{\label{fig:dot_Nion_seed4} {\textit{Top panel:}} the evolution of ionizing emissivity for the fiducial models with black hole seed masses of $10^3\ h^{{-}1}\mathrm{M}_\odot$ ({\color{colorfiducial}{\hdashrule[0.6mm]{6mm}{2.5pt}{}}}) and $10^4\ h^{{-}1}\mathrm{M}_\odot$ ({\color{colorM16}{\hdashrule[0.6mm]{6mm}{2.5pt}{1.5mm 0.4mm 0.4mm 0.4mm}}}), and for the QuasarReion models with $10^3\ h^{{-}1}\mathrm{M}_\odot$  ({\color{colorQuasarReion}{\hdashrule[0.6mm]{6mm}{2.5pt}{1.5mm 0.5mm}}}) and $10^4\ h^{{-}1}\mathrm{M}_\odot$ ({\color{colorstrongQuasar}{\hdashrule[0.6mm]{6mm}{2.5pt}{0.4mm 0.4mm 0.4mm 0.4mm}}}) seed masses. The total emissivity from \citetalias{Becker2013} are indicated with grey triangles. {\textit{Middle panel:}} the evolution of the mass-weighted global neutral hydrogen fraction. {\textit{Bottom panel:}} the Thomson scattering optical depth as a function of redshift. The dotted and dashed lines with shaded regions indicate the \citetalias{PlanckCollaboration2016} measurement (\citetalias{PlanckCollaboration2016}).}
\end{figure}

Our choice of black hole seed mass, $1000\ h^{{-}1}\mathrm{M}_\odot$ lies between the light seed (${\sim}10^2-10^3 \mathrm{M}_\odot$) from a remnant Pop III star and the massive seed (${\sim} 10^3-10^5 \mathrm{M}_\odot$) from the direct collapse of a gas cloud at early times \citep{Greene2012}. The massive seed mass is frequently used to initialize massive haloes ($\gtrsim 10^6-10^{12} \mathrm{M}_\odot$) in hydrodynamic simulations \citep{Springel2005,Vogelsberger2014,Schaye2014,Feng2016} while the ${\sim}10^3 \mathrm{M}_\odot$ seeds are also often adopted in semi-analytic models \citep{Somerville2008,Bonoli2009}. We note that this seed mass assumption only affects the black hole mass at early times and in the least massive galaxies. The main conclusions of this work are not significantly affected by this assumption. For example, with exactly the same adopted parameters (see Table \ref{tab:parameters}) but 10 times larger seed mass, the properties such as the black hole mass function, the Magorrian relation and the UV luminosity function are changed by less than 5 per cent in massive galaxies ($M_* > 10^{9}\mathrm{M}_\odot$). On the other hand, the model predicts a significant number of less massive black holes with masses ${\sim}10^5 \mathrm{M}_\odot$, which is more than ${\sim}1$ order of magnitude larger than the \citet{Shankar2009} sample. However, this has negligible impact on the total instantaneous emissivity and consequently, the reionization history does not change significantly. Fig. \ref{fig:dot_Nion_seed4} presents the evolution of emissivity, neutral hydrogen fraction and optical depth for the fiducial and QuasarReion models with larger black hole seed masses of $10^4\ h^{{-}1}\mathrm{M}_\odot$. Compared to the original models, we see that the quasar emissivity increases with a larger seed mass while the stellar emissivity decreases due to the stronger feedback from black holes. However, the changes are negligible, resulting in a small perturbation to the reionization history and optical depth.

\section{Conclusions}\label{sec:conclusions}
We have updated the {\sc Meraxes} semi-analytic model (\citetalias{Mutch2016}) with a detailed prescription of black hole evolution as part of the Dark-ages Reionization And Galaxy formation Observables from Numerical Simulations (DRAGONS) project to study the role of AGN in reionization and galaxy formation at high redshift. The model is calibrated against the observed stellar mass function ($z\sim7-0.6$), black hole mass function ($z\lesssim0.5$), quasar luminosity function ($z\sim6-0.6$), ionizing emissivity ($z\sim5-2$) and the Thomson scattering optical depth. The model is in agreement with the observed Magorrian relation at low redshift ($z<0.5$) and predicts a decreasing black hole mass towards higher redshifts at a fixed stellar mass. An opening angle of 80 deg, which corresponds to an un-obscured fraction of ${\sim}23.4$ per cent, allows the model to reproduce the observed quasar luminosity function across a large redshift range ($z\sim6-0.6$). 

Our model suggests that the radiation observed from recently discovered faint AGN at high redshift \citetalias{Giallongo2015} may include a significant fraction of UV flux from stars. Previous direct estimates of quasar contributions to reionization based on these observations \citep{Madau2015,Mitra2015} therefore result in an overestimate of the emissivity of quasars by a factor of 3 at $z\sim6$.

When we include the contribution of AGN to reionization, we find that quasars do not dominate the ionizing photon budget at $z>6$. In a quasar-only reionization model, where the escape fractions of ionizing photons are 1 and 0 for quasars and stars, respectively, we find that reionization happens very late, $z{\sim}3$, with a Thomson scattering optical depth of only half of the \citetalias{PlanckCollaboration2016} measurement (\citetalias{PlanckCollaboration2016}). However, at low redshift, quasars are able to provide a large number of ionizing photons. With quasars contributing all of their ionizing photons ($f_\mathrm{esc,q}=1$), our model prefers a redshift-dependent escape fraction for stars, having the form of $f_\mathrm{esc,*}\left(z\right)=\min\left[0.06\times\left(\dfrac{1+z}{6}\right)^{0.5}, 1\right]$. This corresponds to quasars contributing 10 per cent of the total ionizing photons at $z\sim6$.

\section*{Acknowledgements}
We would like to thank the anonymous referees for providing helpful suggestions that improves the paper substantially. This research was supported by the Victorian Life Sciences Computation Initiative (VLSCI), grant no. UOM0005, on its Peak Computing Facility hosted at the University of Melbourne, an initiative of the Victorian Government, Australia. Part of this work was performed on the gSTAR national facility at Swinburne University of Technology. gSTAR is funded by Swinburne and the Australian Governments Education Investment Fund. This research programme is funded by the Australian Research Council through the ARC Laureate Fellowship FL110100072 awarded to JSBW. This work was supported by the Flagship Allocation Scheme of the NCI National Facility at the ANU, generous allocations of time through the iVEC Partner Share and Australian Supercomputer Time Allocation Committee. AM acknowledges support from the European Research Council (ERC) under the European Union's Horizon 2020 research and innovation programme (Grant no. 638809 -- AIDA).
\bibliographystyle{\dir mn2e}
\bibliography{reference}

\appendix
\section{calculating the mean number of ionizing photons produced per black hole}\label{app:Ngamma}

\begin{figure}
	\includegraphics[width=\columnwidth]{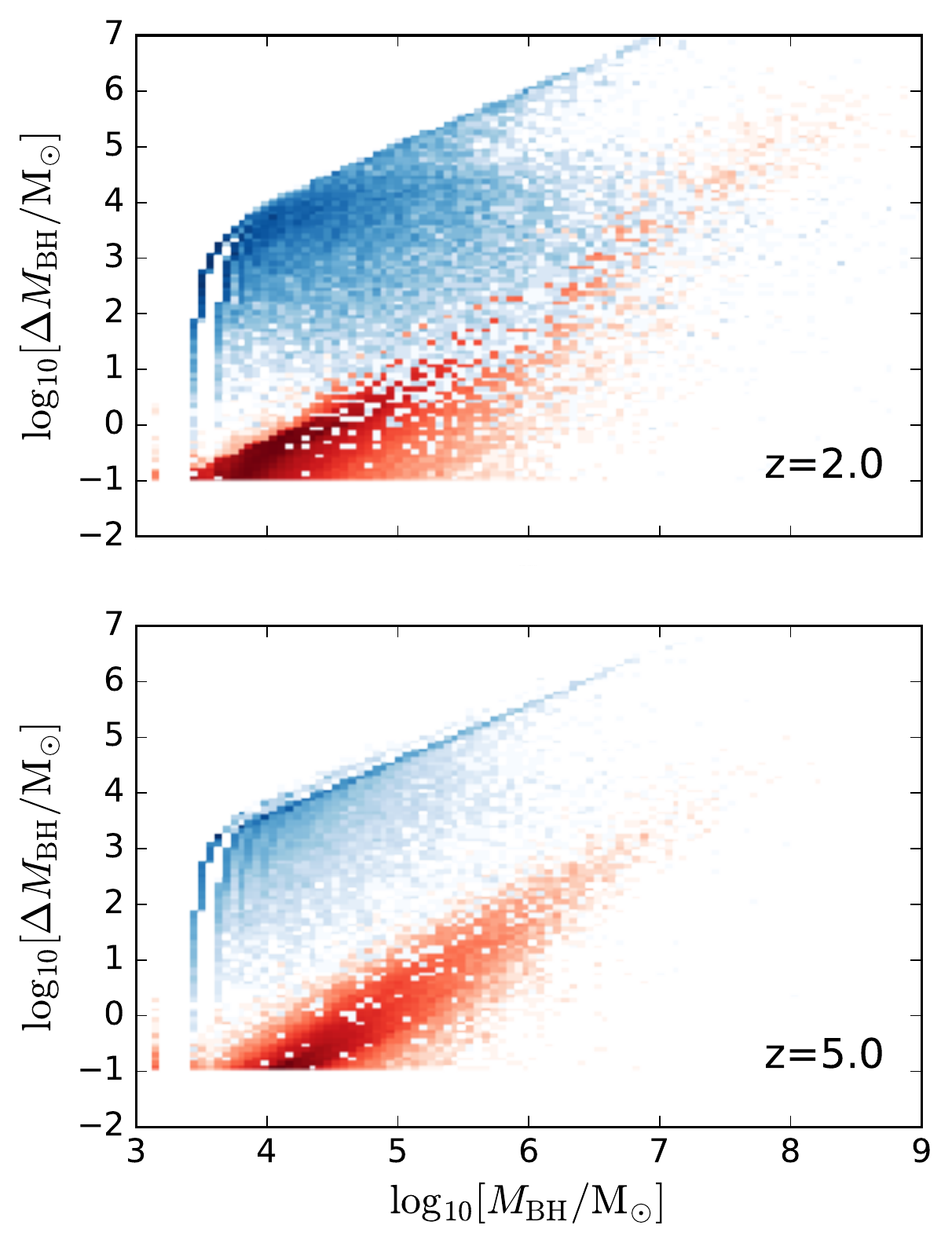}
	\caption{\label{app:fig:accretemass} Accretion mass $\Delta M_{\mathrm{BH}}$ versus black hole mass $M_{\mathrm{BH}}$ at $z{=}2$ and 5 in the \textit{Tiamat} result. The accretion masses from the radio and quasar modes are indicated red and blue, respectively.}
\end{figure}

During one time step, for a black hole with a given initial mass of $M_\mathrm{BH}$, its bolometric luminosity at Eddington rate can be calculated through the right hand of equation (\ref{eq:eddington_limit}). Since the bolometric correction \citep{Hopkins2007} adopted in this work is dependent on the bolometric luminosity (see equation \ref{eq:k_b}), the UV magnitude, $M_{1450}$ of the quasar changes during its accretion, so does the emissivity. In our model, because the accretion mass is always smaller than the black hole mass ($\Delta {M}_\mathrm{BH}<M_\mathrm{BH}$, see Fig. \ref{app:fig:accretemass}), for the sake of calculation speed, we estimate the mean number of ionizing photons produced per black hole, $N_{\gamma,q}$ with the bolometric luminosity at the beginning of accretion. We calculate $N_{\gamma,q}$ as follows:
\begin{enumerate}
	\item We calculate the UV magnitude, $M_{1450}$ using equations (\ref{eq:luminosity})-({\ref{eq:app:M1450}}).
	\item We calculate the UV flux with $M_{1450}$ in units of $\mathrm{erg\ s^{-1} Hz^{-1}}$ through
	\begin{equation}
	F_{1450} = 10^{\left(M_{1450}-48.6\right)/-2.5}\times4\pi\left(\dfrac{10\mathrm{pc}}{1\mathrm{cm}}\right)^2.
	\end{equation}
	\item We calculate the flux at Lyman limit following \citetalias{Giallongo2015}
	\begin{equation}
	F_{912} = F_{1450}\left(\dfrac{1200}{1450}\right)^{\alpha_{q,\mathrm{optical}}}\left(\dfrac{912}{1200}\right)^{\alpha_\mathrm{q}},
	\end{equation}
	where $\alpha_{q,\mathrm{optical}} = 0.44$ and $\alpha_\mathrm{q} = 1.57$ correspond to a double power-law AGN SED.
	\item We calculate the instantaneous emissivity by
	\begin{equation}
	\dot{N}_\mathrm{ion} \equiv \int_{\nu_{912}}^{\infty} 	F_{912}\left(\dfrac{\nu}{\nu_{912}}\right)^{-\alpha_\mathrm{q}} \dfrac{d\nu}{h\nu} = \dfrac{F_{912}}{h\alpha_\mathrm{q}}.
	\end{equation}
	\item The duration of accreting mass, $\Delta M_\mathrm{BH}$ can be calculated through 
	\begin{equation}
	t_\mathrm{acc} = \ln\left(\dfrac{\Delta {M}_\mathrm{BH}}{M_\mathrm{BH}} +1\right)\times\dfrac{\eta t_\mathrm{Edd}}{\epsilon}.
	\end{equation}
	Therefore, the total number of ionizing photons emitted is $\dot{N}_\mathrm{ion}t_\mathrm{acc}$ and the mean number of ionizing photons produced per black hole is
	\begin{equation}\label{eq:N_gamma}
	N_{\gamma,q} = \dfrac{\int_{0}^{t_\mathrm{acc}}\dot{N}_\mathrm{ion}\left(t\right)\mathrm{d}t}{\left(1-\eta\right)\Delta {M}_\mathrm{BH}/m_\mathrm{p}} \approx \dfrac{\dot{N}_\mathrm{ion}|_{t=t_\mathrm{acc}/2}t_\mathrm{acc}}{\left(1-\eta\right)\Delta {M}_\mathrm{BH}/m_\mathrm{p}},
	\end{equation}
	where the last step adopts the instantaneous emissivity at the middle of accretion for the sake of computational speed.
\end{enumerate}

We note that during the accretion, with an exponential increase of black hole mass, the AGN bolometric luminosity, $L_\mathrm{bol}$ increases exponentially (equation \ref{eq:luminosity}). Since the \textit{B}-band bolometric correction, $k_\mathrm{B}$, decreases with increasing luminosity following a double power law (equation \ref{eq:k_b}), $\dot{N}_\mathrm{ion}$ is a convex function of time. Therefore, the approximation in equation (\ref{eq:N_gamma}) underestimates the number of ionizing photons produced by black hole. In order to test whether this has a significant impact on our conclusion, we rerun the QuasarReion model assuming a constant bolometric correction with $k_\mathrm{B}\left(t\right) \approx k_\mathrm{B}|_{t=t_\mathrm{acc}}$. Eliminating the complex dependence of time from $k_\mathrm{B}$, $N_{\gamma,q}$ can be analytically calculated by integrating the AGN light curve. However, we note that since $k_\mathrm{B}\left(t\right) \leq k_\mathrm{B}|_{t=t_\mathrm{acc}}$, this approximation overestimates $N_{\gamma,q}$.

\begin{figure}
	\includegraphics[width=\columnwidth]{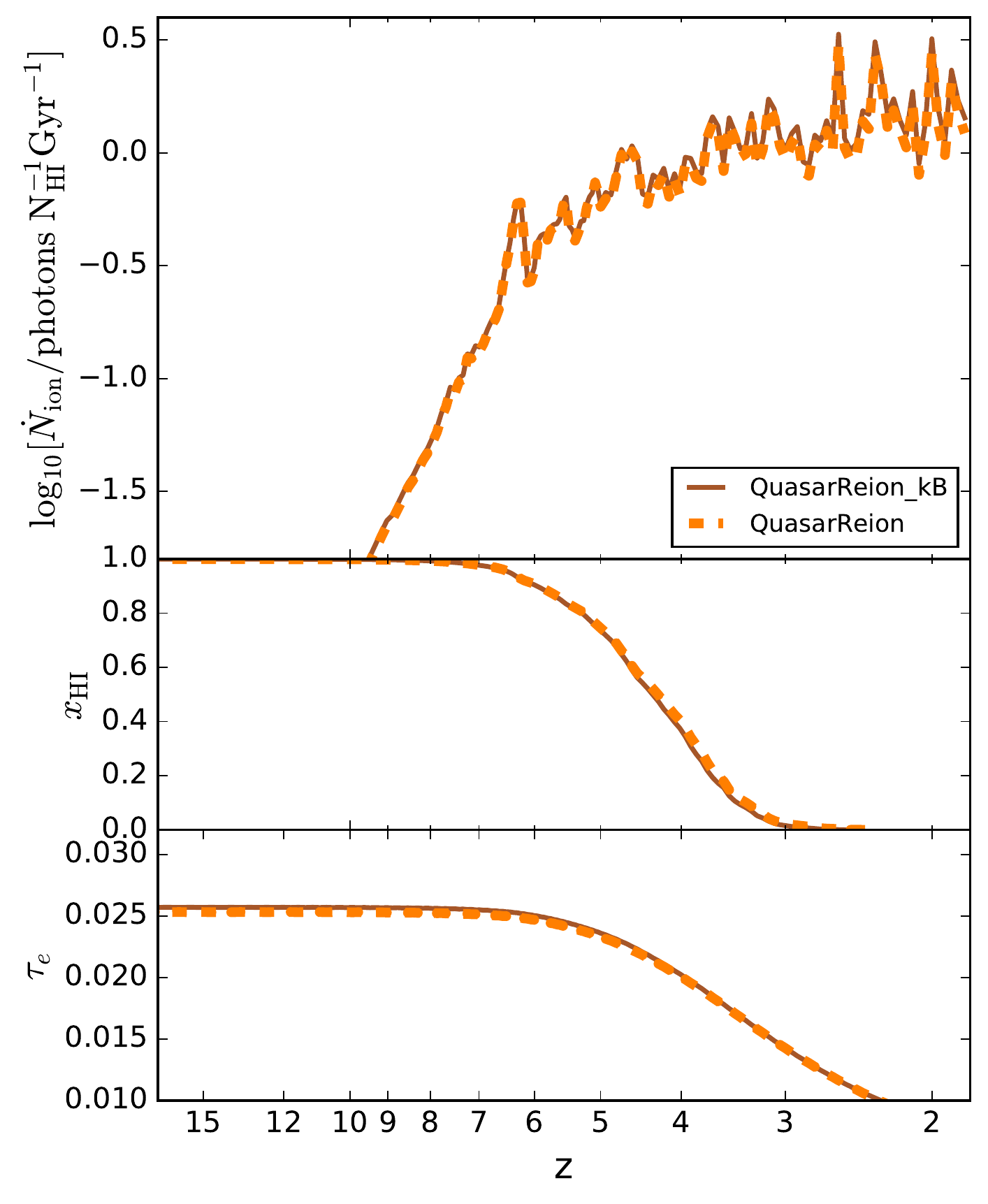}
	\caption{\label{app:fig:Lbolend} {\textit{Top panel:}} the evolution of ionizing emissivity for models assuming constant $\dot{N}_\mathrm{ion}$ (QuasarReion, {\color{colorQuasarReion}{\hdashrule[0.6mm]{6mm}{2.5pt}{1.5mm 0.5mm}}}) and $k_\mathrm{B}$ (QuasarReion\_kB, {\color{colorQuasarReionnodust}{\hdashrule[0.6mm]{6mm}{1.pt}{}}}). {\textit{Middle panel:}} the evolution of the mass-weighted global neutral hydrogen fraction. {\textit{Bottom panel:}} the Thomson scattering optical depth as a function of redshift.}
\end{figure}

Fig. \ref{app:fig:Lbolend} presents the evolution of emissivity, neutral hydrogen fraction and optical depth for different QuasarReion models assuming constant $\dot{N}_\mathrm{ion}$ (QuasarReion) and $k_\mathrm{B}$ (QuasarReion\_kB), respectively. Since the time interval between two snapshots is much smaller than the Eddington accretion time-scale ($t_\mathrm{Edd}\sim450\mathrm{Myr}$), the black hole mass increment is still within the linear regime and therefore we see that the impact from the calculation of $N_{\gamma,q}$ is not significant.

\label{lastpage}
\end{document}